\documentclass[aps, prl,10pt,twocolumn,superscriptaddress,nofootinbib]{revtex4-1}
\usepackage{mathrsfs, amssymb, amsmath}  
\usepackage{epsfig, cancel}
\usepackage{latexsym}
\usepackage{natbib, comment}
\usepackage{url}
\usepackage{dcolumn}
\usepackage{multirow}
\usepackage{color}
\usepackage{cancel}
\usepackage{soul}
\usepackage[normalem]{ulem}
\usepackage{amsfonts,amssymb,amsmath, txfonts}
\usepackage{graphicx,epsfig}
\usepackage{psfrag}
\usepackage{hyperref}
\hypersetup{colorlinks=true}
\usepackage{mathtools}
\usepackage{enumitem}
\usepackage{float}
\usepackage[dvipsnames]{xcolor}
\usepackage{xcolor}
\hypersetup{ linktoc=all,
    colorlinks, linkcolor={brightpink},
    citecolor={blue}, urlcolor={blue}
}
\definecolor{rosy}{RGB}{230,235,252}
\definecolor{myframetitle}{RGB}{90,89,170}
\definecolor{myblocktitle}{RGB}{140,185,249}
\definecolor{mytitle}{RGB}{10,80,26}

\definecolor{darkgreen}{RGB}{27,130,45}
\definecolor{darkblue}{rgb}{0,0,0.3}
\definecolor{darkred}{rgb}{0.7,0,0}

\definecolor{light gray}{RGB}{220,220,220}
\definecolor{dark purple}{RGB}{108,0,217}
\definecolor{pink}{RGB}{190,20,100}
\definecolor{orang}{RGB}{193,63,0}
\definecolor{green}{RGB}{11,98,17}
\definecolor{darkpink}{RGB}{153,0,76}
\definecolor{bluegreen}{RGB}{0,102,102}
\definecolor{greenlagan}{RGB}{0,102,0}
\definecolor{redgreen}{RGB}{102,102,0}
\definecolor{Redgreen}{RGB}{153,76,0}
\definecolor{vividviolet}{rgb}{0.62, 0.0, 1.0}
\definecolor{amaranth}{rgb}{0.9, 0.17, 0.31}
\definecolor{palatinateblue}{rgb}{0.15, 0.23, 0.89}
\definecolor{brightpink}{rgb}{1.0, 0.0, 0.5}
\definecolor{cornflowerblue}{rgb}{0.39, 0.58, 0.93}
\definecolor{deepcarminepink}{rgb}{0.94, 0.19, 0.22}
\definecolor{radicalred}{rgb}{1.0, 0.21, 0.37}

\def\H0{{\text{H}\hspace*{-2.05mm}\text{H} 0\hspace*{-1.35mm}0\ }}

\def\be{\begin{equation}}
\def\ee{\end{equation}}
\def\beq{\begin{equation}}
\def\eeq{\end{equation}}
\def\bea{\begin{eqnarray}}
\def\eea{\end{eqnarray}}

\begin{document}

\title{Further support for \texorpdfstring{$S_8$}{S8} increasing with effective redshift}

\author{\"{O}zg\"ur Akarsu}
\affiliation{Department of Physics, Istanbul Technical University, Maslak 34469 Istanbul, Turkey}
\author{Eoin \'O Colg\'ain}
\affiliation{Atlantic Technological University, Ash Lane, Sligo, Ireland}
\author{Anjan A. Sen}
\affiliation{Centre for Theoretical Physics, Jamia Millia Islamia, New Delhi - 110025, India}
\author{M. M. Sheikh-Jabbari} 
\affiliation{School of Physics, Institute for Research in Fundamental Sciences (IPM), P.O.Box 19395-5531, Tehran, Iran}

\begin{abstract}
In Adil et al.~(2023)~\cite{Adil:2023jtu}, we reported an increasing trend in $S_8$ with effective redshift $z_{\textrm{eff}}$ based on $f \sigma_8(z)$ constraints over the redshift range $0 \lesssim z \lesssim 2$, and predicted that this trend would be observable in independent datasets. Recently, the studies by Artis et al.~\cite{Artis:2024zag} and the ACT/DESI collaboration~\cite{ACT:2024nrz} appeared, presenting data that aligns with the expected trends. In this letter, we quantify the statistical significance of the increasing $S_8$ trends in~\cite{Artis:2024zag,ACT:2024nrz} by fitting a linear model to estimate the slope $\Delta\,S_8/\Delta\, z_{\textrm{eff}}$, and comparing the results to mock simulations. We find probabilities of $p = 0.0163$ and $p = 0.0185$, corresponding to approximately $2.1\sigma$ for each dataset. Using Fisher’s method to combine the independent probabilities, we obtain $p = 0.0027$ ($2.8\sigma$). When we incorporate our earlier findings~\cite{Adil:2023jtu}, the combined statistical significance reaches between $3\sigma$ and $3.7\sigma$. Finally, we discuss how ``scatter" in $\sigma_8$/$S_8$ constraints from recent DESI full-shape galaxy clustering fits this picture at low statistical significance. This letter continues a series of studies initiated in 2020 that explore redshift-dependent $\Lambda$CDM parameters as an indication of a breakdown in the standard cosmological model.

\end{abstract}

\maketitle

\section{Introduction}
In science, and particularly in physics, dynamical models break down when their parameters acquire time dependence when fitted to observational data. One may, of course, persist in studying the model, but if the fitting parameters vary across epochs, there are no meaningful predictions one can make. In cosmology, redshift $z$ is a proxy for time. In the era of cosmological tensions~\cite{DiValentino:2021izs, Perivolaropoulos:2021jda, Abdalla:2022yfr}, it is imperative to study $\Lambda$CDM cosmological parameters to identify the redshift ranges where physics is missing~\cite{Akarsu:2024qiq}. This is a key prerequisite to informed model building.

Naturally, such tests begin with $H_0$, a universal (integration) constant in any Friedmann-Lema\^itre-Robertson-Walker (FLRW) cosmology. The motivation is clear: the Hubble tension~\cite{Planck:2018vyg, Riess:2021jrx, Freedman:2021ahq, Pesce:2020xfe, Blakeslee:2021rqi, Kourkchi:2020iyz} suggests that $H_0$ may not be a constant under the assumption of the $\Lambda$CDM model. Corroborating support can be found in claims of descending trends in $H_0$ with effective redshift $z_{\rm eff}$ in the literature~\cite{Wong:2019kwg, Millon:2019slk, Krishnan:2020obg, Dainotti:2021pqg, Dainotti:2022bzg, Colgain:2022nlb, Colgain:2022rxy, Malekjani:2023dky, Hu:2022kes, Jia:2022ycc, Vagnozzi:2023nrq}. Within $\Lambda$CDM, the matter density today ($z=0$), $\Omega_m$, is the next most relevant parameter in the late Universe (at low redshifts). Noting that the late Universe Hubble parameter $H(z)$ at $z \gtrsim 1$ is essentially governed by the combination $H_0^2 \Omega_m$, one expects an increasing trend in $\Omega_m$ with effective redshift. This complementary signal has been reported in various forms~\cite{Colgain:2022nlb, Colgain:2022rxy, Malekjani:2023dky, Colgain:2024ksa, Risaliti:2018reu, Lusso:2020pdb, Yang:2019vgk, Khadka:2020vlh, Khadka:2020tlm, Khadka:2021xcc, Pourojaghi:2022zrh, Pasten:2023rpc}.\footnote{Alternatively, if $\Omega_m$ is a genuine constant in line with the prevailing assumption, it is important to calibrate it model independently \cite{Pedrotti:2024kpn}.}

Besides the background parameters $H_0, \Omega_m$, one may explore parameters describing perturbations, most notably the weighted amplitude of matter fluctuations $S_8 := \sigma_8 \sqrt{\Omega_m/0.3}$,  where a milder tension has persisted~\cite{Planck:2018vyg, Heymans:2013fya, Joudaki:2016mvz, DES:2017qwj, HSC:2018mrq, KiDS:2020suj, DES:2021wwk}. Here too, if the model is breaking down, one does not expect $S_8$ to remain constant, especially since it explicitly depends on $\Omega_m$.  In~\cite{Adil:2023jtu}, based on redshift-binned $f \sigma_8(z)$ constraints, it was demonstrated that $S_8$ increases with effective redshift. It is worth noting that~\cite{White:2021yvw} reports an earlier increase of $\sigma_8$ with redshift for fixed $\Omega_m$ (see Fig. 11 of~\cite{White:2021yvw})\footnote{Results are typically presented as $\sigma_8(z)$ or $S_8(z)$ versus $z$ \cite{White:2021yvw, Garcia-Garcia:2021unp, DES:2022ccp, SPT:2024qbr}.} albeit with large uncertainties, in a cross-correlation of luminous red galaxies with CMB lensing, but attributes the effect to a statistical fluctuation. The purpose of this note is to reinforce the claim by demonstrating the trend in recent independent studies of SRG/eROSITA number counts of massive halos~\cite{Artis:2024zag} and cross-correlations of CMB lensing with DESI Legacy imaging galaxies~\cite{ACT:2024nrz}, elucidating the combined statistical significance of these trends.  Systematics aside, all observables point to a breakdown of the $\Lambda$CDM model \textit{in the linear regime}. We note in passing that~\cite{Esposito:2022plo, Tutusaus:2023aux, Manna:2024wak, Sailer:2024coh} also indicate an increasing   $S_8$ or $\sigma_8$ parameter with effective redshift, though we do not incorporate these trends.\footnote{The anomaly in the late-time integrated Sachs-Wolfe effect in supervoids~\cite{DES:2018nlb, Kovacs:2021mnf} also suggests weaker growth at lower and stronger growth at higher redshifts, thereby interpolating Planck expectations at intermediate redshifts. This weaker growth may also be interpreted in terms of the $\gamma$CDM model, e.g.,~\cite{Nguyen:2023fip}, but as remarked in \cite{Artis:2024zag} with which we agree, the $\gamma$CDM model \textit{lacks an immediate physical interpretation}. Retracing its origins to \cite{Wang:1998gt}, it is, at best, a simple and effective approximation.}

\section{Analysis}

Our methodology is straightforward and follows~\cite{Wong:2019kwg}. We fit a line to the observed $S_8$ constraints and their effective redshifts $z_{\textrm{eff}}$ to extract a slope, $m=\Delta\,S_8/\Delta\, z_{\textrm{eff}}$. We then reset the $S_8$ constraints to an average value and generate new data points randomly in normal distributions, using the $S_8$ error as the standard deviation for each new data point. We then construct $10^5$ mock datasets and count the number of mocks where a larger slope is found by chance, converting this into a corresponding $p$-value. Note that the average value is unimportant here, as we are only concerned with the slope; however, where possible, we respect the combined values reported in the literature.

From Table~\ref{tab:artis}, which is reproduced from Table~2 of~\cite{Artis:2024zag}, a study based on the SRG/eROSITA catalogues~\cite{Bulbul:2024mfj, Kluge:2024ghp}, one should choose a  $z_{\textrm{eff}}$. In the absence of further guidance, we adopt the mean redshift in each redshift bin as $z_{\textrm{eff}}$. Furthermore, for the average $S_8$ value, we adopt~$S_8 = 0.86$ \cite{Ghirardini:2024yni} from the first column of Table~2 in~\cite{Artis:2024zag}. Note that the choice made here is irrelevant for determining the slope. Fitting the observed data in Table~\ref{tab:artis}, with $z_{\textrm{eff}} \in \{ 0.138, 0.210, 0.287, 0.391, 0.626\}$, results in a slope of $m = 0.18^{+0.08}_{-0.09}$ from Markov Chain Monte Carlo (MCMC) analysis, which is approximately $2\sigma$ away from a horizontal line with slope $m=0$, corresponding to a constant, non-evolving $S_8$.

\begin{table}[t]
\centering
\begin{tabular}{|c|c|}
\hline
\rule{0pt}{3ex} \textbf{Redshift Range ($z$)} & \textbf{$S_8$} \\
\hline\hline
\rule{0pt}{3ex} $0.100 \, < \, z \, < \, 0.175$  & $0.83 \pm 0.02$ \\
\hline
\rule{0pt}{3ex} $0.175 \, < \, z \, < \, 0.244$  & $0.85 \pm 0.07$ \\
\hline
\rule{0pt}{3ex} $0.244 \, < \, z \, < \, 0.330$  & $0.86 \pm 0.06$ \\
\hline
\rule{0pt}{3ex} $0.330 \, < \, z \, < \, 0.452$  & $0.84 \pm 0.03$ \\
\hline
\rule{0pt}{3ex} $0.452 \, < \, z \, < \, 0.800$  & $0.94 \pm 0.04$ \\
\hline
\end{tabular}
\caption{Redshift ranges and $S_8$ constraints from Table 2 of \cite{Artis:2024zag}.}
\label{tab:artis}
\end{table}

\begin{figure}[ht]
\includegraphics[width=75mm]{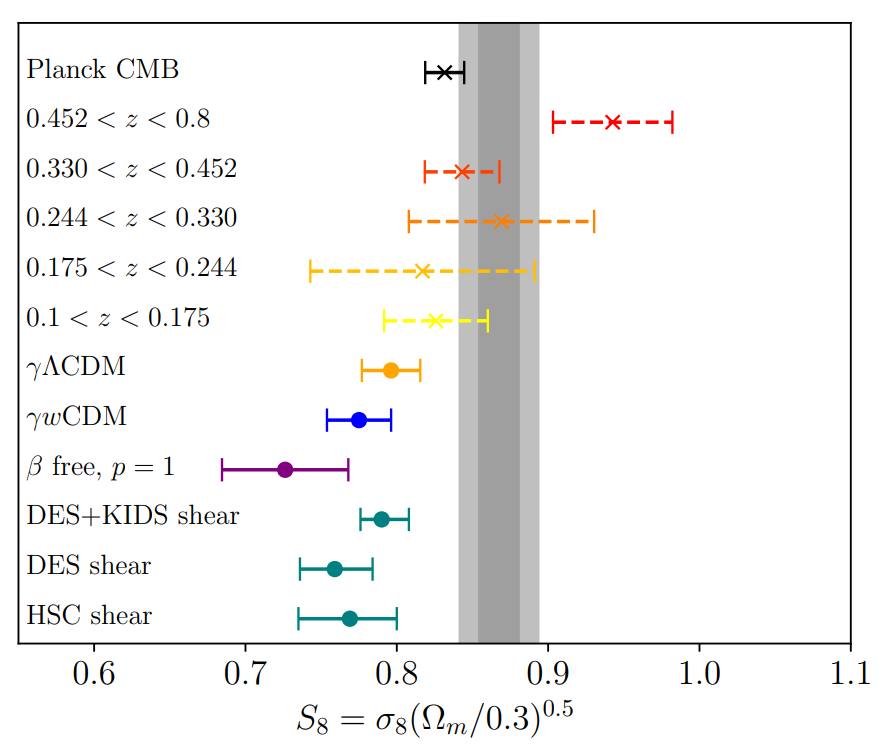} 
\caption{Reproduced from Fig. 11 of \cite{Artis:2024zag}. The constraints of interest are dashed and the grey bands denote the combined $1 \sigma$ and $2 \sigma$ constraint of all the binned data.}
\label{fig:artis}
\end{figure}

It is evident from Fig.~\ref{fig:artis}, that this trend is essentially driven by the highest redshift bin. We have confirmed that removing this bin reduces the slope to $m = 0.05 \pm 0.14$. The resulting slope is consistent with no trend. It is worth bearing in mind that the first four bins collectively span a redshift range similar to that of the highest redshift bin, so it may be unsurprising that little evolution is observed across the narrower bins when the highest bin is excluded. Nevertheless, excluding the highest redshift bin effectively eliminates the trend, restoring consistency with the Planck-$\Lambda$CDM prediction for $S_8$~\cite{Planck:2018vyg}.

As explained, to better assess the statistical significance for all bins, we set the $S_8$ constraints to the average value $S_8 = 0.86$, generate  $10^5$ mock realizations, and fit a slope to each one. We find a probability of $p = 0.0163$---or more precisely, 1627 out of $10^5$ realizations---that a larger slope could arise by chance. Translated into a one-sided Gaussian, this corresponds to a statistical significance of approximately $2.1 \sigma$, which agrees well with the MCMC analysis of the observed data.

\begin{figure*}[htb]
\includegraphics[width=140mm]{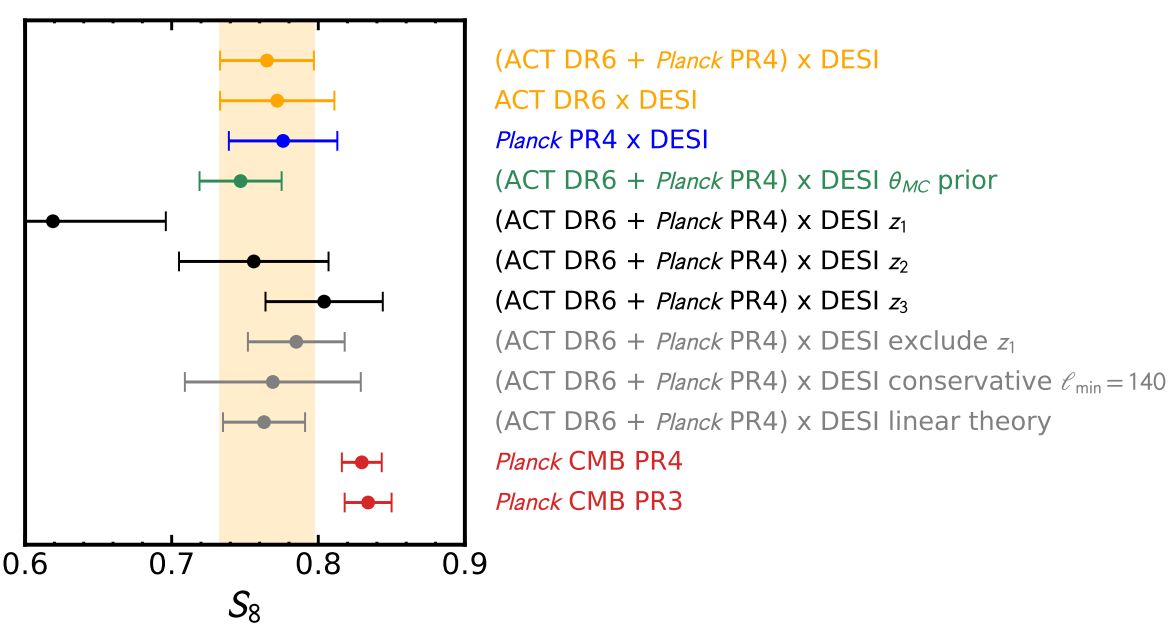} 
\caption{Reproduced from Fig. 13 of \cite{ACT:2024nrz}. The constraints of interest are in black. Orange band denotes the combined $1 \sigma$ constraint.}
\label{fig:ACT_DESI}
\end{figure*}

We now turn to the ACT DR6~\cite{ACT:2023dou, ACT:2023kun}, Planck PR4~\cite{Carron:2022eyg}, DESI~\cite{DESI:2018ymu} constraints in Table~\ref{tab:ACT_DESI}, reproduced from Table 4 of \cite{ACT:2024nrz}. The effective redshifts $z_{\textrm{eff}} \in \{0.37,\, 0.51,\, 0.65 \}$ can be found in Table~1 of~\cite{ACT:2024nrz}. The average or ``joint" constraint is $S_8 = 0.765$. The increasing trend of $S_8$ with $z_{\textrm{eff}}$ is evident in both Table~\ref{tab:ACT_DESI} and Fig.~\ref{fig:ACT_DESI}. Note that, with reference to Fig.~\ref{fig:artis}, despite the studies disagreeing on whether the results are closer to Planck at lower or higher redshifts---suggesting potential systematics in the observables---the same increasing trend of $S_8$ with $z_{\textrm{eff}}$ remains evident. For completeness, we isolate the constraints from Table~4 of \cite{ACT:2024nrz} that correspond to Fig.~\ref{fig:ACT_DESI}. One could extend this analysis to include BAO data, but doing so would not further develop the point being made. Repeating the same steps as above, we find a slope of $m = 0.59^{+0.29}_{-0.28}$ from MCMC analysis, which suggests a discrepancy of approximately $\sim 2.1 \sigma$ from a horizontal line. Once again, we compare to mock simulations and find that the probability of obtaining a larger slope is $p = 0.0185$ ($1851$ from $10^5$ realisations), corresponding to $\sim 2.1 \sigma$ for a one-sided Gaussian, thus confirming the MCMC analysis. Given that the probabilities are independent, we can combine them using Fisher's method to obtain $p = 0.0027$. This confirms the increasing trend of $S_8$ with $z_{\textrm{eff}}$ across two independent data sets, with a statistical significance of approximately $2.8\sigma$.

\begin{table}[htb]
\centering 
\begin{tabular}{|c|c|c|}
\hline
\rule{0pt}{3ex} \textbf{Redshift ($z$)} & \textbf{$\Omega_m$} & \textbf{$S_8$} \\
\hline\hline
\rule{0pt}{3ex} $0.37$ & $0.237 \pm 0.067$ & $0.619 \pm 0.077$ \\
\hline
\rule{0pt}{3ex} $0.51$ & $0.247 \pm 0.060$ & $0.756 \pm 0.051$ \\
\hline
\rule{0pt}{3ex} $0.65$ & $0.304 \pm 0.080$ & $0.807 \pm 0.039$ \\
\hline
\end{tabular}
\caption{Effective redshifts, $\Omega_m$, and $S_8$ constraints reproduced from Table 1 and Table 4 of \cite{ACT:2024nrz}. }
\label{tab:ACT_DESI}
\end{table}

We now return to the results of \cite{Adil:2023jtu}, where an increasing trend in $S_8$ was reported for both a small $f \sigma_8(z)$ dataset and a larger archival $f \sigma_8(z)$ dataset, with statistical significances of $\sim 1.6 \sigma$ and $\sim 2.8 \sigma$, respectively. We note that the second dataset comprises overlapping data constraints, so it should be interpreted with caution. An additional caveat is that these results assume $\Omega_m$ is constant, constrained to $\Omega_m = 0.3111 \pm 0.0056$ based on CMB and BAO data~\cite{Planck:2018vyg, BOSS:2016wmc, Hou:2020rse, Neveux:2020voa}. Given the higher-quality data from recent studies~\cite{Artis:2024zag, ACT:2024nrz}, the trend can be observed without additional priors or external input, though the large fractional errors in $f \sigma_8(z)$ constraints necessitate more stringent assumptions.\footnote{The additional assumption of a constant $\Omega_m$ is relatively mild; however, it should be noted that $\Omega_m$ may now vary between datasets \cite{DESI:2024mwx} (Fig.~3).} Nevertheless, as the data is independent of~\cite{Artis:2024zag, ACT:2024nrz}, we can calculate a combined $p$-value. This yields a $p$-value ranging from $p = 0.0012$ ($\sim 3 \sigma$) to $p = 0.00009$ ($\sim 3.7 \sigma$). As explained above, we are not performing the same test across datasets, so these results should be interpreted with caution; nevertheless, we can confidently place the increasing $S_8$ trend with $z_{\textrm{eff}}$ at approximately $3\sigma$ based on three independent datasets.

Before concluding, we discuss another interesting point. In Table~\ref{tab:ACT_DESI}, it is evident that  $\Omega_m$ also increases with effective redshift, echoing independent claims in the literature that the $\Lambda$CDM fitting parameter $\Omega_m$ increases with redshift~\cite{Colgain:2022nlb, Colgain:2022rxy, Malekjani:2023dky, Colgain:2024ksa}. Repeating the steps above with $S_8$ replaced by $\Omega_m$ and using the joint result $\Omega_m = 0.271$, we find a slope of $m = 0.23^{+0.37}_{-0.36}$, which is naively consistent with $m = 0$ at $\sim 0.6 \sigma$. Moreover, from the simulations, we find a probability of $p = 0.268$, for obtaining a larger slope, which corresponds to a statistical significance of only $\sim 0.6 \sigma$ for a one-sided Gaussian. The key takeaway from this exercise is that, although an increasing $\Omega_m$ trend is evident in the central values in Table~\ref{tab:ACT_DESI}, the associated errors are much larger than those for $S_8$. Ultimately, this is unsurprising, as it is well documented in the literature that the $S_8$ parameter is better constrained than $\sigma_8$ or $\Omega_m$, thereby justifying the focus on $S_8$.

\section{DESI Full-Shape Galaxy Clustering}
DESI recently released cosmological constraints from the full-shape modelling of galaxy clusters \cite{DESI:2024hhd, DESI:2024jis}. While the statistical significance of any signal for a $\Lambda$CDM deviation is low without external data, we explain how results are consistent with our claims. Given the overlap in DESI data with ACT/DESI observations \cite{ACT:2024nrz}, one cannot combine $p$-values using Fisher's method.

DESI analysed a bright galaxy sample (BGS) with redshift $0.1 < z < 0.4$, three luminous red galaxy (LRG) samples with redshifts $0.4 < z < 0.6$, $0.6 < z < 0.8$ and $0.8 < z < 1.1$, respectively, an emission line galaxy (ELG) sample in the range $1.1 < z < 1.6$ and a quasar (QSO) sample with $0.8 < z < 2.1$. In Table \ref{tab:full_shape} we reproduce the effective redshift $z_{\textrm{eff}}$ from Table 1 \cite{DESI:2024jis} and the $\Omega_m$ and $\sigma_8$ constraints from Table 10  \cite{DESI:2024jis} for each tracer. We focus on full-modelling without BAO. As is evident from Table \ref{tab:full_shape}, the low redshift BGS $\Omega_m$ and $\sigma_8$ values are lower than the corresponding high redshift QSO values by $\sim 0.8 \sigma$ and $\sim 2.2 \sigma$, respectively. In contrast, the tracers probing the redshifts in between exhibit scatter about intermediate values with no definite trend.

\begin{table}[htpb!]
\centering 
\begin{tabular}{|c|c|c|c|c|}
\hline
\rule{0pt}{3ex} \textbf{Tracer} & \textbf{Redshift ($z$)} & \textbf{$\Omega_m$} & \textbf{$\sigma_8$} & \textbf{$S_8$} \\
\hline\hline
\rule{0pt}{3ex} BGS & $0.295$ & $0.272 \pm 0.027$ & $0.664^{+0.098}_{-0.14}$ & $0.63^{+0.12}_{-0.16}$ \\
\hline
\rule{0pt}{3ex} LRG1 & $0.510$ & $0.297 \pm 0.021$ & $0.844^{+0.083}_{-0.095}$ & $0.84^{+0.11}_{-0.12}$ \\
\hline
\rule{0pt}{3ex} LRG2 & $0.706$ & $0.280 \pm 0.020$ & $0.888^{+0.071}_{-0.084}$ & $0.858^{+0.096}_{-0.11}$ \\
\hline
\rule{0pt}{3ex} LRG3 & $0.930$ & $0.294 \pm 0.025$ & $0.810^{+0.069}_{-0.078}$ & $0.802^{+0.099}_{-0.11}$ \\
\hline
\rule{0pt}{3ex} ELG2 & $1.317$ & $0.297^{+0.028}_{-0.036}$ & $0.749^{+0.055}_{-0.064}$ & $0.745^{+0.087}_{-0.10}$ \\
\hline
\rule{0pt}{3ex} QSO & $1.491$ & $0.310^{+0.030}_{-0.041}$ & $0.950^{+0.070}_{-0.084}$ & $0.97^{+0.11}_{-0.14}$ \\
\hline
\end{tabular}
\caption{Effective redshifts, $\Omega_m$, and $\sigma_8$ constraints reproduced from Table 1 and Table 10 of \cite{DESI:2024jis}. We estimate the $S_8$ constraints as outlined in the text.}
\label{tab:full_shape}
\end{table}

\begin{figure*}[t]
\includegraphics[width=80mm]{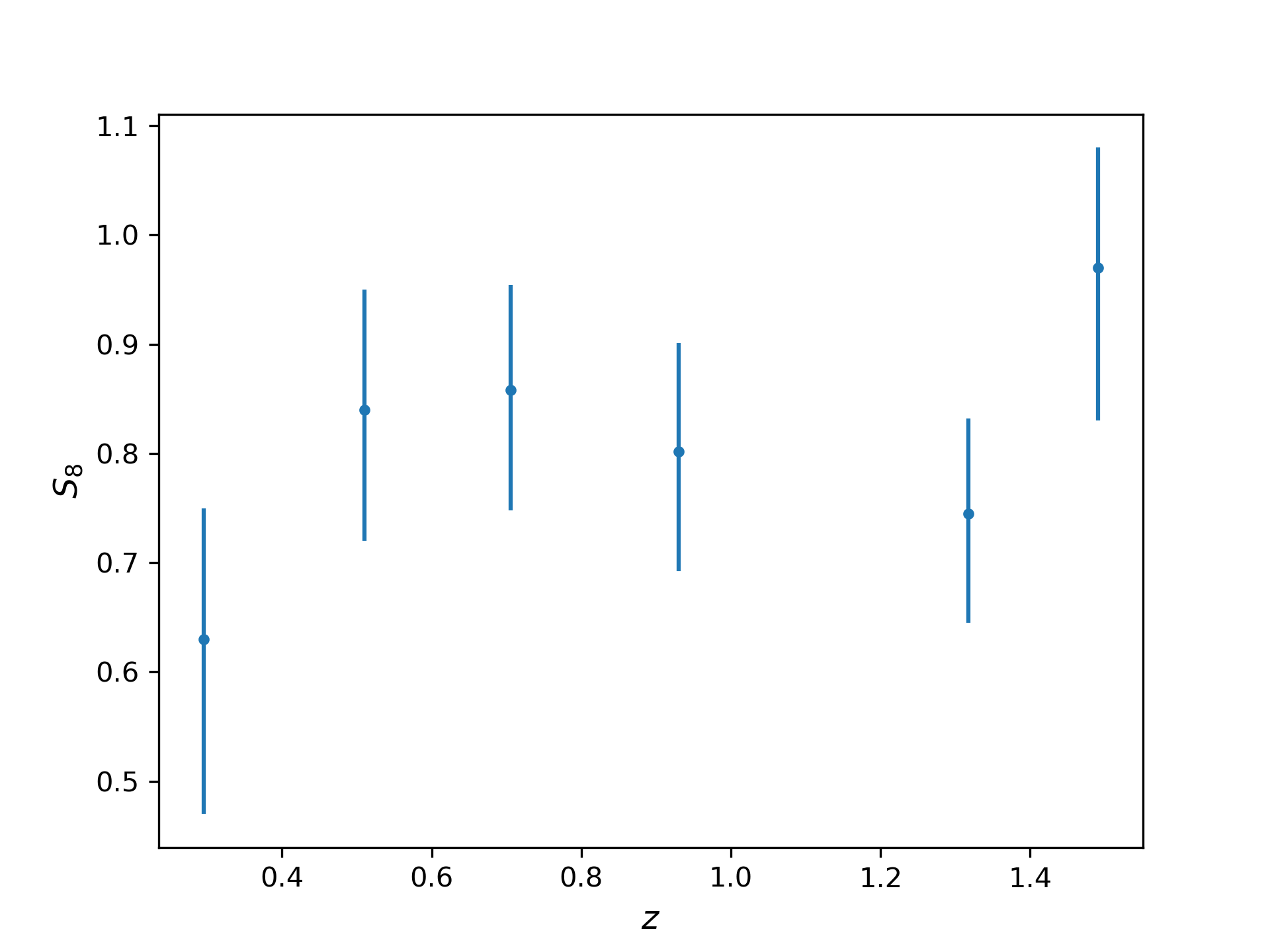} 
\caption{$S_8$ versus effective redshift for DESI full-shape modelling of galaxy clusters.}
\label{fig:full_shape}
\end{figure*}

That being said, it is plausible that the $S_8 = \sigma_8 \sqrt{\Omega_m/0.3}$ values, which are omitted from \cite{DESI:2024jis}, show a more definite trend, because $\Omega_m$ and $\sigma_8$ from the low redshift BGS sample are evidently both lower than the corresponding constraints from the QSO sample. See \cite{Colgain:2024xqj} for speculation of an increasing trend of $\Omega_m$ with $z_{\textrm{eff}}$ in DESI BAO constraints \cite{DESI:2024mwx}. Here we will estimate the $S_8$ constraints. To do so, we approximate the $\Omega_m$ and $\sigma_8$ constraints as antisymmetric Gaussian distributions. More precisely, we assume that the distributions for values larger and smaller than the median ( $50^{\textrm{th}}$ percentile) are Gaussian with standard deviations dictated by the upper and lower errors, respectively. In the case of symmetric errors, the distribution reduces to a standard Gaussian. For each ``half" of the Gaussian, we randomly generate $2 \times 10^5$ odd configurations in a normal distribution, which we restrict to exactly $1.5 \times 10^5$ configurations before combining to get $3 \times 10^5$ configurations in the full distribution. By construction, this leads to $16^{\textrm{th}}, 50^{\textrm{th}}$ and $84^{\textrm{th}}$ percentiles that recover the central values and the errors up to small numbers for both $\Omega_m$ and $\sigma_8$. Given that our distribution for $\Omega_m$ and $\sigma_8$ has the same number of configurations, we combine them to get an $S_8$ distribution and extract the $16^{\textrm{th}}, 50^{\textrm{th}}$ and $84^{\textrm{th}}$ to record them in Table \ref{tab:full_shape}, thus providing us an estimate of the $S_8$ constraints from each tracer. The $S_8$ values appear against $z_{\textrm{eff}}$ in Fig. \ref{fig:full_shape}.

As is evident from Fig. \ref{fig:full_shape}, if there is an increasing trend of $S_8$ with $z_{\textrm{eff}}$, it is not significant. At this juncture, we assume the errors are Gaussian by adopting the largest error and repeat the earlier analysis. From MCMC analysis, we find a slope of $m = 0.07^{+0.13}_{-0.12}$ for all tracers and a slope of $m=0.50^{+0.46}_{-0.45}$ for the first three data points in Fig. \ref{fig:full_shape}, where one should note that these constraints overlap in redshift range with the SRG/eROSITA and ACT/DESI observations. This motivates restricting the redshift range. Regardless, the statistical significance is low at $\sim 0.6 \sigma$ to $\sim 1.1 \sigma$ in DESI full-shape results \cite{DESI:2024hhd, DESI:2024jis}. We further remark that the BGS and QSO $S_8$ values now differ at $\sim 1.8 \sigma$. In switching from $\sigma_8$ to $S_8$, the errors inflate, but this effect is countered by some movement in the $\Omega_m$ central values.\footnote{Under similar assumptions, replacing $S_8$ with $\sigma_8$ marginally decreases the statistical significance of an increasing trend with $z_{\textrm{eff}}$ $\sim 0.6 \sigma \rightarrow \sim 0.5 \sigma$ for all redshifts and marginally increases the statistical significance of the trend $\sim 1.1 \sigma \rightarrow \sim 1.3 \sigma$ if one restricts to the lowest three effective redshifts.}

\section{Discussion}
When confronted with $\Lambda$CDM tensions---in essence, the $H_0$~\cite{Planck:2018vyg, Riess:2021jrx, Freedman:2021ahq, Pesce:2020xfe, Blakeslee:2021rqi, Kourkchi:2020iyz} and $S_8$ tensions~\cite{Planck:2018vyg, Heymans:2013fya, Joudaki:2016mvz, DES:2017qwj, HSC:2018mrq, KiDS:2020suj, DES:2021wwk}---it is natural to first consider potential systematics. When systematics fail to resolve the issue, it is natural to look for minimal resolutions as the next step. Minimal resolutions start with sample variance~\cite{Wu:2017fpr}, though suggestions for new recombination physics \cite{Knox:2019rjx} and adjustments in baryonic physics~\cite{Amon:2022azi, Preston:2023uup} can also be viewed as minimal concessions. The key point is that all of these proposals are designed to retain as much of the $\Lambda$CDM framework as possible. However, none of these resolutions may be considered natural from a fundamental physics perspective.

What \textit{is} natural---and what one can confidently assert---is that the $\Lambda$CDM model represents a measure-zero subset within the space of cosmological models, even when restricted to FLRW. For this reason, it is far more likely that we are simply using an incorrect $H(z)$ function and that Nature prefers a different one. If this is the case, one would expect the $\Lambda$CDM parameters to exhibit redshift dependence, marking a breakdown of the model~\cite{Krishnan:2020obg}. $\Lambda$CDM is a phenomenological model with limited theoretical backing, and in light of persistent tensions, it is imperative to conduct consistency checks across different subsectors of the model~\cite{Akarsu:2024qiq}. Model breakdown is a very natural outcome. As discussed in~\cite{Akarsu:2024qiq}, it is essential to first ``localize'' the problem by epoch/redshift ---made possible by specifying the evolution of parameters---before attempting to resolve the tensions or propose alternative models. Within this context, the point of this letter is that~\cite{Artis:2024zag, ACT:2024nrz} now independently show the same increasing $S_8$ trend with effective redshift that was originally highlighted in \cite{Adil:2023jtu}. 

Our work departs from reconstructions of $S_8(z)/\sigma_8(z)$ \cite{White:2021yvw, Garcia-Garcia:2021unp, DES:2022ccp, SPT:2024qbr} in two key ways. First, while $S_8(z)/\sigma_8(z)$ reconstructions can show deviations from Planck-$\Lambda$CDM behaviour, and identify redshift ranges where physics is missing, one still needs to separate missing scale-dependent physics from missing redshift-dependent physics. Here, this is less of a concern as ACT/DESI cross-correlations, SRG/eROSITA observations and $f \sigma_8(z)$ constraints from redshift space distortions typically only probe linear scales. Secondly, $S_{8}(z)/\sigma_8(z)$ reconstructions fail to map the problem back into constant $\Lambda$CDM fitting parameters showing that the $\Lambda$CDM model has broken down. With $S_8(z)/\sigma_8(z)$ reconstructions one has the weaker statement that the $\Lambda$CDM model is not tracking the data, not the stronger statement that the $\Lambda$CDM model fails to make unique predictions.

This is not to suggest that systematics are unimportant. As highlighted in the text, the fact that the Planck $S_8$ constraint aligns more closely with~\cite{Artis:2024zag} at lower redshifts but with~\cite{ACT:2024nrz} at higher redshifts suggests that these observables are not tracking the same $S_8$. This points to the presence of systematics or unknown physics in these observables that is not fully understood. It is also worth noting that the late-time integrated Sachs-Wolfe (ISW) effect anomaly hints at stronger growth than predicted by Planck-$\Lambda$CDM at higher redshifts~\cite{DES:2018nlb, Kovacs:2021mnf}, suggesting that it may be premature to conclude that growth is only suppressed at lower redshifts ($z\lesssim1$). That being said, each observable has its own quirks; nevertheless, they must ultimately align on the underlying physics. Redshift-dependent $\Lambda$CDM fitting parameters are a signature of missing physics, provided that the same trend is observed across different observables and datasets. The studies \cite{Adil:2023jtu, Artis:2024zag, ACT:2024nrz} independently show an increasing trend of $S_8$ with $z_{\textrm{eff}}$ that is difficult to overlook.  As we have argued, the combined significance of this trend is approximately $3\sigma$.

An analogy may help to illustrate this perspective. Part of what motivated Einstein to rewrite Newtonian gravity was the contradiction between the finite speed of light and Newtonian gravity’s instantaneous action at a distance. Naturally, one could perturb around Newton's theory to construct a phenomenologically viable theory of gravity, but this fails to address fundamental theoretical contradictions. In a similar vein, the $\Lambda$CDM model introduces the long-standing cosmological constant problem~\cite{Weinberg:1988cp,Weinberg:2000yb,Peebles:2002gy}, and the sign of the cosmological constant conflicts with generic string theory frameworks~\cite{Dvali:2014gua, Dvali:2018fqu, Obied:2018sgi}. Minimal resolutions to $\Lambda$CDM tensions, such as sample variance~\cite{Wu:2017fpr}, new recombination physics~\cite{Knox:2019rjx}, or baryonic feedback~\cite{Amon:2022azi, Preston:2023uup}, do little to address these theoretical issues. Nevertheless, from a theoretical perspective, it is entirely natural for dynamical models to reach their limits when fitting parameters begin to vary with time (or, in this case, redshift) in light of observational data. Our claims, which stretch back to~\cite{Krishnan:2020obg}\footnote{This research line was inspired by~\cite{Wong:2019kwg, Millon:2019slk}, though TDCOSMO's current research program is primarily focused on determining $H_0$ to high precision within the $\Lambda$CDM framework. Over time, TDCOSMO’s findings may either recover a unique $H_0$ value or suggest that $H_0$ cannot be uniquely determined from cosmological scales.}, are best appreciated in such a light.

\section*{Acknowledgements}
We thank Emmanuel Artis, Ido Ben-Dayan, Eleonora Di Valentino, Sunny Vagnozzi and Martin White for correspondence on related topics. \"{O}A acknowledges the support by the Turkish Academy of Sciences in the scheme of the Outstanding Young Scientist Award (T\"{U}BA-GEB\.{I}P). \"{O}A is supported in part by TUBITAK grant 122F124.   MMShJ is supported in part by  Iran National Science Foundation (INSF) Grant No.  4026712 and in part by the ICTP through the senior Associates Programme (2023-2028).  AAS acknowledges the funding from SERB, Govt of India under the research grant no: CRG/2023/003984. This article/publication is based upon work from COST Action CA21136 – “Addressing observational tensions in cosmology with systematics and fundamental physics (CosmoVerse)”, supported by COST (European Cooperation in Science and Technology).

\bibliography{refs}

\begin{thebibliography}{76}%
\makeatletter
\providecommand \@ifxundefined [1]{%
 \@ifx{#1\undefined}
}%
\providecommand \@ifnum [1]{%
 \ifnum #1\expandafter \@firstoftwo
 \else \expandafter \@secondoftwo
 \fi
}%
\providecommand \@ifx [1]{%
 \ifx #1\expandafter \@firstoftwo
 \else \expandafter \@secondoftwo
 \fi
}%
\providecommand \natexlab [1]{#1}%
\providecommand \enquote  [1]{``#1''}%
\providecommand \bibnamefont  [1]{#1}%
\providecommand \bibfnamefont [1]{#1}%
\providecommand \citenamefont [1]{#1}%
\providecommand \href@noop [0]{\@secondoftwo}%
\providecommand \href [0]{\begingroup \@sanitize@url \@href}%
\providecommand \@href[1]{\@@startlink{#1}\@@href}%
\providecommand \@@href[1]{\endgroup#1\@@endlink}%
\providecommand \@sanitize@url [0]{\catcode `\\12\catcode `\$12\catcode
  `\&12\catcode `\#12\catcode `\^12\catcode `\_12\catcode `\%12\relax}%
\providecommand \@@startlink[1]{}%
\providecommand \@@endlink[0]{}%
\providecommand \url  [0]{\begingroup\@sanitize@url \@url }%
\providecommand \@url [1]{\endgroup\@href {#1}{\urlprefix }}%
\providecommand \urlprefix  [0]{URL }%
\providecommand \Eprint [0]{\href }%
\providecommand \doibase [0]{http://dx.doi.org/}%
\providecommand \selectlanguage [0]{\@gobble}%
\providecommand \bibinfo  [0]{\@secondoftwo}%
\providecommand \bibfield  [0]{\@secondoftwo}%
\providecommand \translation [1]{[#1]}%
\providecommand \BibitemOpen [0]{}%
\providecommand \bibitemStop [0]{}%
\providecommand \bibitemNoStop [0]{.\EOS\space}%
\providecommand \EOS [0]{\spacefactor3000\relax}%
\providecommand \BibitemShut  [1]{\csname bibitem#1\endcsname}%
\let\auto@bib@innerbib\@empty
\bibitem [{\citenamefont {Adil}\ \emph {et~al.}(2023)\citenamefont {Adil},
  \citenamefont {Akarsu}, \citenamefont {Malekjani}, \citenamefont
  {\'O~Colg\'ain}, \citenamefont {Pourojaghi}, \citenamefont {Sen},\ and\
  \citenamefont {Sheikh-Jabbari}}]{Adil:2023jtu}%
  \BibitemOpen
  \bibfield  {author} {\bibinfo {author} {\bibfnamefont {S.~A.}\ \bibnamefont
  {Adil}}, \bibinfo {author} {\bibfnamefont {O.}~\bibnamefont {Akarsu}},
  \bibinfo {author} {\bibfnamefont {M.}~\bibnamefont {Malekjani}}, \bibinfo
  {author} {\bibfnamefont {E.}~\bibnamefont {\'O~Colg\'ain}}, \bibinfo {author}
  {\bibfnamefont {S.}~\bibnamefont {Pourojaghi}}, \bibinfo {author}
  {\bibfnamefont {A.~A.}\ \bibnamefont {Sen}}, \ and\ \bibinfo {author}
  {\bibfnamefont {M.~M.}\ \bibnamefont {Sheikh-Jabbari}},\ }\href {\doibase
  10.1093/mnrasl/slad165} {\bibfield  {journal} {\bibinfo  {journal} {Mon. Not.
  Roy. Astron. Soc.}\ }\textbf {\bibinfo {volume} {528}},\ \bibinfo {pages}
  {L20} (\bibinfo {year} {2023})},\ \Eprint {http://arxiv.org/abs/2303.06928}
  {arXiv:2303.06928 [astro-ph.CO]} \BibitemShut {NoStop}%
\bibitem [{\citenamefont {{Artis}}\ \emph {et~al.}(2024)\citenamefont
  {{Artis}}, \citenamefont {{Bulbul}}, \citenamefont {{Grandis}}, \citenamefont
  {{Ghirardini}}, \citenamefont {{Clerc}}, \citenamefont {{Seppi}},
  \citenamefont {{Comparat}}, \citenamefont {{Cataneo}}, \citenamefont {{von
  der Linden}}, \citenamefont {{Bahar}}, \citenamefont {{Balzer}},
  \citenamefont {{Chiu}}, \citenamefont {{Gruen}}, \citenamefont
  {{Kleinebreil}}, \citenamefont {{Kluge}}, \citenamefont {{Krippendorf}},
  \citenamefont {{Li}}, \citenamefont {{Liu}}, \citenamefont {{Malavasi}},
  \citenamefont {{Merloni}}, \citenamefont {{Miyatake}}, \citenamefont
  {{Miyazaki}}, \citenamefont {{Nandra}}, \citenamefont {{Okabe}},
  \citenamefont {{Pacaud}}, \citenamefont {{Predehl}}, \citenamefont
  {{Ramos-Ceja}}, \citenamefont {{Reiprich}}, \citenamefont {{Sanders}},
  \citenamefont {{Schrabback}}, \citenamefont {{Zelmer}},\ and\ \citenamefont
  {{Zhang}}}]{Artis:2024zag}%
  \BibitemOpen
  \bibfield  {author} {\bibinfo {author} {\bibfnamefont {E.}~\bibnamefont
  {{Artis}}}, \bibinfo {author} {\bibfnamefont {E.}~\bibnamefont {{Bulbul}}},
  \bibinfo {author} {\bibfnamefont {S.}~\bibnamefont {{Grandis}}}, \bibinfo
  {author} {\bibfnamefont {V.}~\bibnamefont {{Ghirardini}}}, \bibinfo {author}
  {\bibfnamefont {N.}~\bibnamefont {{Clerc}}}, \bibinfo {author} {\bibfnamefont
  {R.}~\bibnamefont {{Seppi}}}, \bibinfo {author} {\bibfnamefont
  {J.}~\bibnamefont {{Comparat}}}, \bibinfo {author} {\bibfnamefont
  {M.}~\bibnamefont {{Cataneo}}}, \bibinfo {author} {\bibfnamefont
  {A.}~\bibnamefont {{von der Linden}}}, \bibinfo {author} {\bibfnamefont
  {Y.~E.}\ \bibnamefont {{Bahar}}}, \bibinfo {author} {\bibfnamefont
  {F.}~\bibnamefont {{Balzer}}}, \bibinfo {author} {\bibfnamefont
  {I.}~\bibnamefont {{Chiu}}}, \bibinfo {author} {\bibfnamefont
  {D.}~\bibnamefont {{Gruen}}}, \bibinfo {author} {\bibfnamefont
  {F.}~\bibnamefont {{Kleinebreil}}}, \bibinfo {author} {\bibfnamefont
  {M.}~\bibnamefont {{Kluge}}}, \bibinfo {author} {\bibfnamefont
  {S.}~\bibnamefont {{Krippendorf}}}, \bibinfo {author} {\bibfnamefont
  {X.}~\bibnamefont {{Li}}}, \bibinfo {author} {\bibfnamefont {A.}~\bibnamefont
  {{Liu}}}, \bibinfo {author} {\bibfnamefont {N.}~\bibnamefont {{Malavasi}}},
  \bibinfo {author} {\bibfnamefont {A.}~\bibnamefont {{Merloni}}}, \bibinfo
  {author} {\bibfnamefont {H.}~\bibnamefont {{Miyatake}}}, \bibinfo {author}
  {\bibfnamefont {S.}~\bibnamefont {{Miyazaki}}}, \bibinfo {author}
  {\bibfnamefont {K.}~\bibnamefont {{Nandra}}}, \bibinfo {author}
  {\bibfnamefont {N.}~\bibnamefont {{Okabe}}}, \bibinfo {author} {\bibfnamefont
  {F.}~\bibnamefont {{Pacaud}}}, \bibinfo {author} {\bibfnamefont
  {P.}~\bibnamefont {{Predehl}}}, \bibinfo {author} {\bibfnamefont {M.~E.}\
  \bibnamefont {{Ramos-Ceja}}}, \bibinfo {author} {\bibfnamefont {T.~H.}\
  \bibnamefont {{Reiprich}}}, \bibinfo {author} {\bibfnamefont {J.~S.}\
  \bibnamefont {{Sanders}}}, \bibinfo {author} {\bibfnamefont {T.}~\bibnamefont
  {{Schrabback}}}, \bibinfo {author} {\bibfnamefont {S.}~\bibnamefont
  {{Zelmer}}}, \ and\ \bibinfo {author} {\bibfnamefont {X.}~\bibnamefont
  {{Zhang}}},\ }\href {\doibase 10.48550/arXiv.2410.09499} {\bibfield
  {journal} {\bibinfo  {journal} {arXiv e-prints}\ ,\ \bibinfo {eid}
  {arXiv:2410.09499}} (\bibinfo {year} {2024})},\ \Eprint
  {http://arxiv.org/abs/2410.09499} {arXiv:2410.09499 [astro-ph.CO]}
  \BibitemShut {NoStop}%
\bibitem [{\citenamefont {{Qu}}\ \emph {et~al.}(2024)\citenamefont {{Qu}},
  \citenamefont {{Hang}}, \citenamefont {{Farren}}, \citenamefont {{Bolliet}},
  \citenamefont {{Aguilar}}, \citenamefont {{Ahlen}}, \citenamefont {{Alam}},
  \citenamefont {{Brooks}}, \citenamefont {{Cai}}, \citenamefont {{Calabrese}},
  \citenamefont {{Claybaugh}}, \citenamefont {{de la Macorra}}, \citenamefont
  {{Devlin}}, \citenamefont {{Doel}}, \citenamefont {{Embil-Villagra}},
  \citenamefont {{Ferraro}}, \citenamefont {{Font-Ribera}}, \citenamefont
  {{Forero-Romero}}, \citenamefont {{Gazta{\~n}aga}}, \citenamefont
  {{Gluscevic}}, \citenamefont {{Gontcho}}, \citenamefont {{Gutierrez}},
  \citenamefont {{Howlett}}, \citenamefont {{Kehoe}}, \citenamefont {{Kim}},
  \citenamefont {{Kremin}}, \citenamefont {{Lambert}}, \citenamefont
  {{Landriau}}, \citenamefont {{Le Guillou}}, \citenamefont {{Levi}},
  \citenamefont {{Louis}}, \citenamefont {{Meisner}}, \citenamefont {{Miquel}},
  \citenamefont {{Moustakas}}, \citenamefont {{Newman}}, \citenamefont {{Niz}},
  \citenamefont {{Peacock}}, \citenamefont {{Percival}}, \citenamefont
  {{Poppett}}, \citenamefont {{Prada}}, \citenamefont {{P{\'e}rez-R{\`a}fols}},
  \citenamefont {{Rossi}}, \citenamefont {{Sanchez}}, \citenamefont
  {{Schlegel}}, \citenamefont {{Sehgal}}, \citenamefont {{Shaikh}},
  \citenamefont {{Sherwin}}, \citenamefont {{Sif{\'o}n}}, \citenamefont
  {{Schubnell}}, \citenamefont {{Sprayberry}}, \citenamefont {{Tarl{\'e}}},
  \citenamefont {{Weaver}}, \citenamefont {{Wollack}},\ and\ \citenamefont
  {{Zou}}}]{ACT:2024nrz}%
  \BibitemOpen
  \bibfield  {author} {\bibinfo {author} {\bibfnamefont {F.~J.}\ \bibnamefont
  {{Qu}}}, \bibinfo {author} {\bibfnamefont {Q.}~\bibnamefont {{Hang}}},
  \bibinfo {author} {\bibfnamefont {G.}~\bibnamefont {{Farren}}}, \bibinfo
  {author} {\bibfnamefont {B.}~\bibnamefont {{Bolliet}}}, \bibinfo {author}
  {\bibfnamefont {J.~N.}\ \bibnamefont {{Aguilar}}}, \bibinfo {author}
  {\bibfnamefont {S.}~\bibnamefont {{Ahlen}}}, \bibinfo {author} {\bibfnamefont
  {S.}~\bibnamefont {{Alam}}}, \bibinfo {author} {\bibfnamefont
  {D.}~\bibnamefont {{Brooks}}}, \bibinfo {author} {\bibfnamefont {Y.-C.}\
  \bibnamefont {{Cai}}}, \bibinfo {author} {\bibfnamefont {E.}~\bibnamefont
  {{Calabrese}}}, \bibinfo {author} {\bibfnamefont {T.}~\bibnamefont
  {{Claybaugh}}}, \bibinfo {author} {\bibfnamefont {A.}~\bibnamefont {{de la
  Macorra}}}, \bibinfo {author} {\bibfnamefont {M.~J.}\ \bibnamefont
  {{Devlin}}}, \bibinfo {author} {\bibfnamefont {P.}~\bibnamefont {{Doel}}},
  \bibinfo {author} {\bibfnamefont {C.}~\bibnamefont {{Embil-Villagra}}},
  \bibinfo {author} {\bibfnamefont {S.}~\bibnamefont {{Ferraro}}}, \bibinfo
  {author} {\bibfnamefont {A.}~\bibnamefont {{Font-Ribera}}}, \bibinfo {author}
  {\bibfnamefont {J.~E.}\ \bibnamefont {{Forero-Romero}}}, \bibinfo {author}
  {\bibfnamefont {E.}~\bibnamefont {{Gazta{\~n}aga}}}, \bibinfo {author}
  {\bibfnamefont {V.}~\bibnamefont {{Gluscevic}}}, \bibinfo {author}
  {\bibfnamefont {S.~G.~A.}\ \bibnamefont {{Gontcho}}}, \bibinfo {author}
  {\bibfnamefont {G.}~\bibnamefont {{Gutierrez}}}, \bibinfo {author}
  {\bibfnamefont {C.}~\bibnamefont {{Howlett}}}, \bibinfo {author}
  {\bibfnamefont {R.}~\bibnamefont {{Kehoe}}}, \bibinfo {author} {\bibfnamefont
  {J.}~\bibnamefont {{Kim}}}, \bibinfo {author} {\bibfnamefont
  {A.}~\bibnamefont {{Kremin}}}, \bibinfo {author} {\bibfnamefont
  {A.}~\bibnamefont {{Lambert}}}, \bibinfo {author} {\bibfnamefont
  {M.}~\bibnamefont {{Landriau}}}, \bibinfo {author} {\bibfnamefont
  {L.}~\bibnamefont {{Le Guillou}}}, \bibinfo {author} {\bibfnamefont
  {M.}~\bibnamefont {{Levi}}}, \bibinfo {author} {\bibfnamefont
  {T.}~\bibnamefont {{Louis}}}, \bibinfo {author} {\bibfnamefont
  {A.}~\bibnamefont {{Meisner}}}, \bibinfo {author} {\bibfnamefont
  {R.}~\bibnamefont {{Miquel}}}, \bibinfo {author} {\bibfnamefont
  {J.}~\bibnamefont {{Moustakas}}}, \bibinfo {author} {\bibfnamefont {J.~A.}\
  \bibnamefont {{Newman}}}, \bibinfo {author} {\bibfnamefont {G.}~\bibnamefont
  {{Niz}}}, \bibinfo {author} {\bibfnamefont {J.}~\bibnamefont {{Peacock}}},
  \bibinfo {author} {\bibfnamefont {W.}~\bibnamefont {{Percival}}}, \bibinfo
  {author} {\bibfnamefont {C.}~\bibnamefont {{Poppett}}}, \bibinfo {author}
  {\bibfnamefont {F.}~\bibnamefont {{Prada}}}, \bibinfo {author} {\bibfnamefont
  {I.}~\bibnamefont {{P{\'e}rez-R{\`a}fols}}}, \bibinfo {author} {\bibfnamefont
  {G.}~\bibnamefont {{Rossi}}}, \bibinfo {author} {\bibfnamefont
  {E.}~\bibnamefont {{Sanchez}}}, \bibinfo {author} {\bibfnamefont
  {D.}~\bibnamefont {{Schlegel}}}, \bibinfo {author} {\bibfnamefont
  {N.}~\bibnamefont {{Sehgal}}}, \bibinfo {author} {\bibfnamefont
  {S.}~\bibnamefont {{Shaikh}}}, \bibinfo {author} {\bibfnamefont
  {B.}~\bibnamefont {{Sherwin}}}, \bibinfo {author} {\bibfnamefont
  {C.}~\bibnamefont {{Sif{\'o}n}}}, \bibinfo {author} {\bibfnamefont
  {M.}~\bibnamefont {{Schubnell}}}, \bibinfo {author} {\bibfnamefont
  {D.}~\bibnamefont {{Sprayberry}}}, \bibinfo {author} {\bibfnamefont
  {G.}~\bibnamefont {{Tarl{\'e}}}}, \bibinfo {author} {\bibfnamefont {B.~A.}\
  \bibnamefont {{Weaver}}}, \bibinfo {author} {\bibfnamefont {E.~J.}\
  \bibnamefont {{Wollack}}}, \ and\ \bibinfo {author} {\bibfnamefont
  {H.}~\bibnamefont {{Zou}}},\ }\href {\doibase 10.48550/arXiv.2410.10808}
  {\bibfield  {journal} {\bibinfo  {journal} {arXiv e-prints}\ ,\ \bibinfo
  {eid} {arXiv:2410.10808}} (\bibinfo {year} {2024})},\ \Eprint
  {http://arxiv.org/abs/2410.10808} {arXiv:2410.10808 [astro-ph.CO]}
  \BibitemShut {NoStop}%
\bibitem [{\citenamefont {Di~Valentino}\ \emph {et~al.}(2021)\citenamefont
  {Di~Valentino}, \citenamefont {Mena}, \citenamefont {Pan}, \citenamefont
  {Visinelli}, \citenamefont {Yang}, \citenamefont {Melchiorri}, \citenamefont
  {Mota}, \citenamefont {Riess},\ and\ \citenamefont
  {Silk}}]{DiValentino:2021izs}%
  \BibitemOpen
  \bibfield  {author} {\bibinfo {author} {\bibfnamefont {E.}~\bibnamefont
  {Di~Valentino}}, \bibinfo {author} {\bibfnamefont {O.}~\bibnamefont {Mena}},
  \bibinfo {author} {\bibfnamefont {S.}~\bibnamefont {Pan}}, \bibinfo {author}
  {\bibfnamefont {L.}~\bibnamefont {Visinelli}}, \bibinfo {author}
  {\bibfnamefont {W.}~\bibnamefont {Yang}}, \bibinfo {author} {\bibfnamefont
  {A.}~\bibnamefont {Melchiorri}}, \bibinfo {author} {\bibfnamefont {D.~F.}\
  \bibnamefont {Mota}}, \bibinfo {author} {\bibfnamefont {A.~G.}\ \bibnamefont
  {Riess}}, \ and\ \bibinfo {author} {\bibfnamefont {J.}~\bibnamefont {Silk}},\
  }\href {\doibase 10.1088/1361-6382/ac086d} {\bibfield  {journal} {\bibinfo
  {journal} {Class. Quant. Grav.}\ }\textbf {\bibinfo {volume} {38}},\ \bibinfo
  {pages} {153001} (\bibinfo {year} {2021})},\ \Eprint
  {http://arxiv.org/abs/2103.01183} {arXiv:2103.01183 [astro-ph.CO]}
  \BibitemShut {NoStop}%
\bibitem [{\citenamefont {Perivolaropoulos}\ and\ \citenamefont
  {Skara}(2022)}]{Perivolaropoulos:2021jda}%
  \BibitemOpen
  \bibfield  {author} {\bibinfo {author} {\bibfnamefont {L.}~\bibnamefont
  {Perivolaropoulos}}\ and\ \bibinfo {author} {\bibfnamefont {F.}~\bibnamefont
  {Skara}},\ }\href {\doibase 10.1016/j.newar.2022.101659} {\bibfield
  {journal} {\bibinfo  {journal} {New Astron. Rev.}\ }\textbf {\bibinfo
  {volume} {95}},\ \bibinfo {pages} {101659} (\bibinfo {year} {2022})},\
  \Eprint {http://arxiv.org/abs/2105.05208} {arXiv:2105.05208 [astro-ph.CO]}
  \BibitemShut {NoStop}%
\bibitem [{\citenamefont {Abdalla}\ \emph {et~al.}(2022)\citenamefont {Abdalla}
  \emph {et~al.}}]{Abdalla:2022yfr}%
  \BibitemOpen
  \bibfield  {author} {\bibinfo {author} {\bibfnamefont {E.}~\bibnamefont
  {Abdalla}} \emph {et~al.},\ }\href {\doibase 10.1016/j.jheap.2022.04.002}
  {\bibfield  {journal} {\bibinfo  {journal} {JHEAp}\ }\textbf {\bibinfo
  {volume} {34}},\ \bibinfo {pages} {49} (\bibinfo {year} {2022})},\ \Eprint
  {http://arxiv.org/abs/2203.06142} {arXiv:2203.06142 [astro-ph.CO]}
  \BibitemShut {NoStop}%
\bibitem [{\citenamefont {Akarsu}\ \emph {et~al.}(2024)\citenamefont {Akarsu},
  \citenamefont {\'O~Colg\'ain}, \citenamefont {Sen},\ and\ \citenamefont
  {Sheikh-Jabbari}}]{Akarsu:2024qiq}%
  \BibitemOpen
  \bibfield  {author} {\bibinfo {author} {\bibfnamefont {O.}~\bibnamefont
  {Akarsu}}, \bibinfo {author} {\bibfnamefont {E.}~\bibnamefont
  {\'O~Colg\'ain}}, \bibinfo {author} {\bibfnamefont {A.~A.}\ \bibnamefont
  {Sen}}, \ and\ \bibinfo {author} {\bibfnamefont {M.~M.}\ \bibnamefont
  {Sheikh-Jabbari}},\ }\href {\doibase 10.3390/universe10080305} {\bibfield
  {journal} {\bibinfo  {journal} {Universe}\ }\textbf {\bibinfo {volume}
  {10}},\ \bibinfo {pages} {305} (\bibinfo {year} {2024})},\ \Eprint
  {http://arxiv.org/abs/2402.04767} {arXiv:2402.04767 [astro-ph.CO]}
  \BibitemShut {NoStop}%
\bibitem [{\citenamefont {Aghanim}\ \emph {et~al.}(2020)\citenamefont {Aghanim}
  \emph {et~al.}}]{Planck:2018vyg}%
  \BibitemOpen
  \bibfield  {author} {\bibinfo {author} {\bibfnamefont {N.}~\bibnamefont
  {Aghanim}} \emph {et~al.} (\bibinfo {collaboration} {Planck}),\ }\href
  {\doibase 10.1051/0004-6361/201833910} {\bibfield  {journal} {\bibinfo
  {journal} {Astron. Astrophys.}\ }\textbf {\bibinfo {volume} {641}},\ \bibinfo
  {pages} {A6} (\bibinfo {year} {2020})},\ \bibinfo {note} {[Erratum:
  Astron.Astrophys. 652, C4 (2021)]},\ \Eprint
  {http://arxiv.org/abs/1807.06209} {arXiv:1807.06209 [astro-ph.CO]}
  \BibitemShut {NoStop}%
\bibitem [{\citenamefont {Riess}\ \emph {et~al.}(2022)\citenamefont {Riess}
  \emph {et~al.}}]{Riess:2021jrx}%
  \BibitemOpen
  \bibfield  {author} {\bibinfo {author} {\bibfnamefont {A.~G.}\ \bibnamefont
  {Riess}} \emph {et~al.},\ }\href {\doibase 10.3847/2041-8213/ac5c5b}
  {\bibfield  {journal} {\bibinfo  {journal} {Astrophys. J. Lett.}\ }\textbf
  {\bibinfo {volume} {934}},\ \bibinfo {pages} {L7} (\bibinfo {year} {2022})},\
  \Eprint {http://arxiv.org/abs/2112.04510} {arXiv:2112.04510 [astro-ph.CO]}
  \BibitemShut {NoStop}%
\bibitem [{\citenamefont {Freedman}(2021)}]{Freedman:2021ahq}%
  \BibitemOpen
  \bibfield  {author} {\bibinfo {author} {\bibfnamefont {W.~L.}\ \bibnamefont
  {Freedman}},\ }\href {\doibase 10.3847/1538-4357/ac0e95} {\bibfield
  {journal} {\bibinfo  {journal} {Astrophys. J.}\ }\textbf {\bibinfo {volume}
  {919}},\ \bibinfo {pages} {16} (\bibinfo {year} {2021})},\ \Eprint
  {http://arxiv.org/abs/2106.15656} {arXiv:2106.15656 [astro-ph.CO]}
  \BibitemShut {NoStop}%
\bibitem [{\citenamefont {Pesce}\ \emph {et~al.}(2020)\citenamefont {Pesce}
  \emph {et~al.}}]{Pesce:2020xfe}%
  \BibitemOpen
  \bibfield  {author} {\bibinfo {author} {\bibfnamefont {D.~W.}\ \bibnamefont
  {Pesce}} \emph {et~al.},\ }\href {\doibase 10.3847/2041-8213/ab75f0}
  {\bibfield  {journal} {\bibinfo  {journal} {Astrophys. J. Lett.}\ }\textbf
  {\bibinfo {volume} {891}},\ \bibinfo {pages} {L1} (\bibinfo {year} {2020})},\
  \Eprint {http://arxiv.org/abs/2001.09213} {arXiv:2001.09213 [astro-ph.CO]}
  \BibitemShut {NoStop}%
\bibitem [{\citenamefont {Blakeslee}\ \emph {et~al.}(2021)\citenamefont
  {Blakeslee}, \citenamefont {Jensen}, \citenamefont {Ma}, \citenamefont
  {Milne},\ and\ \citenamefont {Greene}}]{Blakeslee:2021rqi}%
  \BibitemOpen
  \bibfield  {author} {\bibinfo {author} {\bibfnamefont {J.~P.}\ \bibnamefont
  {Blakeslee}}, \bibinfo {author} {\bibfnamefont {J.~B.}\ \bibnamefont
  {Jensen}}, \bibinfo {author} {\bibfnamefont {C.-P.}\ \bibnamefont {Ma}},
  \bibinfo {author} {\bibfnamefont {P.~A.}\ \bibnamefont {Milne}}, \ and\
  \bibinfo {author} {\bibfnamefont {J.~E.}\ \bibnamefont {Greene}},\ }\href
  {\doibase 10.3847/1538-4357/abe86a} {\bibfield  {journal} {\bibinfo
  {journal} {Astrophys. J.}\ }\textbf {\bibinfo {volume} {911}},\ \bibinfo
  {pages} {65} (\bibinfo {year} {2021})},\ \Eprint
  {http://arxiv.org/abs/2101.02221} {arXiv:2101.02221 [astro-ph.CO]}
  \BibitemShut {NoStop}%
\bibitem [{\citenamefont {Kourkchi}\ \emph {et~al.}(2020)\citenamefont
  {Kourkchi}, \citenamefont {Tully}, \citenamefont {Anand}, \citenamefont
  {Courtois}, \citenamefont {Dupuy}, \citenamefont {Neill}, \citenamefont
  {Rizzi},\ and\ \citenamefont {Seibert}}]{Kourkchi:2020iyz}%
  \BibitemOpen
  \bibfield  {author} {\bibinfo {author} {\bibfnamefont {E.}~\bibnamefont
  {Kourkchi}}, \bibinfo {author} {\bibfnamefont {R.~B.}\ \bibnamefont {Tully}},
  \bibinfo {author} {\bibfnamefont {G.~S.}\ \bibnamefont {Anand}}, \bibinfo
  {author} {\bibfnamefont {H.~M.}\ \bibnamefont {Courtois}}, \bibinfo {author}
  {\bibfnamefont {A.}~\bibnamefont {Dupuy}}, \bibinfo {author} {\bibfnamefont
  {J.~D.}\ \bibnamefont {Neill}}, \bibinfo {author} {\bibfnamefont
  {L.}~\bibnamefont {Rizzi}}, \ and\ \bibinfo {author} {\bibfnamefont
  {M.}~\bibnamefont {Seibert}},\ }\href {\doibase 10.3847/1538-4357/ab901c}
  {\bibfield  {journal} {\bibinfo  {journal} {Astrophys. J.}\ }\textbf
  {\bibinfo {volume} {896}},\ \bibinfo {pages} {3} (\bibinfo {year} {2020})},\
  \Eprint {http://arxiv.org/abs/2004.14499} {arXiv:2004.14499 [astro-ph.GA]}
  \BibitemShut {NoStop}%
\bibitem [{\citenamefont {Wong}\ \emph {et~al.}(2020)\citenamefont {Wong} \emph
  {et~al.}}]{Wong:2019kwg}%
  \BibitemOpen
  \bibfield  {author} {\bibinfo {author} {\bibfnamefont {K.~C.}\ \bibnamefont
  {Wong}} \emph {et~al.},\ }\href {\doibase 10.1093/mnras/stz3094} {\bibfield
  {journal} {\bibinfo  {journal} {Mon. Not. Roy. Astron. Soc.}\ }\textbf
  {\bibinfo {volume} {498}},\ \bibinfo {pages} {1420} (\bibinfo {year}
  {2020})},\ \Eprint {http://arxiv.org/abs/1907.04869} {arXiv:1907.04869
  [astro-ph.CO]} \BibitemShut {NoStop}%
\bibitem [{\citenamefont {Millon}\ \emph {et~al.}(2020)\citenamefont {Millon}
  \emph {et~al.}}]{Millon:2019slk}%
  \BibitemOpen
  \bibfield  {author} {\bibinfo {author} {\bibfnamefont {M.}~\bibnamefont
  {Millon}} \emph {et~al.},\ }\href {\doibase 10.1051/0004-6361/201937351}
  {\bibfield  {journal} {\bibinfo  {journal} {Astron. Astrophys.}\ }\textbf
  {\bibinfo {volume} {639}},\ \bibinfo {pages} {A101} (\bibinfo {year}
  {2020})},\ \Eprint {http://arxiv.org/abs/1912.08027} {arXiv:1912.08027
  [astro-ph.CO]} \BibitemShut {NoStop}%
\bibitem [{\citenamefont {Krishnan}\ \emph {et~al.}(2020)\citenamefont
  {Krishnan}, \citenamefont {\'O~Colg\'ain}, \citenamefont {Ruchika},
  \citenamefont {Sen}, \citenamefont {Sheikh-Jabbari},\ and\ \citenamefont
  {Yang}}]{Krishnan:2020obg}%
  \BibitemOpen
  \bibfield  {author} {\bibinfo {author} {\bibfnamefont {C.}~\bibnamefont
  {Krishnan}}, \bibinfo {author} {\bibfnamefont {E.}~\bibnamefont
  {\'O~Colg\'ain}}, \bibinfo {author} {\bibnamefont {Ruchika}}, \bibinfo
  {author} {\bibfnamefont {A.~A.}\ \bibnamefont {Sen}}, \bibinfo {author}
  {\bibfnamefont {M.~M.}\ \bibnamefont {Sheikh-Jabbari}}, \ and\ \bibinfo
  {author} {\bibfnamefont {T.}~\bibnamefont {Yang}},\ }\href {\doibase
  10.1103/PhysRevD.102.103525} {\bibfield  {journal} {\bibinfo  {journal}
  {Phys. Rev. D}\ }\textbf {\bibinfo {volume} {102}},\ \bibinfo {pages}
  {103525} (\bibinfo {year} {2020})},\ \Eprint
  {http://arxiv.org/abs/2002.06044} {arXiv:2002.06044 [astro-ph.CO]}
  \BibitemShut {NoStop}%
\bibitem [{\citenamefont {Dainotti}\ \emph {et~al.}(2021)\citenamefont
  {Dainotti}, \citenamefont {De~Simone}, \citenamefont {Schiavone},
  \citenamefont {Montani}, \citenamefont {Rinaldi},\ and\ \citenamefont
  {Lambiase}}]{Dainotti:2021pqg}%
  \BibitemOpen
  \bibfield  {author} {\bibinfo {author} {\bibfnamefont {M.~G.}\ \bibnamefont
  {Dainotti}}, \bibinfo {author} {\bibfnamefont {B.}~\bibnamefont {De~Simone}},
  \bibinfo {author} {\bibfnamefont {T.}~\bibnamefont {Schiavone}}, \bibinfo
  {author} {\bibfnamefont {G.}~\bibnamefont {Montani}}, \bibinfo {author}
  {\bibfnamefont {E.}~\bibnamefont {Rinaldi}}, \ and\ \bibinfo {author}
  {\bibfnamefont {G.}~\bibnamefont {Lambiase}},\ }\href {\doibase
  10.3847/1538-4357/abeb73} {\bibfield  {journal} {\bibinfo  {journal}
  {Astrophys. J.}\ }\textbf {\bibinfo {volume} {912}},\ \bibinfo {pages} {150}
  (\bibinfo {year} {2021})},\ \Eprint {http://arxiv.org/abs/2103.02117}
  {arXiv:2103.02117 [astro-ph.CO]} \BibitemShut {NoStop}%
\bibitem [{\citenamefont {Dainotti}\ \emph {et~al.}(2022)\citenamefont
  {Dainotti}, \citenamefont {De~Simone}, \citenamefont {Schiavone},
  \citenamefont {Montani}, \citenamefont {Rinaldi}, \citenamefont {Lambiase},
  \citenamefont {Bogdan},\ and\ \citenamefont {Ugale}}]{Dainotti:2022bzg}%
  \BibitemOpen
  \bibfield  {author} {\bibinfo {author} {\bibfnamefont {M.~G.}\ \bibnamefont
  {Dainotti}}, \bibinfo {author} {\bibfnamefont {B.}~\bibnamefont {De~Simone}},
  \bibinfo {author} {\bibfnamefont {T.}~\bibnamefont {Schiavone}}, \bibinfo
  {author} {\bibfnamefont {G.}~\bibnamefont {Montani}}, \bibinfo {author}
  {\bibfnamefont {E.}~\bibnamefont {Rinaldi}}, \bibinfo {author} {\bibfnamefont
  {G.}~\bibnamefont {Lambiase}}, \bibinfo {author} {\bibfnamefont
  {M.}~\bibnamefont {Bogdan}}, \ and\ \bibinfo {author} {\bibfnamefont
  {S.}~\bibnamefont {Ugale}},\ }\href {\doibase 10.3390/galaxies10010024}
  {\bibfield  {journal} {\bibinfo  {journal} {Galaxies}\ }\textbf {\bibinfo
  {volume} {10}},\ \bibinfo {pages} {24} (\bibinfo {year} {2022})},\ \Eprint
  {http://arxiv.org/abs/2201.09848} {arXiv:2201.09848 [astro-ph.CO]}
  \BibitemShut {NoStop}%
\bibitem [{\citenamefont {\'O~Colg\'ain}\ \emph {et~al.}(2022)\citenamefont
  {\'O~Colg\'ain}, \citenamefont {Sheikh-Jabbari}, \citenamefont {Solomon},
  \citenamefont {Bargiacchi}, \citenamefont {Capozziello}, \citenamefont
  {Dainotti},\ and\ \citenamefont {Stojkovic}}]{Colgain:2022nlb}%
  \BibitemOpen
  \bibfield  {author} {\bibinfo {author} {\bibfnamefont {E.}~\bibnamefont
  {\'O~Colg\'ain}}, \bibinfo {author} {\bibfnamefont {M.~M.}\ \bibnamefont
  {Sheikh-Jabbari}}, \bibinfo {author} {\bibfnamefont {R.}~\bibnamefont
  {Solomon}}, \bibinfo {author} {\bibfnamefont {G.}~\bibnamefont {Bargiacchi}},
  \bibinfo {author} {\bibfnamefont {S.}~\bibnamefont {Capozziello}}, \bibinfo
  {author} {\bibfnamefont {M.~G.}\ \bibnamefont {Dainotti}}, \ and\ \bibinfo
  {author} {\bibfnamefont {D.}~\bibnamefont {Stojkovic}},\ }\href {\doibase
  10.1103/PhysRevD.106.L041301} {\bibfield  {journal} {\bibinfo  {journal}
  {Phys. Rev. D}\ }\textbf {\bibinfo {volume} {106}},\ \bibinfo {pages}
  {L041301} (\bibinfo {year} {2022})},\ \Eprint
  {http://arxiv.org/abs/2203.10558} {arXiv:2203.10558 [astro-ph.CO]}
  \BibitemShut {NoStop}%
\bibitem [{\citenamefont {Colg\'ain}\ \emph {et~al.}(2024)\citenamefont
  {Colg\'ain}, \citenamefont {Sheikh-Jabbari}, \citenamefont {Solomon},
  \citenamefont {Dainotti},\ and\ \citenamefont {Stojkovic}}]{Colgain:2022rxy}%
  \BibitemOpen
  \bibfield  {author} {\bibinfo {author} {\bibfnamefont {E.~O.}\ \bibnamefont
  {Colg\'ain}}, \bibinfo {author} {\bibfnamefont {M.~M.}\ \bibnamefont
  {Sheikh-Jabbari}}, \bibinfo {author} {\bibfnamefont {R.}~\bibnamefont
  {Solomon}}, \bibinfo {author} {\bibfnamefont {M.~G.}\ \bibnamefont
  {Dainotti}}, \ and\ \bibinfo {author} {\bibfnamefont {D.}~\bibnamefont
  {Stojkovic}},\ }\href {\doibase 10.1016/j.dark.2024.101464} {\bibfield
  {journal} {\bibinfo  {journal} {Phys. Dark Univ.}\ }\textbf {\bibinfo
  {volume} {44}},\ \bibinfo {pages} {101464} (\bibinfo {year} {2024})},\
  \Eprint {http://arxiv.org/abs/2206.11447} {arXiv:2206.11447 [astro-ph.CO]}
  \BibitemShut {NoStop}%
\bibitem [{\citenamefont {Malekjani}\ \emph {et~al.}(2024)\citenamefont
  {Malekjani}, \citenamefont {Conville}, \citenamefont {Colg\'ain},
  \citenamefont {Pourojaghi},\ and\ \citenamefont
  {Sheikh-Jabbari}}]{Malekjani:2023dky}%
  \BibitemOpen
  \bibfield  {author} {\bibinfo {author} {\bibfnamefont {M.}~\bibnamefont
  {Malekjani}}, \bibinfo {author} {\bibfnamefont {R.~M.}\ \bibnamefont
  {Conville}}, \bibinfo {author} {\bibfnamefont {E.~O.}\ \bibnamefont
  {Colg\'ain}}, \bibinfo {author} {\bibfnamefont {S.}~\bibnamefont
  {Pourojaghi}}, \ and\ \bibinfo {author} {\bibfnamefont {M.~M.}\ \bibnamefont
  {Sheikh-Jabbari}},\ }\href {\doibase 10.1140/epjc/s10052-024-12667-z}
  {\bibfield  {journal} {\bibinfo  {journal} {Eur. Phys. J. C}\ }\textbf
  {\bibinfo {volume} {84}},\ \bibinfo {pages} {317} (\bibinfo {year} {2024})},\
  \Eprint {http://arxiv.org/abs/2301.12725} {arXiv:2301.12725 [astro-ph.CO]}
  \BibitemShut {NoStop}%
\bibitem [{\citenamefont {Hu}\ and\ \citenamefont {Wang}(2022)}]{Hu:2022kes}%
  \BibitemOpen
  \bibfield  {author} {\bibinfo {author} {\bibfnamefont {J.-P.}\ \bibnamefont
  {Hu}}\ and\ \bibinfo {author} {\bibfnamefont {F.~Y.}\ \bibnamefont {Wang}},\
  }\href {\doibase 10.1093/mnras/stac2728} {\bibfield  {journal} {\bibinfo
  {journal} {Mon. Not. Roy. Astron. Soc.}\ }\textbf {\bibinfo {volume} {517}},\
  \bibinfo {pages} {576} (\bibinfo {year} {2022})},\ \Eprint
  {http://arxiv.org/abs/2203.13037} {arXiv:2203.13037 [astro-ph.CO]}
  \BibitemShut {NoStop}%
\bibitem [{\citenamefont {Jia}\ \emph {et~al.}(2023)\citenamefont {Jia},
  \citenamefont {Hu},\ and\ \citenamefont {Wang}}]{Jia:2022ycc}%
  \BibitemOpen
  \bibfield  {author} {\bibinfo {author} {\bibfnamefont {X.~D.}\ \bibnamefont
  {Jia}}, \bibinfo {author} {\bibfnamefont {J.~P.}\ \bibnamefont {Hu}}, \ and\
  \bibinfo {author} {\bibfnamefont {F.~Y.}\ \bibnamefont {Wang}},\ }\href
  {\doibase 10.1051/0004-6361/202346356} {\bibfield  {journal} {\bibinfo
  {journal} {Astron. Astrophys.}\ }\textbf {\bibinfo {volume} {674}},\ \bibinfo
  {pages} {A45} (\bibinfo {year} {2023})},\ \Eprint
  {http://arxiv.org/abs/2212.00238} {arXiv:2212.00238 [astro-ph.CO]}
  \BibitemShut {NoStop}%
\bibitem [{\citenamefont {Vagnozzi}(2023)}]{Vagnozzi:2023nrq}%
  \BibitemOpen
  \bibfield  {author} {\bibinfo {author} {\bibfnamefont {S.}~\bibnamefont
  {Vagnozzi}},\ }\href {\doibase 10.3390/universe9090393} {\bibfield  {journal}
  {\bibinfo  {journal} {Universe}\ }\textbf {\bibinfo {volume} {9}},\ \bibinfo
  {pages} {393} (\bibinfo {year} {2023})},\ \Eprint
  {http://arxiv.org/abs/2308.16628} {arXiv:2308.16628 [astro-ph.CO]}
  \BibitemShut {NoStop}%
\bibitem [{\citenamefont {{\'O Colg{\'a}in}}\ \emph {et~al.}(2024)\citenamefont
  {{\'O Colg{\'a}in}}, \citenamefont {{Pourojaghi}},\ and\ \citenamefont
  {{Sheikh-Jabbari}}}]{Colgain:2024ksa}%
  \BibitemOpen
  \bibfield  {author} {\bibinfo {author} {\bibfnamefont {E.}~\bibnamefont {{\'O
  Colg{\'a}in}}}, \bibinfo {author} {\bibfnamefont {S.}~\bibnamefont
  {{Pourojaghi}}}, \ and\ \bibinfo {author} {\bibfnamefont {M.~M.}\
  \bibnamefont {{Sheikh-Jabbari}}},\ }\href {\doibase
  10.48550/arXiv.2406.06389} {\bibfield  {journal} {\bibinfo  {journal} {arXiv
  e-prints}\ ,\ \bibinfo {eid} {arXiv:2406.06389}} (\bibinfo {year} {2024})},\
  \Eprint {http://arxiv.org/abs/2406.06389} {arXiv:2406.06389 [astro-ph.CO]}
  \BibitemShut {NoStop}%
\bibitem [{\citenamefont {Risaliti}\ and\ \citenamefont
  {Lusso}(2019)}]{Risaliti:2018reu}%
  \BibitemOpen
  \bibfield  {author} {\bibinfo {author} {\bibfnamefont {G.}~\bibnamefont
  {Risaliti}}\ and\ \bibinfo {author} {\bibfnamefont {E.}~\bibnamefont
  {Lusso}},\ }\href {\doibase 10.1038/s41550-018-0657-z} {\bibfield  {journal}
  {\bibinfo  {journal} {Nature Astron.}\ }\textbf {\bibinfo {volume} {3}},\
  \bibinfo {pages} {272} (\bibinfo {year} {2019})},\ \Eprint
  {http://arxiv.org/abs/1811.02590} {arXiv:1811.02590 [astro-ph.CO]}
  \BibitemShut {NoStop}%
\bibitem [{\citenamefont {Lusso}\ \emph {et~al.}(2020)\citenamefont {Lusso}
  \emph {et~al.}}]{Lusso:2020pdb}%
  \BibitemOpen
  \bibfield  {author} {\bibinfo {author} {\bibfnamefont {E.}~\bibnamefont
  {Lusso}} \emph {et~al.},\ }\href {\doibase 10.1051/0004-6361/202038899}
  {\bibfield  {journal} {\bibinfo  {journal} {Astron. Astrophys.}\ }\textbf
  {\bibinfo {volume} {642}},\ \bibinfo {pages} {A150} (\bibinfo {year}
  {2020})},\ \Eprint {http://arxiv.org/abs/2008.08586} {arXiv:2008.08586
  [astro-ph.GA]} \BibitemShut {NoStop}%
\bibitem [{\citenamefont {Yang}\ \emph {et~al.}(2020)\citenamefont {Yang},
  \citenamefont {Banerjee},\ and\ \citenamefont
  {\'O~Colg\'ain}}]{Yang:2019vgk}%
  \BibitemOpen
  \bibfield  {author} {\bibinfo {author} {\bibfnamefont {T.}~\bibnamefont
  {Yang}}, \bibinfo {author} {\bibfnamefont {A.}~\bibnamefont {Banerjee}}, \
  and\ \bibinfo {author} {\bibfnamefont {E.}~\bibnamefont {\'O~Colg\'ain}},\
  }\href {\doibase 10.1103/PhysRevD.102.123532} {\bibfield  {journal} {\bibinfo
   {journal} {Phys. Rev. D}\ }\textbf {\bibinfo {volume} {102}},\ \bibinfo
  {pages} {123532} (\bibinfo {year} {2020})},\ \Eprint
  {http://arxiv.org/abs/1911.01681} {arXiv:1911.01681 [astro-ph.CO]}
  \BibitemShut {NoStop}%
\bibitem [{\citenamefont {Khadka}\ and\ \citenamefont
  {Ratra}(2020)}]{Khadka:2020vlh}%
  \BibitemOpen
  \bibfield  {author} {\bibinfo {author} {\bibfnamefont {N.}~\bibnamefont
  {Khadka}}\ and\ \bibinfo {author} {\bibfnamefont {B.}~\bibnamefont {Ratra}},\
  }\href {\doibase 10.1093/mnras/staa1855} {\bibfield  {journal} {\bibinfo
  {journal} {Mon. Not. Roy. Astron. Soc.}\ }\textbf {\bibinfo {volume} {497}},\
  \bibinfo {pages} {263} (\bibinfo {year} {2020})},\ \Eprint
  {http://arxiv.org/abs/2004.09979} {arXiv:2004.09979 [astro-ph.CO]}
  \BibitemShut {NoStop}%
\bibitem [{\citenamefont {Khadka}\ and\ \citenamefont
  {Ratra}(2021)}]{Khadka:2020tlm}%
  \BibitemOpen
  \bibfield  {author} {\bibinfo {author} {\bibfnamefont {N.}~\bibnamefont
  {Khadka}}\ and\ \bibinfo {author} {\bibfnamefont {B.}~\bibnamefont {Ratra}},\
  }\href {\doibase 10.1093/mnras/stab486} {\bibfield  {journal} {\bibinfo
  {journal} {Mon. Not. Roy. Astron. Soc.}\ }\textbf {\bibinfo {volume} {502}},\
  \bibinfo {pages} {6140} (\bibinfo {year} {2021})},\ \Eprint
  {http://arxiv.org/abs/2012.09291} {arXiv:2012.09291 [astro-ph.CO]}
  \BibitemShut {NoStop}%
\bibitem [{\citenamefont {Khadka}\ and\ \citenamefont
  {Ratra}(2022)}]{Khadka:2021xcc}%
  \BibitemOpen
  \bibfield  {author} {\bibinfo {author} {\bibfnamefont {N.}~\bibnamefont
  {Khadka}}\ and\ \bibinfo {author} {\bibfnamefont {B.}~\bibnamefont {Ratra}},\
  }\href {\doibase 10.1093/mnras/stab3678} {\bibfield  {journal} {\bibinfo
  {journal} {Mon. Not. Roy. Astron. Soc.}\ }\textbf {\bibinfo {volume} {510}},\
  \bibinfo {pages} {2753} (\bibinfo {year} {2022})},\ \Eprint
  {http://arxiv.org/abs/2107.07600} {arXiv:2107.07600 [astro-ph.CO]}
  \BibitemShut {NoStop}%
\bibitem [{\citenamefont {Pourojaghi}\ \emph {et~al.}(2022)\citenamefont
  {Pourojaghi}, \citenamefont {Zabihi},\ and\ \citenamefont
  {Malekjani}}]{Pourojaghi:2022zrh}%
  \BibitemOpen
  \bibfield  {author} {\bibinfo {author} {\bibfnamefont {S.}~\bibnamefont
  {Pourojaghi}}, \bibinfo {author} {\bibfnamefont {N.~F.}\ \bibnamefont
  {Zabihi}}, \ and\ \bibinfo {author} {\bibfnamefont {M.}~\bibnamefont
  {Malekjani}},\ }\href {\doibase 10.1103/PhysRevD.106.123523} {\bibfield
  {journal} {\bibinfo  {journal} {Phys. Rev. D}\ }\textbf {\bibinfo {volume}
  {106}},\ \bibinfo {pages} {123523} (\bibinfo {year} {2022})},\ \Eprint
  {http://arxiv.org/abs/2212.04118} {arXiv:2212.04118 [astro-ph.CO]}
  \BibitemShut {NoStop}%
\bibitem [{\citenamefont {Past\'en}\ and\ \citenamefont
  {C\'ardenas}(2023)}]{Pasten:2023rpc}%
  \BibitemOpen
  \bibfield  {author} {\bibinfo {author} {\bibfnamefont {E.}~\bibnamefont
  {Past\'en}}\ and\ \bibinfo {author} {\bibfnamefont {V.~H.}\ \bibnamefont
  {C\'ardenas}},\ }\href {\doibase 10.1016/j.dark.2023.101224} {\bibfield
  {journal} {\bibinfo  {journal} {Phys. Dark Univ.}\ }\textbf {\bibinfo
  {volume} {40}},\ \bibinfo {pages} {101224} (\bibinfo {year} {2023})},\
  \Eprint {http://arxiv.org/abs/2301.10740} {arXiv:2301.10740 [astro-ph.CO]}
  \BibitemShut {NoStop}%
\bibitem [{\citenamefont {{Pedrotti}}\ \emph {et~al.}(2024)\citenamefont
  {{Pedrotti}}, \citenamefont {{Jiang}}, \citenamefont {{Escamilla}},
  \citenamefont {{Santos da Costa}},\ and\ \citenamefont
  {{Vagnozzi}}}]{Pedrotti:2024kpn}%
  \BibitemOpen
  \bibfield  {author} {\bibinfo {author} {\bibfnamefont {D.}~\bibnamefont
  {{Pedrotti}}}, \bibinfo {author} {\bibfnamefont {J.-Q.}\ \bibnamefont
  {{Jiang}}}, \bibinfo {author} {\bibfnamefont {L.~A.}\ \bibnamefont
  {{Escamilla}}}, \bibinfo {author} {\bibfnamefont {S.}~\bibnamefont {{Santos
  da Costa}}}, \ and\ \bibinfo {author} {\bibfnamefont {S.}~\bibnamefont
  {{Vagnozzi}}},\ }\href {\doibase 10.48550/arXiv.2408.04530} {\bibfield
  {journal} {\bibinfo  {journal} {arXiv e-prints}\ ,\ \bibinfo {eid}
  {arXiv:2408.04530}} (\bibinfo {year} {2024})},\ \Eprint
  {http://arxiv.org/abs/2408.04530} {arXiv:2408.04530 [astro-ph.CO]}
  \BibitemShut {NoStop}%
\bibitem [{\citenamefont {Heymans}\ \emph {et~al.}(2013)\citenamefont {Heymans}
  \emph {et~al.}}]{Heymans:2013fya}%
  \BibitemOpen
  \bibfield  {author} {\bibinfo {author} {\bibfnamefont {C.}~\bibnamefont
  {Heymans}} \emph {et~al.},\ }\href {\doibase 10.1093/mnras/stt601} {\bibfield
   {journal} {\bibinfo  {journal} {Mon. Not. Roy. Astron. Soc.}\ }\textbf
  {\bibinfo {volume} {432}},\ \bibinfo {pages} {2433} (\bibinfo {year}
  {2013})},\ \Eprint {http://arxiv.org/abs/1303.1808} {arXiv:1303.1808
  [astro-ph.CO]} \BibitemShut {NoStop}%
\bibitem [{\citenamefont {Joudaki}\ \emph {et~al.}(2017)\citenamefont {Joudaki}
  \emph {et~al.}}]{Joudaki:2016mvz}%
  \BibitemOpen
  \bibfield  {author} {\bibinfo {author} {\bibfnamefont {S.}~\bibnamefont
  {Joudaki}} \emph {et~al.},\ }\href {\doibase 10.1093/mnras/stw2665}
  {\bibfield  {journal} {\bibinfo  {journal} {Mon. Not. Roy. Astron. Soc.}\
  }\textbf {\bibinfo {volume} {465}},\ \bibinfo {pages} {2033} (\bibinfo {year}
  {2017})},\ \Eprint {http://arxiv.org/abs/1601.05786} {arXiv:1601.05786
  [astro-ph.CO]} \BibitemShut {NoStop}%
\bibitem [{\citenamefont {Troxel}\ \emph {et~al.}(2018)\citenamefont {Troxel}
  \emph {et~al.}}]{DES:2017qwj}%
  \BibitemOpen
  \bibfield  {author} {\bibinfo {author} {\bibfnamefont {M.~A.}\ \bibnamefont
  {Troxel}} \emph {et~al.} (\bibinfo {collaboration} {DES}),\ }\href {\doibase
  10.1103/PhysRevD.98.043528} {\bibfield  {journal} {\bibinfo  {journal} {Phys.
  Rev. D}\ }\textbf {\bibinfo {volume} {98}},\ \bibinfo {pages} {043528}
  (\bibinfo {year} {2018})},\ \Eprint {http://arxiv.org/abs/1708.01538}
  {arXiv:1708.01538 [astro-ph.CO]} \BibitemShut {NoStop}%
\bibitem [{\citenamefont {Hikage}\ \emph {et~al.}(2019)\citenamefont {Hikage}
  \emph {et~al.}}]{HSC:2018mrq}%
  \BibitemOpen
  \bibfield  {author} {\bibinfo {author} {\bibfnamefont {C.}~\bibnamefont
  {Hikage}} \emph {et~al.} (\bibinfo {collaboration} {HSC}),\ }\href {\doibase
  10.1093/pasj/psz010} {\bibfield  {journal} {\bibinfo  {journal} {Publ.
  Astron. Soc. Jap.}\ }\textbf {\bibinfo {volume} {71}},\ \bibinfo {pages} {43}
  (\bibinfo {year} {2019})},\ \Eprint {http://arxiv.org/abs/1809.09148}
  {arXiv:1809.09148 [astro-ph.CO]} \BibitemShut {NoStop}%
\bibitem [{\citenamefont {Asgari}\ \emph {et~al.}(2021)\citenamefont {Asgari}
  \emph {et~al.}}]{KiDS:2020suj}%
  \BibitemOpen
  \bibfield  {author} {\bibinfo {author} {\bibfnamefont {M.}~\bibnamefont
  {Asgari}} \emph {et~al.} (\bibinfo {collaboration} {KiDS}),\ }\href {\doibase
  10.1051/0004-6361/202039070} {\bibfield  {journal} {\bibinfo  {journal}
  {Astron. Astrophys.}\ }\textbf {\bibinfo {volume} {645}},\ \bibinfo {pages}
  {A104} (\bibinfo {year} {2021})},\ \Eprint {http://arxiv.org/abs/2007.15633}
  {arXiv:2007.15633 [astro-ph.CO]} \BibitemShut {NoStop}%
\bibitem [{\citenamefont {Abbott}\ \emph {et~al.}(2022)\citenamefont {Abbott}
  \emph {et~al.}}]{DES:2021wwk}%
  \BibitemOpen
  \bibfield  {author} {\bibinfo {author} {\bibfnamefont {T.~M.~C.}\
  \bibnamefont {Abbott}} \emph {et~al.} (\bibinfo {collaboration} {DES}),\
  }\href {\doibase 10.1103/PhysRevD.105.023520} {\bibfield  {journal} {\bibinfo
   {journal} {Phys. Rev. D}\ }\textbf {\bibinfo {volume} {105}},\ \bibinfo
  {pages} {023520} (\bibinfo {year} {2022})},\ \Eprint
  {http://arxiv.org/abs/2105.13549} {arXiv:2105.13549 [astro-ph.CO]}
  \BibitemShut {NoStop}%
\bibitem [{\citenamefont {White}\ \emph {et~al.}(2022)\citenamefont {White}
  \emph {et~al.}}]{White:2021yvw}%
  \BibitemOpen
  \bibfield  {author} {\bibinfo {author} {\bibfnamefont {M.}~\bibnamefont
  {White}} \emph {et~al.},\ }\href {\doibase 10.1088/1475-7516/2022/02/007}
  {\bibfield  {journal} {\bibinfo  {journal} {JCAP}\ }\textbf {\bibinfo
  {volume} {02}},\ \bibinfo {pages} {007} (\bibinfo {year} {2022})},\ \Eprint
  {http://arxiv.org/abs/2111.09898} {arXiv:2111.09898 [astro-ph.CO]}
  \BibitemShut {NoStop}%
\bibitem [{\citenamefont {Garc\'\i{}a-Garc\'\i{}a}\ \emph
  {et~al.}(2021)\citenamefont {Garc\'\i{}a-Garc\'\i{}a}, \citenamefont
  {Zapatero}, \citenamefont {Alonso}, \citenamefont {Bellini}, \citenamefont
  {Ferreira}, \citenamefont {Mueller}, \citenamefont {Nicola},\ and\
  \citenamefont {Ruiz-Lapuente}}]{Garcia-Garcia:2021unp}%
  \BibitemOpen
  \bibfield  {author} {\bibinfo {author} {\bibfnamefont {C.}~\bibnamefont
  {Garc\'\i{}a-Garc\'\i{}a}}, \bibinfo {author} {\bibfnamefont {J.~R.}\
  \bibnamefont {Zapatero}}, \bibinfo {author} {\bibfnamefont {D.}~\bibnamefont
  {Alonso}}, \bibinfo {author} {\bibfnamefont {E.}~\bibnamefont {Bellini}},
  \bibinfo {author} {\bibfnamefont {P.~G.}\ \bibnamefont {Ferreira}}, \bibinfo
  {author} {\bibfnamefont {E.-M.}\ \bibnamefont {Mueller}}, \bibinfo {author}
  {\bibfnamefont {A.}~\bibnamefont {Nicola}}, \ and\ \bibinfo {author}
  {\bibfnamefont {P.}~\bibnamefont {Ruiz-Lapuente}},\ }\href {\doibase
  10.1088/1475-7516/2021/10/030} {\bibfield  {journal} {\bibinfo  {journal}
  {JCAP}\ }\textbf {\bibinfo {volume} {10}},\ \bibinfo {pages} {030} (\bibinfo
  {year} {2021})},\ \Eprint {http://arxiv.org/abs/2105.12108} {arXiv:2105.12108
  [astro-ph.CO]} \BibitemShut {NoStop}%
\bibitem [{\citenamefont {Abbott}\ \emph {et~al.}(2023)\citenamefont {Abbott}
  \emph {et~al.}}]{DES:2022ccp}%
  \BibitemOpen
  \bibfield  {author} {\bibinfo {author} {\bibfnamefont {T.~M.~C.}\
  \bibnamefont {Abbott}} \emph {et~al.} (\bibinfo {collaboration} {DES}),\
  }\href {\doibase 10.1103/PhysRevD.107.083504} {\bibfield  {journal} {\bibinfo
   {journal} {Phys. Rev. D}\ }\textbf {\bibinfo {volume} {107}},\ \bibinfo
  {pages} {083504} (\bibinfo {year} {2023})},\ \Eprint
  {http://arxiv.org/abs/2207.05766} {arXiv:2207.05766 [astro-ph.CO]}
  \BibitemShut {NoStop}%
\bibitem [{\citenamefont {Bocquet}\ \emph {et~al.}(2024)\citenamefont {Bocquet}
  \emph {et~al.}}]{SPT:2024qbr}%
  \BibitemOpen
  \bibfield  {author} {\bibinfo {author} {\bibfnamefont {S.}~\bibnamefont
  {Bocquet}} \emph {et~al.} (\bibinfo {collaboration} {SPT, DES}),\ }\href
  {\doibase 10.1103/PhysRevD.110.083510} {\bibfield  {journal} {\bibinfo
  {journal} {Phys. Rev. D}\ }\textbf {\bibinfo {volume} {110}},\ \bibinfo
  {pages} {083510} (\bibinfo {year} {2024})},\ \Eprint
  {http://arxiv.org/abs/2401.02075} {arXiv:2401.02075 [astro-ph.CO]}
  \BibitemShut {NoStop}%
\bibitem [{\citenamefont {Esposito}\ \emph {et~al.}(2022)\citenamefont
  {Esposito}, \citenamefont {Ir\v{s}i\v{c}}, \citenamefont {Costanzi},
  \citenamefont {Borgani}, \citenamefont {Saro},\ and\ \citenamefont
  {Viel}}]{Esposito:2022plo}%
  \BibitemOpen
  \bibfield  {author} {\bibinfo {author} {\bibfnamefont {M.}~\bibnamefont
  {Esposito}}, \bibinfo {author} {\bibfnamefont {V.}~\bibnamefont
  {Ir\v{s}i\v{c}}}, \bibinfo {author} {\bibfnamefont {M.}~\bibnamefont
  {Costanzi}}, \bibinfo {author} {\bibfnamefont {S.}~\bibnamefont {Borgani}},
  \bibinfo {author} {\bibfnamefont {A.}~\bibnamefont {Saro}}, \ and\ \bibinfo
  {author} {\bibfnamefont {M.}~\bibnamefont {Viel}},\ }\href {\doibase
  10.1093/mnras/stac1825} {\bibfield  {journal} {\bibinfo  {journal} {Mon. Not.
  Roy. Astron. Soc.}\ }\textbf {\bibinfo {volume} {515}},\ \bibinfo {pages}
  {857} (\bibinfo {year} {2022})},\ \Eprint {http://arxiv.org/abs/2202.00974}
  {arXiv:2202.00974 [astro-ph.CO]} \BibitemShut {NoStop}%
\bibitem [{\citenamefont {{Tutusaus}}\ \emph {et~al.}(2023)\citenamefont
  {{Tutusaus}}, \citenamefont {{Bonvin}},\ and\ \citenamefont
  {{Grimm}}}]{Tutusaus:2023aux}%
  \BibitemOpen
  \bibfield  {author} {\bibinfo {author} {\bibfnamefont {I.}~\bibnamefont
  {{Tutusaus}}}, \bibinfo {author} {\bibfnamefont {C.}~\bibnamefont
  {{Bonvin}}}, \ and\ \bibinfo {author} {\bibfnamefont {N.}~\bibnamefont
  {{Grimm}}},\ }\href {\doibase 10.48550/arXiv.2312.06434} {\bibfield
  {journal} {\bibinfo  {journal} {arXiv e-prints}\ ,\ \bibinfo {eid}
  {arXiv:2312.06434}} (\bibinfo {year} {2023})},\ \Eprint
  {http://arxiv.org/abs/2312.06434} {arXiv:2312.06434 [astro-ph.CO]}
  \BibitemShut {NoStop}%
\bibitem [{\citenamefont {Manna}\ and\ \citenamefont
  {Desai}(2024)}]{Manna:2024wak}%
  \BibitemOpen
  \bibfield  {author} {\bibinfo {author} {\bibfnamefont {S.}~\bibnamefont
  {Manna}}\ and\ \bibinfo {author} {\bibfnamefont {S.}~\bibnamefont {Desai}},\
  }\href {\doibase 10.1140/epjc/s10052-024-13031-x} {\bibfield  {journal}
  {\bibinfo  {journal} {Eur. Phys. J. C}\ }\textbf {\bibinfo {volume} {84}},\
  \bibinfo {pages} {661} (\bibinfo {year} {2024})},\ \Eprint
  {http://arxiv.org/abs/2406.10931} {arXiv:2406.10931 [astro-ph.CO]}
  \BibitemShut {NoStop}%
\bibitem [{\citenamefont {{Sailer}}\ \emph {et~al.}(2024)\citenamefont
  {{Sailer}}, \citenamefont {{Kim}}, \citenamefont {{Ferraro}}, \citenamefont
  {{Madhavacheril}}, \citenamefont {{White}}, \citenamefont {{Abril-Cabezas}},
  \citenamefont {{Aguilar}}, \citenamefont {{Ahlen}}, \citenamefont {{Bond}},
  \citenamefont {{Brooks}}, \citenamefont {{Burtin}}, \citenamefont
  {{Calabrese}}, \citenamefont {{Chen}}, \citenamefont {{Choi}}, \citenamefont
  {{Claybaugh}}, \citenamefont {{Dawson}}, \citenamefont {{de la Macorra}},
  \citenamefont {{DeRose}}, \citenamefont {{Dey}}, \citenamefont {{Dey}},
  \citenamefont {{Doel}}, \citenamefont {{Dunkley}}, \citenamefont
  {{Embil-Villagra}}, \citenamefont {{Farren}}, \citenamefont {{Font-Ribera}},
  \citenamefont {{Forero-Romero}}, \citenamefont {{Gazta{\~n}aga}},
  \citenamefont {{Gluscevic}}, \citenamefont {{Gontcho}}, \citenamefont
  {{Honscheid}}, \citenamefont {{Howlett}}, \citenamefont {{Juneau}},
  \citenamefont {{Kirkby}}, \citenamefont {{Kisner}}, \citenamefont {{Kremin}},
  \citenamefont {{Landriau}}, \citenamefont {{Le Guillou}}, \citenamefont
  {{Levi}}, \citenamefont {{Manera}}, \citenamefont {{Meisner}}, \citenamefont
  {{Miquel}}, \citenamefont {{Moodley}}, \citenamefont {{Moustakas}},
  \citenamefont {{Niemack}}, \citenamefont {{Niz}}, \citenamefont
  {{Palanque-Delabrouille}}, \citenamefont {{Percival}}, \citenamefont
  {{Prada}}, \citenamefont {{Qu}}, \citenamefont {{Rossi}}, \citenamefont
  {{Sanchez}}, \citenamefont {{Schaan}}, \citenamefont {{Schlafly}},
  \citenamefont {{Schlegel}}, \citenamefont {{Schubnell}}, \citenamefont
  {{Sehgal}}, \citenamefont {{Seo}}, \citenamefont {{Sherwin}}, \citenamefont
  {{Sif{\'o}n}}, \citenamefont {{Sprayberry}}, \citenamefont {{Staggs}},
  \citenamefont {{Tarl{\'e}}}, \citenamefont {{Weaver}}, \citenamefont
  {{Y{\`e}che}}, \citenamefont {{Zhou}},\ and\ \citenamefont
  {{Zou}}}]{Sailer:2024coh}%
  \BibitemOpen
  \bibfield  {author} {\bibinfo {author} {\bibfnamefont {N.}~\bibnamefont
  {{Sailer}}}, \bibinfo {author} {\bibfnamefont {J.}~\bibnamefont {{Kim}}},
  \bibinfo {author} {\bibfnamefont {S.}~\bibnamefont {{Ferraro}}}, \bibinfo
  {author} {\bibfnamefont {M.~S.}\ \bibnamefont {{Madhavacheril}}}, \bibinfo
  {author} {\bibfnamefont {M.}~\bibnamefont {{White}}}, \bibinfo {author}
  {\bibfnamefont {I.}~\bibnamefont {{Abril-Cabezas}}}, \bibinfo {author}
  {\bibfnamefont {J.~N.}\ \bibnamefont {{Aguilar}}}, \bibinfo {author}
  {\bibfnamefont {S.}~\bibnamefont {{Ahlen}}}, \bibinfo {author} {\bibfnamefont
  {J.~R.}\ \bibnamefont {{Bond}}}, \bibinfo {author} {\bibfnamefont
  {D.}~\bibnamefont {{Brooks}}}, \bibinfo {author} {\bibfnamefont
  {E.}~\bibnamefont {{Burtin}}}, \bibinfo {author} {\bibfnamefont
  {E.}~\bibnamefont {{Calabrese}}}, \bibinfo {author} {\bibfnamefont {S.-F.}\
  \bibnamefont {{Chen}}}, \bibinfo {author} {\bibfnamefont {S.~K.}\
  \bibnamefont {{Choi}}}, \bibinfo {author} {\bibfnamefont {T.}~\bibnamefont
  {{Claybaugh}}}, \bibinfo {author} {\bibfnamefont {K.}~\bibnamefont
  {{Dawson}}}, \bibinfo {author} {\bibfnamefont {A.}~\bibnamefont {{de la
  Macorra}}}, \bibinfo {author} {\bibfnamefont {J.}~\bibnamefont {{DeRose}}},
  \bibinfo {author} {\bibfnamefont {A.}~\bibnamefont {{Dey}}}, \bibinfo
  {author} {\bibfnamefont {B.}~\bibnamefont {{Dey}}}, \bibinfo {author}
  {\bibfnamefont {P.}~\bibnamefont {{Doel}}}, \bibinfo {author} {\bibfnamefont
  {J.}~\bibnamefont {{Dunkley}}}, \bibinfo {author} {\bibfnamefont
  {C.}~\bibnamefont {{Embil-Villagra}}}, \bibinfo {author} {\bibfnamefont
  {G.~S.}\ \bibnamefont {{Farren}}}, \bibinfo {author} {\bibfnamefont
  {A.}~\bibnamefont {{Font-Ribera}}}, \bibinfo {author} {\bibfnamefont {J.~E.}\
  \bibnamefont {{Forero-Romero}}}, \bibinfo {author} {\bibfnamefont
  {E.}~\bibnamefont {{Gazta{\~n}aga}}}, \bibinfo {author} {\bibfnamefont
  {V.}~\bibnamefont {{Gluscevic}}}, \bibinfo {author} {\bibfnamefont
  {S.~G.~A.}\ \bibnamefont {{Gontcho}}}, \bibinfo {author} {\bibfnamefont
  {K.}~\bibnamefont {{Honscheid}}}, \bibinfo {author} {\bibfnamefont
  {C.}~\bibnamefont {{Howlett}}}, \bibinfo {author} {\bibfnamefont
  {S.}~\bibnamefont {{Juneau}}}, \bibinfo {author} {\bibfnamefont
  {D.}~\bibnamefont {{Kirkby}}}, \bibinfo {author} {\bibfnamefont
  {T.}~\bibnamefont {{Kisner}}}, \bibinfo {author} {\bibfnamefont
  {A.}~\bibnamefont {{Kremin}}}, \bibinfo {author} {\bibfnamefont
  {M.}~\bibnamefont {{Landriau}}}, \bibinfo {author} {\bibfnamefont
  {L.}~\bibnamefont {{Le Guillou}}}, \bibinfo {author} {\bibfnamefont
  {M.}~\bibnamefont {{Levi}}}, \bibinfo {author} {\bibfnamefont
  {M.}~\bibnamefont {{Manera}}}, \bibinfo {author} {\bibfnamefont
  {A.}~\bibnamefont {{Meisner}}}, \bibinfo {author} {\bibfnamefont
  {R.}~\bibnamefont {{Miquel}}}, \bibinfo {author} {\bibfnamefont
  {K.}~\bibnamefont {{Moodley}}}, \bibinfo {author} {\bibfnamefont
  {J.}~\bibnamefont {{Moustakas}}}, \bibinfo {author} {\bibfnamefont {M.~D.}\
  \bibnamefont {{Niemack}}}, \bibinfo {author} {\bibfnamefont {G.}~\bibnamefont
  {{Niz}}}, \bibinfo {author} {\bibfnamefont {N.}~\bibnamefont
  {{Palanque-Delabrouille}}}, \bibinfo {author} {\bibfnamefont
  {W.}~\bibnamefont {{Percival}}}, \bibinfo {author} {\bibfnamefont
  {F.}~\bibnamefont {{Prada}}}, \bibinfo {author} {\bibfnamefont {F.~J.}\
  \bibnamefont {{Qu}}}, \bibinfo {author} {\bibfnamefont {G.}~\bibnamefont
  {{Rossi}}}, \bibinfo {author} {\bibfnamefont {E.}~\bibnamefont {{Sanchez}}},
  \bibinfo {author} {\bibfnamefont {E.}~\bibnamefont {{Schaan}}}, \bibinfo
  {author} {\bibfnamefont {E.}~\bibnamefont {{Schlafly}}}, \bibinfo {author}
  {\bibfnamefont {D.}~\bibnamefont {{Schlegel}}}, \bibinfo {author}
  {\bibfnamefont {M.}~\bibnamefont {{Schubnell}}}, \bibinfo {author}
  {\bibfnamefont {N.}~\bibnamefont {{Sehgal}}}, \bibinfo {author}
  {\bibfnamefont {H.-J.}\ \bibnamefont {{Seo}}}, \bibinfo {author}
  {\bibfnamefont {B.}~\bibnamefont {{Sherwin}}}, \bibinfo {author}
  {\bibfnamefont {C.}~\bibnamefont {{Sif{\'o}n}}}, \bibinfo {author}
  {\bibfnamefont {D.}~\bibnamefont {{Sprayberry}}}, \bibinfo {author}
  {\bibfnamefont {S.~T.}\ \bibnamefont {{Staggs}}}, \bibinfo {author}
  {\bibfnamefont {G.}~\bibnamefont {{Tarl{\'e}}}}, \bibinfo {author}
  {\bibfnamefont {B.~A.}\ \bibnamefont {{Weaver}}}, \bibinfo {author}
  {\bibfnamefont {C.}~\bibnamefont {{Y{\`e}che}}}, \bibinfo {author}
  {\bibfnamefont {R.}~\bibnamefont {{Zhou}}}, \ and\ \bibinfo {author}
  {\bibfnamefont {H.}~\bibnamefont {{Zou}}},\ }\href {\doibase
  10.48550/arXiv.2407.04607} {\bibfield  {journal} {\bibinfo  {journal} {arXiv
  e-prints}\ ,\ \bibinfo {eid} {arXiv:2407.04607}} (\bibinfo {year} {2024})},\
  \Eprint {http://arxiv.org/abs/2407.04607} {arXiv:2407.04607 [astro-ph.CO]}
  \BibitemShut {NoStop}%
\bibitem [{\citenamefont {Kov\'acs}\ \emph {et~al.}(2019)\citenamefont
  {Kov\'acs} \emph {et~al.}}]{DES:2018nlb}%
  \BibitemOpen
  \bibfield  {author} {\bibinfo {author} {\bibfnamefont {A.}~\bibnamefont
  {Kov\'acs}} \emph {et~al.} (\bibinfo {collaboration} {DES}),\ }\href
  {\doibase 10.1093/mnras/stz341} {\bibfield  {journal} {\bibinfo  {journal}
  {Mon. Not. Roy. Astron. Soc.}\ }\textbf {\bibinfo {volume} {484}},\ \bibinfo
  {pages} {5267} (\bibinfo {year} {2019})},\ \Eprint
  {http://arxiv.org/abs/1811.07812} {arXiv:1811.07812 [astro-ph.CO]}
  \BibitemShut {NoStop}%
\bibitem [{\citenamefont {Kov\'acs}\ \emph {et~al.}(2022)\citenamefont
  {Kov\'acs}, \citenamefont {Beck}, \citenamefont {Smith}, \citenamefont
  {R\'acz}, \citenamefont {Csabai},\ and\ \citenamefont
  {Szapudi}}]{Kovacs:2021mnf}%
  \BibitemOpen
  \bibfield  {author} {\bibinfo {author} {\bibfnamefont {A.}~\bibnamefont
  {Kov\'acs}}, \bibinfo {author} {\bibfnamefont {R.}~\bibnamefont {Beck}},
  \bibinfo {author} {\bibfnamefont {A.}~\bibnamefont {Smith}}, \bibinfo
  {author} {\bibfnamefont {G.}~\bibnamefont {R\'acz}}, \bibinfo {author}
  {\bibfnamefont {I.}~\bibnamefont {Csabai}}, \ and\ \bibinfo {author}
  {\bibfnamefont {I.}~\bibnamefont {Szapudi}},\ }\href {\doibase
  10.1093/mnras/stac903} {\bibfield  {journal} {\bibinfo  {journal} {Mon. Not.
  Roy. Astron. Soc.}\ }\textbf {\bibinfo {volume} {513}},\ \bibinfo {pages}
  {15} (\bibinfo {year} {2022})},\ \Eprint {http://arxiv.org/abs/2107.13038}
  {arXiv:2107.13038 [astro-ph.CO]} \BibitemShut {NoStop}%
\bibitem [{\citenamefont {Nguyen}\ \emph {et~al.}(2023)\citenamefont {Nguyen},
  \citenamefont {Huterer},\ and\ \citenamefont {Wen}}]{Nguyen:2023fip}%
  \BibitemOpen
  \bibfield  {author} {\bibinfo {author} {\bibfnamefont {N.-M.}\ \bibnamefont
  {Nguyen}}, \bibinfo {author} {\bibfnamefont {D.}~\bibnamefont {Huterer}}, \
  and\ \bibinfo {author} {\bibfnamefont {Y.}~\bibnamefont {Wen}},\ }\href
  {\doibase 10.1103/PhysRevLett.131.111001} {\bibfield  {journal} {\bibinfo
  {journal} {Phys. Rev. Lett.}\ }\textbf {\bibinfo {volume} {131}},\ \bibinfo
  {pages} {111001} (\bibinfo {year} {2023})},\ \Eprint
  {http://arxiv.org/abs/2302.01331} {arXiv:2302.01331 [astro-ph.CO]}
  \BibitemShut {NoStop}%
\bibitem [{\citenamefont {Wang}\ and\ \citenamefont
  {Steinhardt}(1998)}]{Wang:1998gt}%
  \BibitemOpen
  \bibfield  {author} {\bibinfo {author} {\bibfnamefont {L.-M.}\ \bibnamefont
  {Wang}}\ and\ \bibinfo {author} {\bibfnamefont {P.~J.}\ \bibnamefont
  {Steinhardt}},\ }\href {\doibase 10.1086/306436} {\bibfield  {journal}
  {\bibinfo  {journal} {Astrophys. J.}\ }\textbf {\bibinfo {volume} {508}},\
  \bibinfo {pages} {483} (\bibinfo {year} {1998})},\ \Eprint
  {http://arxiv.org/abs/astro-ph/9804015} {arXiv:astro-ph/9804015} \BibitemShut
  {NoStop}%
\bibitem [{\citenamefont {Bulbul}\ \emph {et~al.}(2024)\citenamefont {Bulbul}
  \emph {et~al.}}]{Bulbul:2024mfj}%
  \BibitemOpen
  \bibfield  {author} {\bibinfo {author} {\bibfnamefont {E.}~\bibnamefont
  {Bulbul}} \emph {et~al.},\ }\href {\doibase 10.1051/0004-6361/202348264}
  {\bibfield  {journal} {\bibinfo  {journal} {Astron. Astrophys.}\ }\textbf
  {\bibinfo {volume} {685}},\ \bibinfo {pages} {A106} (\bibinfo {year}
  {2024})},\ \Eprint {http://arxiv.org/abs/2402.08452} {arXiv:2402.08452
  [astro-ph.CO]} \BibitemShut {NoStop}%
\bibitem [{\citenamefont {Kluge}\ \emph {et~al.}(2024)\citenamefont {Kluge}
  \emph {et~al.}}]{Kluge:2024ghp}%
  \BibitemOpen
  \bibfield  {author} {\bibinfo {author} {\bibfnamefont {M.}~\bibnamefont
  {Kluge}} \emph {et~al.},\ }\href {\doibase 10.1051/0004-6361/202349031}
  {\bibfield  {journal} {\bibinfo  {journal} {Astron. Astrophys.}\ }\textbf
  {\bibinfo {volume} {688}},\ \bibinfo {pages} {A210} (\bibinfo {year}
  {2024})},\ \Eprint {http://arxiv.org/abs/2402.08453} {arXiv:2402.08453
  [astro-ph.CO]} \BibitemShut {NoStop}%
\bibitem [{\citenamefont {Ghirardini}\ \emph {et~al.}(2024)\citenamefont
  {Ghirardini} \emph {et~al.}}]{Ghirardini:2024yni}%
  \BibitemOpen
  \bibfield  {author} {\bibinfo {author} {\bibfnamefont {V.}~\bibnamefont
  {Ghirardini}} \emph {et~al.},\ }\href {\doibase 10.1051/0004-6361/202348852}
  {\bibfield  {journal} {\bibinfo  {journal} {Astron. Astrophys.}\ }\textbf
  {\bibinfo {volume} {689}},\ \bibinfo {pages} {A298} (\bibinfo {year}
  {2024})},\ \Eprint {http://arxiv.org/abs/2402.08458} {arXiv:2402.08458
  [astro-ph.CO]} \BibitemShut {NoStop}%
\bibitem [{\citenamefont {{Qu}}\ \emph {et~al.}(2023)\citenamefont {{Qu}},
  \citenamefont {{Sherwin}}, \citenamefont {{Madhavacheril}}, \citenamefont
  {{Han}}, \citenamefont {{Crowley}}, \citenamefont {{Abril-Cabezas}},
  \citenamefont {{Ade}}, \citenamefont {{Aiola}}, \citenamefont {{Alford}},
  \citenamefont {{Amiri}}, \citenamefont {{Amodeo}}, \citenamefont {{An}},
  \citenamefont {{Atkins}}, \citenamefont {{Austermann}}, \citenamefont
  {{Battaglia}}, \citenamefont {{Battistelli}}, \citenamefont {{Beall}},
  \citenamefont {{Bean}}, \citenamefont {{Beringue}}, \citenamefont
  {{Bhandarkar}}, \citenamefont {{Biermann}}, \citenamefont {{Bolliet}},
  \citenamefont {{Bond}}, \citenamefont {{Cai}}, \citenamefont {{Calabrese}},
  \citenamefont {{Calafut}}, \citenamefont {{Capalbo}}, \citenamefont
  {{Carrero}}, \citenamefont {{Carron}}, \citenamefont {{Challinor}},
  \citenamefont {{Chesmore}}, \citenamefont {{Cho}}, \citenamefont {{Choi}},
  \citenamefont {{Clark}}, \citenamefont {{C{\'o}rdova Rosado}}, \citenamefont
  {{Cothard}}, \citenamefont {{Coughlin}}, \citenamefont {{Coulton}},
  \citenamefont {{Dalal}}, \citenamefont {{Darwish}}, \citenamefont {{Devlin}},
  \citenamefont {{Dicker}}, \citenamefont {{Doze}}, \citenamefont {{Duell}},
  \citenamefont {{Duff}}, \citenamefont {{Duivenvoorden}}, \citenamefont
  {{Dunkley}}, \citenamefont {{D{\"u}nner}}, \citenamefont {{Fanfani}},
  \citenamefont {{Fankhanel}}, \citenamefont {{Farren}}, \citenamefont
  {{Ferraro}}, \citenamefont {{Freundt}}, \citenamefont {{Fuzia}},
  \citenamefont {{Gallardo}}, \citenamefont {{Garrido}}, \citenamefont
  {{Gluscevic}}, \citenamefont {{Golec}}, \citenamefont {{Guan}}, \citenamefont
  {{Halpern}}, \citenamefont {{Harrison}}, \citenamefont {{Hasselfield}},
  \citenamefont {{Healy}}, \citenamefont {{Henderson}}, \citenamefont
  {{Hensley}}, \citenamefont {{Herv{\'\i}as-Caimapo}}, \citenamefont {{Hill}},
  \citenamefont {{Hilton}}, \citenamefont {{Hilton}}, \citenamefont {{Hincks}},
  \citenamefont {{Hlo{\v{z}}ek}}, \citenamefont {{Ho}}, \citenamefont
  {{Huber}}, \citenamefont {{Hubmayr}}, \citenamefont {{Huffenberger}},
  \citenamefont {{Hughes}}, \citenamefont {{Irwin}}, \citenamefont {{Isopi}},
  \citenamefont {{Jense}}, \citenamefont {{Keller}}, \citenamefont {{Kim}},
  \citenamefont {{Knowles}}, \citenamefont {{Koopman}}, \citenamefont
  {{Kosowsky}}, \citenamefont {{Kramer}}, \citenamefont {{Kusiak}},
  \citenamefont {{La Posta}}, \citenamefont {{Lague}}, \citenamefont {{Lakey}},
  \citenamefont {{Lee}}, \citenamefont {{Li}}, \citenamefont {{Li}},
  \citenamefont {{Limon}}, \citenamefont {{Lokken}}, \citenamefont {{Louis}},
  \citenamefont {{Lungu}}, \citenamefont {{MacCrann}}, \citenamefont
  {{MacInnis}}, \citenamefont {{Maldonado}}, \citenamefont {{Maldonado}},
  \citenamefont {{Mallaby-Kay}}, \citenamefont {{Marques}}, \citenamefont
  {{McMahon}}, \citenamefont {{Mehta}}, \citenamefont {{Menanteau}},
  \citenamefont {{Moodley}}, \citenamefont {{Morris}}, \citenamefont
  {{Mroczkowski}}, \citenamefont {{Naess}}, \citenamefont {{Namikawa}},
  \citenamefont {{Nati}}, \citenamefont {{Newburgh}}, \citenamefont {{Nicola}},
  \citenamefont {{Niemack}}, \citenamefont {{Nolta}}, \citenamefont
  {{Orlowski-Scherer}}, \citenamefont {{Page}}, \citenamefont {{Pandey}},
  \citenamefont {{Partridge}}, \citenamefont {{Prince}}, \citenamefont
  {{Puddu}}, \citenamefont {{Radiconi}}, \citenamefont {{Robertson}},
  \citenamefont {{Rojas}}, \citenamefont {{Sakuma}}, \citenamefont
  {{Salatino}}, \citenamefont {{Schaan}}, \citenamefont {{Schmitt}},
  \citenamefont {{Sehgal}}, \citenamefont {{Shaikh}}, \citenamefont {{Sierra}},
  \citenamefont {{Sievers}}, \citenamefont {{Sif{\'o}n}}, \citenamefont
  {{Simon}}, \citenamefont {{Sonka}}, \citenamefont {{Spergel}}, \citenamefont
  {{Staggs}}, \citenamefont {{Storer}}, \citenamefont {{Switzer}},
  \citenamefont {{Tampier}}, \citenamefont {{Thornton}}, \citenamefont
  {{Trac}}, \citenamefont {{Treu}}, \citenamefont {{Tucker}}, \citenamefont
  {{Ulluom}}, \citenamefont {{Vale}}, \citenamefont {{Van Engelen}},
  \citenamefont {{Van Lanen}}, \citenamefont {{van Marrewijk}}, \citenamefont
  {{Vargas}}, \citenamefont {{Vavagiakis}}, \citenamefont {{Wagoner}},
  \citenamefont {{Wang}}, \citenamefont {{Wenzl}}, \citenamefont {{Wollack}},
  \citenamefont {{Xu}}, \citenamefont {{Zago}},\ and\ \citenamefont
  {{Zhang}}}]{ACT:2023dou}%
  \BibitemOpen
  \bibfield  {author} {\bibinfo {author} {\bibfnamefont {F.~J.}\ \bibnamefont
  {{Qu}}}, \bibinfo {author} {\bibfnamefont {B.~D.}\ \bibnamefont {{Sherwin}}},
  \bibinfo {author} {\bibfnamefont {M.~S.}\ \bibnamefont {{Madhavacheril}}},
  \bibinfo {author} {\bibfnamefont {D.}~\bibnamefont {{Han}}}, \bibinfo
  {author} {\bibfnamefont {K.~T.}\ \bibnamefont {{Crowley}}}, \bibinfo {author}
  {\bibfnamefont {I.}~\bibnamefont {{Abril-Cabezas}}}, \bibinfo {author}
  {\bibfnamefont {P.~A.~R.}\ \bibnamefont {{Ade}}}, \bibinfo {author}
  {\bibfnamefont {S.}~\bibnamefont {{Aiola}}}, \bibinfo {author} {\bibfnamefont
  {T.}~\bibnamefont {{Alford}}}, \bibinfo {author} {\bibfnamefont
  {M.}~\bibnamefont {{Amiri}}}, \bibinfo {author} {\bibfnamefont
  {S.}~\bibnamefont {{Amodeo}}}, \bibinfo {author} {\bibfnamefont
  {R.}~\bibnamefont {{An}}}, \bibinfo {author} {\bibfnamefont {Z.}~\bibnamefont
  {{Atkins}}}, \bibinfo {author} {\bibfnamefont {J.~E.}\ \bibnamefont
  {{Austermann}}}, \bibinfo {author} {\bibfnamefont {N.}~\bibnamefont
  {{Battaglia}}}, \bibinfo {author} {\bibfnamefont {E.~S.}\ \bibnamefont
  {{Battistelli}}}, \bibinfo {author} {\bibfnamefont {J.~A.}\ \bibnamefont
  {{Beall}}}, \bibinfo {author} {\bibfnamefont {R.}~\bibnamefont {{Bean}}},
  \bibinfo {author} {\bibfnamefont {B.}~\bibnamefont {{Beringue}}}, \bibinfo
  {author} {\bibfnamefont {T.}~\bibnamefont {{Bhandarkar}}}, \bibinfo {author}
  {\bibfnamefont {E.}~\bibnamefont {{Biermann}}}, \bibinfo {author}
  {\bibfnamefont {B.}~\bibnamefont {{Bolliet}}}, \bibinfo {author}
  {\bibfnamefont {J.~R.}\ \bibnamefont {{Bond}}}, \bibinfo {author}
  {\bibfnamefont {H.}~\bibnamefont {{Cai}}}, \bibinfo {author} {\bibfnamefont
  {E.}~\bibnamefont {{Calabrese}}}, \bibinfo {author} {\bibfnamefont
  {V.}~\bibnamefont {{Calafut}}}, \bibinfo {author} {\bibfnamefont
  {V.}~\bibnamefont {{Capalbo}}}, \bibinfo {author} {\bibfnamefont
  {F.}~\bibnamefont {{Carrero}}}, \bibinfo {author} {\bibfnamefont
  {J.}~\bibnamefont {{Carron}}}, \bibinfo {author} {\bibfnamefont
  {A.}~\bibnamefont {{Challinor}}}, \bibinfo {author} {\bibfnamefont {G.~E.}\
  \bibnamefont {{Chesmore}}}, \bibinfo {author} {\bibfnamefont {H.-m.}\
  \bibnamefont {{Cho}}}, \bibinfo {author} {\bibfnamefont {S.~K.}\ \bibnamefont
  {{Choi}}}, \bibinfo {author} {\bibfnamefont {S.~E.}\ \bibnamefont {{Clark}}},
  \bibinfo {author} {\bibfnamefont {R.}~\bibnamefont {{C{\'o}rdova Rosado}}},
  \bibinfo {author} {\bibfnamefont {N.~F.}\ \bibnamefont {{Cothard}}}, \bibinfo
  {author} {\bibfnamefont {K.}~\bibnamefont {{Coughlin}}}, \bibinfo {author}
  {\bibfnamefont {W.}~\bibnamefont {{Coulton}}}, \bibinfo {author}
  {\bibfnamefont {R.}~\bibnamefont {{Dalal}}}, \bibinfo {author} {\bibfnamefont
  {O.}~\bibnamefont {{Darwish}}}, \bibinfo {author} {\bibfnamefont {M.~J.}\
  \bibnamefont {{Devlin}}}, \bibinfo {author} {\bibfnamefont {S.}~\bibnamefont
  {{Dicker}}}, \bibinfo {author} {\bibfnamefont {P.}~\bibnamefont {{Doze}}},
  \bibinfo {author} {\bibfnamefont {C.~J.}\ \bibnamefont {{Duell}}}, \bibinfo
  {author} {\bibfnamefont {S.~M.}\ \bibnamefont {{Duff}}}, \bibinfo {author}
  {\bibfnamefont {A.~J.}\ \bibnamefont {{Duivenvoorden}}}, \bibinfo {author}
  {\bibfnamefont {J.}~\bibnamefont {{Dunkley}}}, \bibinfo {author}
  {\bibfnamefont {R.}~\bibnamefont {{D{\"u}nner}}}, \bibinfo {author}
  {\bibfnamefont {V.}~\bibnamefont {{Fanfani}}}, \bibinfo {author}
  {\bibfnamefont {M.}~\bibnamefont {{Fankhanel}}}, \bibinfo {author}
  {\bibfnamefont {G.}~\bibnamefont {{Farren}}}, \bibinfo {author}
  {\bibfnamefont {S.}~\bibnamefont {{Ferraro}}}, \bibinfo {author}
  {\bibfnamefont {R.}~\bibnamefont {{Freundt}}}, \bibinfo {author}
  {\bibfnamefont {B.}~\bibnamefont {{Fuzia}}}, \bibinfo {author} {\bibfnamefont
  {P.~A.}\ \bibnamefont {{Gallardo}}}, \bibinfo {author} {\bibfnamefont
  {X.}~\bibnamefont {{Garrido}}}, \bibinfo {author} {\bibfnamefont
  {V.}~\bibnamefont {{Gluscevic}}}, \bibinfo {author} {\bibfnamefont {J.~E.}\
  \bibnamefont {{Golec}}}, \bibinfo {author} {\bibfnamefont {Y.}~\bibnamefont
  {{Guan}}}, \bibinfo {author} {\bibfnamefont {M.}~\bibnamefont {{Halpern}}},
  \bibinfo {author} {\bibfnamefont {I.}~\bibnamefont {{Harrison}}}, \bibinfo
  {author} {\bibfnamefont {M.}~\bibnamefont {{Hasselfield}}}, \bibinfo {author}
  {\bibfnamefont {E.}~\bibnamefont {{Healy}}}, \bibinfo {author} {\bibfnamefont
  {S.}~\bibnamefont {{Henderson}}}, \bibinfo {author} {\bibfnamefont
  {B.}~\bibnamefont {{Hensley}}}, \bibinfo {author} {\bibfnamefont
  {C.}~\bibnamefont {{Herv{\'\i}as-Caimapo}}}, \bibinfo {author} {\bibfnamefont
  {J.~C.}\ \bibnamefont {{Hill}}}, \bibinfo {author} {\bibfnamefont {G.~C.}\
  \bibnamefont {{Hilton}}}, \bibinfo {author} {\bibfnamefont {M.}~\bibnamefont
  {{Hilton}}}, \bibinfo {author} {\bibfnamefont {A.~D.}\ \bibnamefont
  {{Hincks}}}, \bibinfo {author} {\bibfnamefont {R.}~\bibnamefont
  {{Hlo{\v{z}}ek}}}, \bibinfo {author} {\bibfnamefont {S.-P.~P.}\ \bibnamefont
  {{Ho}}}, \bibinfo {author} {\bibfnamefont {Z.~B.}\ \bibnamefont {{Huber}}},
  \bibinfo {author} {\bibfnamefont {J.}~\bibnamefont {{Hubmayr}}}, \bibinfo
  {author} {\bibfnamefont {K.~M.}\ \bibnamefont {{Huffenberger}}}, \bibinfo
  {author} {\bibfnamefont {J.~P.}\ \bibnamefont {{Hughes}}}, \bibinfo {author}
  {\bibfnamefont {K.}~\bibnamefont {{Irwin}}}, \bibinfo {author} {\bibfnamefont
  {G.}~\bibnamefont {{Isopi}}}, \bibinfo {author} {\bibfnamefont {H.~T.}\
  \bibnamefont {{Jense}}}, \bibinfo {author} {\bibfnamefont {B.}~\bibnamefont
  {{Keller}}}, \bibinfo {author} {\bibfnamefont {J.}~\bibnamefont {{Kim}}},
  \bibinfo {author} {\bibfnamefont {K.}~\bibnamefont {{Knowles}}}, \bibinfo
  {author} {\bibfnamefont {B.~J.}\ \bibnamefont {{Koopman}}}, \bibinfo {author}
  {\bibfnamefont {A.}~\bibnamefont {{Kosowsky}}}, \bibinfo {author}
  {\bibfnamefont {D.}~\bibnamefont {{Kramer}}}, \bibinfo {author}
  {\bibfnamefont {A.}~\bibnamefont {{Kusiak}}}, \bibinfo {author}
  {\bibfnamefont {A.}~\bibnamefont {{La Posta}}}, \bibinfo {author}
  {\bibfnamefont {A.}~\bibnamefont {{Lague}}}, \bibinfo {author} {\bibfnamefont
  {V.}~\bibnamefont {{Lakey}}}, \bibinfo {author} {\bibfnamefont
  {E.}~\bibnamefont {{Lee}}}, \bibinfo {author} {\bibfnamefont
  {Z.}~\bibnamefont {{Li}}}, \bibinfo {author} {\bibfnamefont {Y.}~\bibnamefont
  {{Li}}}, \bibinfo {author} {\bibfnamefont {M.}~\bibnamefont {{Limon}}},
  \bibinfo {author} {\bibfnamefont {M.}~\bibnamefont {{Lokken}}}, \bibinfo
  {author} {\bibfnamefont {T.}~\bibnamefont {{Louis}}}, \bibinfo {author}
  {\bibfnamefont {M.}~\bibnamefont {{Lungu}}}, \bibinfo {author} {\bibfnamefont
  {N.}~\bibnamefont {{MacCrann}}}, \bibinfo {author} {\bibfnamefont
  {A.}~\bibnamefont {{MacInnis}}}, \bibinfo {author} {\bibfnamefont
  {D.}~\bibnamefont {{Maldonado}}}, \bibinfo {author} {\bibfnamefont
  {F.}~\bibnamefont {{Maldonado}}}, \bibinfo {author} {\bibfnamefont
  {M.}~\bibnamefont {{Mallaby-Kay}}}, \bibinfo {author} {\bibfnamefont {G.~A.}\
  \bibnamefont {{Marques}}}, \bibinfo {author} {\bibfnamefont {J.}~\bibnamefont
  {{McMahon}}}, \bibinfo {author} {\bibfnamefont {Y.}~\bibnamefont {{Mehta}}},
  \bibinfo {author} {\bibfnamefont {F.}~\bibnamefont {{Menanteau}}}, \bibinfo
  {author} {\bibfnamefont {K.}~\bibnamefont {{Moodley}}}, \bibinfo {author}
  {\bibfnamefont {T.~W.}\ \bibnamefont {{Morris}}}, \bibinfo {author}
  {\bibfnamefont {T.}~\bibnamefont {{Mroczkowski}}}, \bibinfo {author}
  {\bibfnamefont {S.}~\bibnamefont {{Naess}}}, \bibinfo {author} {\bibfnamefont
  {T.}~\bibnamefont {{Namikawa}}}, \bibinfo {author} {\bibfnamefont
  {F.}~\bibnamefont {{Nati}}}, \bibinfo {author} {\bibfnamefont
  {L.}~\bibnamefont {{Newburgh}}}, \bibinfo {author} {\bibfnamefont
  {A.}~\bibnamefont {{Nicola}}}, \bibinfo {author} {\bibfnamefont {M.~D.}\
  \bibnamefont {{Niemack}}}, \bibinfo {author} {\bibfnamefont {M.~R.}\
  \bibnamefont {{Nolta}}}, \bibinfo {author} {\bibfnamefont {J.}~\bibnamefont
  {{Orlowski-Scherer}}}, \bibinfo {author} {\bibfnamefont {L.~A.}\ \bibnamefont
  {{Page}}}, \bibinfo {author} {\bibfnamefont {S.}~\bibnamefont {{Pandey}}},
  \bibinfo {author} {\bibfnamefont {B.}~\bibnamefont {{Partridge}}}, \bibinfo
  {author} {\bibfnamefont {H.}~\bibnamefont {{Prince}}}, \bibinfo {author}
  {\bibfnamefont {R.}~\bibnamefont {{Puddu}}}, \bibinfo {author} {\bibfnamefont
  {F.}~\bibnamefont {{Radiconi}}}, \bibinfo {author} {\bibfnamefont
  {N.}~\bibnamefont {{Robertson}}}, \bibinfo {author} {\bibfnamefont
  {F.}~\bibnamefont {{Rojas}}}, \bibinfo {author} {\bibfnamefont
  {T.}~\bibnamefont {{Sakuma}}}, \bibinfo {author} {\bibfnamefont
  {M.}~\bibnamefont {{Salatino}}}, \bibinfo {author} {\bibfnamefont
  {E.}~\bibnamefont {{Schaan}}}, \bibinfo {author} {\bibfnamefont {B.~L.}\
  \bibnamefont {{Schmitt}}}, \bibinfo {author} {\bibfnamefont {N.}~\bibnamefont
  {{Sehgal}}}, \bibinfo {author} {\bibfnamefont {S.}~\bibnamefont {{Shaikh}}},
  \bibinfo {author} {\bibfnamefont {C.}~\bibnamefont {{Sierra}}}, \bibinfo
  {author} {\bibfnamefont {J.}~\bibnamefont {{Sievers}}}, \bibinfo {author}
  {\bibfnamefont {C.}~\bibnamefont {{Sif{\'o}n}}}, \bibinfo {author}
  {\bibfnamefont {S.}~\bibnamefont {{Simon}}}, \bibinfo {author} {\bibfnamefont
  {R.}~\bibnamefont {{Sonka}}}, \bibinfo {author} {\bibfnamefont {D.~N.}\
  \bibnamefont {{Spergel}}}, \bibinfo {author} {\bibfnamefont {S.~T.}\
  \bibnamefont {{Staggs}}}, \bibinfo {author} {\bibfnamefont {E.}~\bibnamefont
  {{Storer}}}, \bibinfo {author} {\bibfnamefont {E.~R.}\ \bibnamefont
  {{Switzer}}}, \bibinfo {author} {\bibfnamefont {N.}~\bibnamefont
  {{Tampier}}}, \bibinfo {author} {\bibfnamefont {R.}~\bibnamefont
  {{Thornton}}}, \bibinfo {author} {\bibfnamefont {H.}~\bibnamefont {{Trac}}},
  \bibinfo {author} {\bibfnamefont {J.}~\bibnamefont {{Treu}}}, \bibinfo
  {author} {\bibfnamefont {C.}~\bibnamefont {{Tucker}}}, \bibinfo {author}
  {\bibfnamefont {J.}~\bibnamefont {{Ulluom}}}, \bibinfo {author}
  {\bibfnamefont {L.~R.}\ \bibnamefont {{Vale}}}, \bibinfo {author}
  {\bibfnamefont {A.}~\bibnamefont {{Van Engelen}}}, \bibinfo {author}
  {\bibfnamefont {J.}~\bibnamefont {{Van Lanen}}}, \bibinfo {author}
  {\bibfnamefont {J.}~\bibnamefont {{van Marrewijk}}}, \bibinfo {author}
  {\bibfnamefont {C.}~\bibnamefont {{Vargas}}}, \bibinfo {author}
  {\bibfnamefont {E.~M.}\ \bibnamefont {{Vavagiakis}}}, \bibinfo {author}
  {\bibfnamefont {K.}~\bibnamefont {{Wagoner}}}, \bibinfo {author}
  {\bibfnamefont {Y.}~\bibnamefont {{Wang}}}, \bibinfo {author} {\bibfnamefont
  {L.}~\bibnamefont {{Wenzl}}}, \bibinfo {author} {\bibfnamefont {E.~J.}\
  \bibnamefont {{Wollack}}}, \bibinfo {author} {\bibfnamefont {Z.}~\bibnamefont
  {{Xu}}}, \bibinfo {author} {\bibfnamefont {F.}~\bibnamefont {{Zago}}}, \ and\
  \bibinfo {author} {\bibfnamefont {K.}~\bibnamefont {{Zhang}}},\ }\href
  {\doibase 10.48550/arXiv.2304.05202} {\bibfield  {journal} {\bibinfo
  {journal} {arXiv e-prints}\ ,\ \bibinfo {eid} {arXiv:2304.05202}} (\bibinfo
  {year} {2023})},\ \Eprint {http://arxiv.org/abs/2304.05202} {arXiv:2304.05202
  [astro-ph.CO]} \BibitemShut {NoStop}%
\bibitem [{\citenamefont {{Madhavacheril}}\ \emph {et~al.}(2023)\citenamefont
  {{Madhavacheril}}, \citenamefont {{Qu}}, \citenamefont {{Sherwin}},
  \citenamefont {{MacCrann}}, \citenamefont {{Li}}, \citenamefont
  {{Abril-Cabezas}}, \citenamefont {{Ade}}, \citenamefont {{Aiola}},
  \citenamefont {{Alford}}, \citenamefont {{Amiri}}, \citenamefont {{Amodeo}},
  \citenamefont {{An}}, \citenamefont {{Atkins}}, \citenamefont {{Austermann}},
  \citenamefont {{Battaglia}}, \citenamefont {{Battistelli}}, \citenamefont
  {{Beall}}, \citenamefont {{Bean}}, \citenamefont {{Beringue}}, \citenamefont
  {{Bhandarkar}}, \citenamefont {{Biermann}}, \citenamefont {{Bolliet}},
  \citenamefont {{Bond}}, \citenamefont {{Cai}}, \citenamefont {{Calabrese}},
  \citenamefont {{Calafut}}, \citenamefont {{Capalbo}}, \citenamefont
  {{Carrero}}, \citenamefont {{Challinor}}, \citenamefont {{Chesmore}},
  \citenamefont {{Cho}}, \citenamefont {{Choi}}, \citenamefont {{Clark}},
  \citenamefont {{C{\'o}rdova Rosado}}, \citenamefont {{Cothard}},
  \citenamefont {{Coughlin}}, \citenamefont {{Coulton}}, \citenamefont
  {{Crowley}}, \citenamefont {{Dalal}}, \citenamefont {{Darwish}},
  \citenamefont {{Devlin}}, \citenamefont {{Dicker}}, \citenamefont {{Doze}},
  \citenamefont {{Duell}}, \citenamefont {{Duff}}, \citenamefont
  {{Duivenvoorden}}, \citenamefont {{Dunkley}}, \citenamefont {{D{\"u}nner}},
  \citenamefont {{Fanfani}}, \citenamefont {{Fankhanel}}, \citenamefont
  {{Farren}}, \citenamefont {{Ferraro}}, \citenamefont {{Freundt}},
  \citenamefont {{Fuzia}}, \citenamefont {{Gallardo}}, \citenamefont
  {{Garrido}}, \citenamefont {{Givans}}, \citenamefont {{Gluscevic}},
  \citenamefont {{Golec}}, \citenamefont {{Guan}}, \citenamefont {{Hall}},
  \citenamefont {{Halpern}}, \citenamefont {{Han}}, \citenamefont {{Harrison}},
  \citenamefont {{Hasselfield}}, \citenamefont {{Healy}}, \citenamefont
  {{Henderson}}, \citenamefont {{Hensley}}, \citenamefont
  {{Herv{\'\i}as-Caimapo}}, \citenamefont {{Hill}}, \citenamefont {{Hilton}},
  \citenamefont {{Hilton}}, \citenamefont {{Hincks}}, \citenamefont
  {{Hlo{\v{z}}ek}}, \citenamefont {{Ho}}, \citenamefont {{Huber}},
  \citenamefont {{Hubmayr}}, \citenamefont {{Huffenberger}}, \citenamefont
  {{Hughes}}, \citenamefont {{Irwin}}, \citenamefont {{Isopi}}, \citenamefont
  {{Jense}}, \citenamefont {{Keller}}, \citenamefont {{Kim}}, \citenamefont
  {{Knowles}}, \citenamefont {{Koopman}}, \citenamefont {{Kosowsky}},
  \citenamefont {{Kramer}}, \citenamefont {{Kusiak}}, \citenamefont {{La
  Posta}}, \citenamefont {{Lague}}, \citenamefont {{Lakey}}, \citenamefont
  {{Lee}}, \citenamefont {{Li}}, \citenamefont {{Limon}}, \citenamefont
  {{Lokken}}, \citenamefont {{Louis}}, \citenamefont {{Lungu}}, \citenamefont
  {{MacInnis}}, \citenamefont {{Maldonado}}, \citenamefont {{Maldonado}},
  \citenamefont {{Mallaby-Kay}}, \citenamefont {{Marques}}, \citenamefont
  {{McMahon}}, \citenamefont {{Mehta}}, \citenamefont {{Menanteau}},
  \citenamefont {{Moodley}}, \citenamefont {{Morris}}, \citenamefont
  {{Mroczkowski}}, \citenamefont {{Naess}}, \citenamefont {{Namikawa}},
  \citenamefont {{Nati}}, \citenamefont {{Newburgh}}, \citenamefont {{Nicola}},
  \citenamefont {{Niemack}}, \citenamefont {{Nolta}}, \citenamefont
  {{Orlowski-Scherer}}, \citenamefont {{Page}}, \citenamefont {{Pandey}},
  \citenamefont {{Partridge}}, \citenamefont {{Prince}}, \citenamefont
  {{Puddu}}, \citenamefont {{Radiconi}}, \citenamefont {{Robertson}},
  \citenamefont {{Rojas}}, \citenamefont {{Sakuma}}, \citenamefont
  {{Salatino}}, \citenamefont {{Schaan}}, \citenamefont {{Schmitt}},
  \citenamefont {{Sehgal}}, \citenamefont {{Shaikh}}, \citenamefont {{Sierra}},
  \citenamefont {{Sievers}}, \citenamefont {{Sif{\'o}n}}, \citenamefont
  {{Simon}}, \citenamefont {{Sonka}}, \citenamefont {{Spergel}}, \citenamefont
  {{Staggs}}, \citenamefont {{Storer}}, \citenamefont {{Switzer}},
  \citenamefont {{Tampier}}, \citenamefont {{Thornton}}, \citenamefont
  {{Trac}}, \citenamefont {{Treu}}, \citenamefont {{Tucker}}, \citenamefont
  {{Ulluom}}, \citenamefont {{Vale}}, \citenamefont {{Van Engelen}},
  \citenamefont {{Van Lanen}}, \citenamefont {{van Marrewijk}}, \citenamefont
  {{Vargas}}, \citenamefont {{Vavagiakis}}, \citenamefont {{Wagoner}},
  \citenamefont {{Wang}}, \citenamefont {{Wenzl}}, \citenamefont {{Wollack}},
  \citenamefont {{Xu}}, \citenamefont {{Zago}},\ and\ \citenamefont
  {{Zhang}}}]{ACT:2023kun}%
  \BibitemOpen
  \bibfield  {author} {\bibinfo {author} {\bibfnamefont {M.~S.}\ \bibnamefont
  {{Madhavacheril}}}, \bibinfo {author} {\bibfnamefont {F.~J.}\ \bibnamefont
  {{Qu}}}, \bibinfo {author} {\bibfnamefont {B.~D.}\ \bibnamefont {{Sherwin}}},
  \bibinfo {author} {\bibfnamefont {N.}~\bibnamefont {{MacCrann}}}, \bibinfo
  {author} {\bibfnamefont {Y.}~\bibnamefont {{Li}}}, \bibinfo {author}
  {\bibfnamefont {I.}~\bibnamefont {{Abril-Cabezas}}}, \bibinfo {author}
  {\bibfnamefont {P.~A.~R.}\ \bibnamefont {{Ade}}}, \bibinfo {author}
  {\bibfnamefont {S.}~\bibnamefont {{Aiola}}}, \bibinfo {author} {\bibfnamefont
  {T.}~\bibnamefont {{Alford}}}, \bibinfo {author} {\bibfnamefont
  {M.}~\bibnamefont {{Amiri}}}, \bibinfo {author} {\bibfnamefont
  {S.}~\bibnamefont {{Amodeo}}}, \bibinfo {author} {\bibfnamefont
  {R.}~\bibnamefont {{An}}}, \bibinfo {author} {\bibfnamefont {Z.}~\bibnamefont
  {{Atkins}}}, \bibinfo {author} {\bibfnamefont {J.~E.}\ \bibnamefont
  {{Austermann}}}, \bibinfo {author} {\bibfnamefont {N.}~\bibnamefont
  {{Battaglia}}}, \bibinfo {author} {\bibfnamefont {E.~S.}\ \bibnamefont
  {{Battistelli}}}, \bibinfo {author} {\bibfnamefont {J.~A.}\ \bibnamefont
  {{Beall}}}, \bibinfo {author} {\bibfnamefont {R.}~\bibnamefont {{Bean}}},
  \bibinfo {author} {\bibfnamefont {B.}~\bibnamefont {{Beringue}}}, \bibinfo
  {author} {\bibfnamefont {T.}~\bibnamefont {{Bhandarkar}}}, \bibinfo {author}
  {\bibfnamefont {E.}~\bibnamefont {{Biermann}}}, \bibinfo {author}
  {\bibfnamefont {B.}~\bibnamefont {{Bolliet}}}, \bibinfo {author}
  {\bibfnamefont {J.~R.}\ \bibnamefont {{Bond}}}, \bibinfo {author}
  {\bibfnamefont {H.}~\bibnamefont {{Cai}}}, \bibinfo {author} {\bibfnamefont
  {E.}~\bibnamefont {{Calabrese}}}, \bibinfo {author} {\bibfnamefont
  {V.}~\bibnamefont {{Calafut}}}, \bibinfo {author} {\bibfnamefont
  {V.}~\bibnamefont {{Capalbo}}}, \bibinfo {author} {\bibfnamefont
  {F.}~\bibnamefont {{Carrero}}}, \bibinfo {author} {\bibfnamefont
  {A.}~\bibnamefont {{Challinor}}}, \bibinfo {author} {\bibfnamefont {G.~E.}\
  \bibnamefont {{Chesmore}}}, \bibinfo {author} {\bibfnamefont {H.-m.}\
  \bibnamefont {{Cho}}}, \bibinfo {author} {\bibfnamefont {S.~K.}\ \bibnamefont
  {{Choi}}}, \bibinfo {author} {\bibfnamefont {S.~E.}\ \bibnamefont {{Clark}}},
  \bibinfo {author} {\bibfnamefont {R.}~\bibnamefont {{C{\'o}rdova Rosado}}},
  \bibinfo {author} {\bibfnamefont {N.~F.}\ \bibnamefont {{Cothard}}}, \bibinfo
  {author} {\bibfnamefont {K.}~\bibnamefont {{Coughlin}}}, \bibinfo {author}
  {\bibfnamefont {W.}~\bibnamefont {{Coulton}}}, \bibinfo {author}
  {\bibfnamefont {K.~T.}\ \bibnamefont {{Crowley}}}, \bibinfo {author}
  {\bibfnamefont {R.}~\bibnamefont {{Dalal}}}, \bibinfo {author} {\bibfnamefont
  {O.}~\bibnamefont {{Darwish}}}, \bibinfo {author} {\bibfnamefont {M.~J.}\
  \bibnamefont {{Devlin}}}, \bibinfo {author} {\bibfnamefont {S.}~\bibnamefont
  {{Dicker}}}, \bibinfo {author} {\bibfnamefont {P.}~\bibnamefont {{Doze}}},
  \bibinfo {author} {\bibfnamefont {C.~J.}\ \bibnamefont {{Duell}}}, \bibinfo
  {author} {\bibfnamefont {S.~M.}\ \bibnamefont {{Duff}}}, \bibinfo {author}
  {\bibfnamefont {A.~J.}\ \bibnamefont {{Duivenvoorden}}}, \bibinfo {author}
  {\bibfnamefont {J.}~\bibnamefont {{Dunkley}}}, \bibinfo {author}
  {\bibfnamefont {R.}~\bibnamefont {{D{\"u}nner}}}, \bibinfo {author}
  {\bibfnamefont {V.}~\bibnamefont {{Fanfani}}}, \bibinfo {author}
  {\bibfnamefont {M.}~\bibnamefont {{Fankhanel}}}, \bibinfo {author}
  {\bibfnamefont {G.}~\bibnamefont {{Farren}}}, \bibinfo {author}
  {\bibfnamefont {S.}~\bibnamefont {{Ferraro}}}, \bibinfo {author}
  {\bibfnamefont {R.}~\bibnamefont {{Freundt}}}, \bibinfo {author}
  {\bibfnamefont {B.}~\bibnamefont {{Fuzia}}}, \bibinfo {author} {\bibfnamefont
  {P.~A.}\ \bibnamefont {{Gallardo}}}, \bibinfo {author} {\bibfnamefont
  {X.}~\bibnamefont {{Garrido}}}, \bibinfo {author} {\bibfnamefont
  {J.}~\bibnamefont {{Givans}}}, \bibinfo {author} {\bibfnamefont
  {V.}~\bibnamefont {{Gluscevic}}}, \bibinfo {author} {\bibfnamefont {J.~E.}\
  \bibnamefont {{Golec}}}, \bibinfo {author} {\bibfnamefont {Y.}~\bibnamefont
  {{Guan}}}, \bibinfo {author} {\bibfnamefont {K.~R.}\ \bibnamefont {{Hall}}},
  \bibinfo {author} {\bibfnamefont {M.}~\bibnamefont {{Halpern}}}, \bibinfo
  {author} {\bibfnamefont {D.}~\bibnamefont {{Han}}}, \bibinfo {author}
  {\bibfnamefont {I.}~\bibnamefont {{Harrison}}}, \bibinfo {author}
  {\bibfnamefont {M.}~\bibnamefont {{Hasselfield}}}, \bibinfo {author}
  {\bibfnamefont {E.}~\bibnamefont {{Healy}}}, \bibinfo {author} {\bibfnamefont
  {S.}~\bibnamefont {{Henderson}}}, \bibinfo {author} {\bibfnamefont
  {B.}~\bibnamefont {{Hensley}}}, \bibinfo {author} {\bibfnamefont
  {C.}~\bibnamefont {{Herv{\'\i}as-Caimapo}}}, \bibinfo {author} {\bibfnamefont
  {J.~C.}\ \bibnamefont {{Hill}}}, \bibinfo {author} {\bibfnamefont {G.~C.}\
  \bibnamefont {{Hilton}}}, \bibinfo {author} {\bibfnamefont {M.}~\bibnamefont
  {{Hilton}}}, \bibinfo {author} {\bibfnamefont {A.~D.}\ \bibnamefont
  {{Hincks}}}, \bibinfo {author} {\bibfnamefont {R.}~\bibnamefont
  {{Hlo{\v{z}}ek}}}, \bibinfo {author} {\bibfnamefont {S.-P.~P.}\ \bibnamefont
  {{Ho}}}, \bibinfo {author} {\bibfnamefont {Z.~B.}\ \bibnamefont {{Huber}}},
  \bibinfo {author} {\bibfnamefont {J.}~\bibnamefont {{Hubmayr}}}, \bibinfo
  {author} {\bibfnamefont {K.~M.}\ \bibnamefont {{Huffenberger}}}, \bibinfo
  {author} {\bibfnamefont {J.~P.}\ \bibnamefont {{Hughes}}}, \bibinfo {author}
  {\bibfnamefont {K.}~\bibnamefont {{Irwin}}}, \bibinfo {author} {\bibfnamefont
  {G.}~\bibnamefont {{Isopi}}}, \bibinfo {author} {\bibfnamefont {H.~T.}\
  \bibnamefont {{Jense}}}, \bibinfo {author} {\bibfnamefont {B.}~\bibnamefont
  {{Keller}}}, \bibinfo {author} {\bibfnamefont {J.}~\bibnamefont {{Kim}}},
  \bibinfo {author} {\bibfnamefont {K.}~\bibnamefont {{Knowles}}}, \bibinfo
  {author} {\bibfnamefont {B.~J.}\ \bibnamefont {{Koopman}}}, \bibinfo {author}
  {\bibfnamefont {A.}~\bibnamefont {{Kosowsky}}}, \bibinfo {author}
  {\bibfnamefont {D.}~\bibnamefont {{Kramer}}}, \bibinfo {author}
  {\bibfnamefont {A.}~\bibnamefont {{Kusiak}}}, \bibinfo {author}
  {\bibfnamefont {A.}~\bibnamefont {{La Posta}}}, \bibinfo {author}
  {\bibfnamefont {A.}~\bibnamefont {{Lague}}}, \bibinfo {author} {\bibfnamefont
  {V.}~\bibnamefont {{Lakey}}}, \bibinfo {author} {\bibfnamefont
  {E.}~\bibnamefont {{Lee}}}, \bibinfo {author} {\bibfnamefont
  {Z.}~\bibnamefont {{Li}}}, \bibinfo {author} {\bibfnamefont {M.}~\bibnamefont
  {{Limon}}}, \bibinfo {author} {\bibfnamefont {M.}~\bibnamefont {{Lokken}}},
  \bibinfo {author} {\bibfnamefont {T.}~\bibnamefont {{Louis}}}, \bibinfo
  {author} {\bibfnamefont {M.}~\bibnamefont {{Lungu}}}, \bibinfo {author}
  {\bibfnamefont {A.}~\bibnamefont {{MacInnis}}}, \bibinfo {author}
  {\bibfnamefont {D.}~\bibnamefont {{Maldonado}}}, \bibinfo {author}
  {\bibfnamefont {F.}~\bibnamefont {{Maldonado}}}, \bibinfo {author}
  {\bibfnamefont {M.}~\bibnamefont {{Mallaby-Kay}}}, \bibinfo {author}
  {\bibfnamefont {G.~A.}\ \bibnamefont {{Marques}}}, \bibinfo {author}
  {\bibfnamefont {J.}~\bibnamefont {{McMahon}}}, \bibinfo {author}
  {\bibfnamefont {Y.}~\bibnamefont {{Mehta}}}, \bibinfo {author} {\bibfnamefont
  {F.}~\bibnamefont {{Menanteau}}}, \bibinfo {author} {\bibfnamefont
  {K.}~\bibnamefont {{Moodley}}}, \bibinfo {author} {\bibfnamefont {T.~W.}\
  \bibnamefont {{Morris}}}, \bibinfo {author} {\bibfnamefont {T.}~\bibnamefont
  {{Mroczkowski}}}, \bibinfo {author} {\bibfnamefont {S.}~\bibnamefont
  {{Naess}}}, \bibinfo {author} {\bibfnamefont {T.}~\bibnamefont {{Namikawa}}},
  \bibinfo {author} {\bibfnamefont {F.}~\bibnamefont {{Nati}}}, \bibinfo
  {author} {\bibfnamefont {L.}~\bibnamefont {{Newburgh}}}, \bibinfo {author}
  {\bibfnamefont {A.}~\bibnamefont {{Nicola}}}, \bibinfo {author}
  {\bibfnamefont {M.~D.}\ \bibnamefont {{Niemack}}}, \bibinfo {author}
  {\bibfnamefont {M.~R.}\ \bibnamefont {{Nolta}}}, \bibinfo {author}
  {\bibfnamefont {J.}~\bibnamefont {{Orlowski-Scherer}}}, \bibinfo {author}
  {\bibfnamefont {L.~A.}\ \bibnamefont {{Page}}}, \bibinfo {author}
  {\bibfnamefont {S.}~\bibnamefont {{Pandey}}}, \bibinfo {author}
  {\bibfnamefont {B.}~\bibnamefont {{Partridge}}}, \bibinfo {author}
  {\bibfnamefont {H.}~\bibnamefont {{Prince}}}, \bibinfo {author}
  {\bibfnamefont {R.}~\bibnamefont {{Puddu}}}, \bibinfo {author} {\bibfnamefont
  {F.}~\bibnamefont {{Radiconi}}}, \bibinfo {author} {\bibfnamefont
  {N.}~\bibnamefont {{Robertson}}}, \bibinfo {author} {\bibfnamefont
  {F.}~\bibnamefont {{Rojas}}}, \bibinfo {author} {\bibfnamefont
  {T.}~\bibnamefont {{Sakuma}}}, \bibinfo {author} {\bibfnamefont
  {M.}~\bibnamefont {{Salatino}}}, \bibinfo {author} {\bibfnamefont
  {E.}~\bibnamefont {{Schaan}}}, \bibinfo {author} {\bibfnamefont {B.~L.}\
  \bibnamefont {{Schmitt}}}, \bibinfo {author} {\bibfnamefont {N.}~\bibnamefont
  {{Sehgal}}}, \bibinfo {author} {\bibfnamefont {S.}~\bibnamefont {{Shaikh}}},
  \bibinfo {author} {\bibfnamefont {C.}~\bibnamefont {{Sierra}}}, \bibinfo
  {author} {\bibfnamefont {J.}~\bibnamefont {{Sievers}}}, \bibinfo {author}
  {\bibfnamefont {C.}~\bibnamefont {{Sif{\'o}n}}}, \bibinfo {author}
  {\bibfnamefont {S.}~\bibnamefont {{Simon}}}, \bibinfo {author} {\bibfnamefont
  {R.}~\bibnamefont {{Sonka}}}, \bibinfo {author} {\bibfnamefont {D.~N.}\
  \bibnamefont {{Spergel}}}, \bibinfo {author} {\bibfnamefont {S.~T.}\
  \bibnamefont {{Staggs}}}, \bibinfo {author} {\bibfnamefont {E.}~\bibnamefont
  {{Storer}}}, \bibinfo {author} {\bibfnamefont {E.~R.}\ \bibnamefont
  {{Switzer}}}, \bibinfo {author} {\bibfnamefont {N.}~\bibnamefont
  {{Tampier}}}, \bibinfo {author} {\bibfnamefont {R.}~\bibnamefont
  {{Thornton}}}, \bibinfo {author} {\bibfnamefont {H.}~\bibnamefont {{Trac}}},
  \bibinfo {author} {\bibfnamefont {J.}~\bibnamefont {{Treu}}}, \bibinfo
  {author} {\bibfnamefont {C.}~\bibnamefont {{Tucker}}}, \bibinfo {author}
  {\bibfnamefont {J.}~\bibnamefont {{Ulluom}}}, \bibinfo {author}
  {\bibfnamefont {L.~R.}\ \bibnamefont {{Vale}}}, \bibinfo {author}
  {\bibfnamefont {A.}~\bibnamefont {{Van Engelen}}}, \bibinfo {author}
  {\bibfnamefont {J.}~\bibnamefont {{Van Lanen}}}, \bibinfo {author}
  {\bibfnamefont {J.}~\bibnamefont {{van Marrewijk}}}, \bibinfo {author}
  {\bibfnamefont {C.}~\bibnamefont {{Vargas}}}, \bibinfo {author}
  {\bibfnamefont {E.~M.}\ \bibnamefont {{Vavagiakis}}}, \bibinfo {author}
  {\bibfnamefont {K.}~\bibnamefont {{Wagoner}}}, \bibinfo {author}
  {\bibfnamefont {Y.}~\bibnamefont {{Wang}}}, \bibinfo {author} {\bibfnamefont
  {L.}~\bibnamefont {{Wenzl}}}, \bibinfo {author} {\bibfnamefont {E.~J.}\
  \bibnamefont {{Wollack}}}, \bibinfo {author} {\bibfnamefont {Z.}~\bibnamefont
  {{Xu}}}, \bibinfo {author} {\bibfnamefont {F.}~\bibnamefont {{Zago}}}, \ and\
  \bibinfo {author} {\bibfnamefont {K.}~\bibnamefont {{Zhang}}},\ }\href
  {\doibase 10.48550/arXiv.2304.05203} {\bibfield  {journal} {\bibinfo
  {journal} {arXiv e-prints}\ ,\ \bibinfo {eid} {arXiv:2304.05203}} (\bibinfo
  {year} {2023})},\ \Eprint {http://arxiv.org/abs/2304.05203} {arXiv:2304.05203
  [astro-ph.CO]} \BibitemShut {NoStop}%
\bibitem [{\citenamefont {Carron}\ \emph {et~al.}(2022)\citenamefont {Carron},
  \citenamefont {Mirmelstein},\ and\ \citenamefont {Lewis}}]{Carron:2022eyg}%
  \BibitemOpen
  \bibfield  {author} {\bibinfo {author} {\bibfnamefont {J.}~\bibnamefont
  {Carron}}, \bibinfo {author} {\bibfnamefont {M.}~\bibnamefont {Mirmelstein}},
  \ and\ \bibinfo {author} {\bibfnamefont {A.}~\bibnamefont {Lewis}},\ }\href
  {\doibase 10.1088/1475-7516/2022/09/039} {\bibfield  {journal} {\bibinfo
  {journal} {JCAP}\ }\textbf {\bibinfo {volume} {09}},\ \bibinfo {pages} {039}
  (\bibinfo {year} {2022})},\ \Eprint {http://arxiv.org/abs/2206.07773}
  {arXiv:2206.07773 [astro-ph.CO]} \BibitemShut {NoStop}%
\bibitem [{\citenamefont {Dey}\ \emph {et~al.}(2019)\citenamefont {Dey} \emph
  {et~al.}}]{DESI:2018ymu}%
  \BibitemOpen
  \bibfield  {author} {\bibinfo {author} {\bibfnamefont {A.}~\bibnamefont
  {Dey}} \emph {et~al.} (\bibinfo {collaboration} {DESI}),\ }\href {\doibase
  10.3847/1538-3881/ab089d} {\bibfield  {journal} {\bibinfo  {journal} {Astron.
  J.}\ }\textbf {\bibinfo {volume} {157}},\ \bibinfo {pages} {168} (\bibinfo
  {year} {2019})},\ \Eprint {http://arxiv.org/abs/1804.08657} {arXiv:1804.08657
  [astro-ph.IM]} \BibitemShut {NoStop}%
\bibitem [{\citenamefont {Alam}\ \emph {et~al.}(2017)\citenamefont {Alam} \emph
  {et~al.}}]{BOSS:2016wmc}%
  \BibitemOpen
  \bibfield  {author} {\bibinfo {author} {\bibfnamefont {S.}~\bibnamefont
  {Alam}} \emph {et~al.} (\bibinfo {collaboration} {BOSS}),\ }\href {\doibase
  10.1093/mnras/stx721} {\bibfield  {journal} {\bibinfo  {journal} {Mon. Not.
  Roy. Astron. Soc.}\ }\textbf {\bibinfo {volume} {470}},\ \bibinfo {pages}
  {2617} (\bibinfo {year} {2017})},\ \Eprint {http://arxiv.org/abs/1607.03155}
  {arXiv:1607.03155 [astro-ph.CO]} \BibitemShut {NoStop}%
\bibitem [{\citenamefont {Hou}\ \emph {et~al.}(2020)\citenamefont {Hou} \emph
  {et~al.}}]{Hou:2020rse}%
  \BibitemOpen
  \bibfield  {author} {\bibinfo {author} {\bibfnamefont {J.}~\bibnamefont
  {Hou}} \emph {et~al.},\ }\href {\doibase 10.1093/mnras/staa3234} {\bibfield
  {journal} {\bibinfo  {journal} {Mon. Not. Roy. Astron. Soc.}\ }\textbf
  {\bibinfo {volume} {500}},\ \bibinfo {pages} {1201} (\bibinfo {year}
  {2020})},\ \Eprint {http://arxiv.org/abs/2007.08998} {arXiv:2007.08998
  [astro-ph.CO]} \BibitemShut {NoStop}%
\bibitem [{\citenamefont {Neveux}\ \emph {et~al.}(2020)\citenamefont {Neveux}
  \emph {et~al.}}]{Neveux:2020voa}%
  \BibitemOpen
  \bibfield  {author} {\bibinfo {author} {\bibfnamefont {R.}~\bibnamefont
  {Neveux}} \emph {et~al.},\ }\href {\doibase 10.1093/mnras/staa2780}
  {\bibfield  {journal} {\bibinfo  {journal} {Mon. Not. Roy. Astron. Soc.}\
  }\textbf {\bibinfo {volume} {499}},\ \bibinfo {pages} {210} (\bibinfo {year}
  {2020})},\ \Eprint {http://arxiv.org/abs/2007.08999} {arXiv:2007.08999
  [astro-ph.CO]} \BibitemShut {NoStop}%
\bibitem [{\citenamefont {{DESI Collaboration}}\ \emph
  {et~al.}(2024)\citenamefont {{DESI Collaboration}}, \citenamefont {{Adame}},
  \citenamefont {{Aguilar}}, \citenamefont {{Ahlen}}, \citenamefont {{Alam}},
  \citenamefont {{Alexander}}, \citenamefont {{Alvarez}}, \citenamefont
  {{Alves}}, \citenamefont {{Anand}}, \citenamefont {{Andrade}}, \citenamefont
  {{Armengaud}}, \citenamefont {{Avila}}, \citenamefont {{Aviles}},
  \citenamefont {{Awan}}, \citenamefont {{Bahr-Kalus}}, \citenamefont
  {{Bailey}}, \citenamefont {{Baltay}}, \citenamefont {{Bault}}, \citenamefont
  {{Behera}}, \citenamefont {{BenZvi}}, \citenamefont {{Bera}}, \citenamefont
  {{Beutler}}, \citenamefont {{Bianchi}}, \citenamefont {{Blake}},
  \citenamefont {{Blum}}, \citenamefont {{Brieden}}, \citenamefont
  {{Brodzeller}}, \citenamefont {{Brooks}}, \citenamefont {{Buckley-Geer}},
  \citenamefont {{Burtin}}, \citenamefont {{Calderon}}, \citenamefont
  {{Canning}}, \citenamefont {{Carnero Rosell}}, \citenamefont {{Cereskaite}},
  \citenamefont {{Cervantes-Cota}}, \citenamefont {{Chabanier}}, \citenamefont
  {{Chaussidon}}, \citenamefont {{Chaves-Montero}}, \citenamefont {{Chen}},
  \citenamefont {{Chen}}, \citenamefont {{Claybaugh}}, \citenamefont {{Cole}},
  \citenamefont {{Cuceu}}, \citenamefont {{Davis}}, \citenamefont {{Dawson}},
  \citenamefont {{de la Macorra}}, \citenamefont {{de Mattia}}, \citenamefont
  {{Deiosso}}, \citenamefont {{Dey}}, \citenamefont {{Dey}}, \citenamefont
  {{Ding}}, \citenamefont {{Doel}}, \citenamefont {{Edelstein}}, \citenamefont
  {{Eftekharzadeh}}, \citenamefont {{Eisenstein}}, \citenamefont {{Elliott}},
  \citenamefont {{Fagrelius}}, \citenamefont {{Fanning}}, \citenamefont
  {{Ferraro}}, \citenamefont {{Ereza}}, \citenamefont {{Findlay}},
  \citenamefont {{Flaugher}}, \citenamefont {{Font-Ribera}}, \citenamefont
  {{Forero-S{\'a}nchez}}, \citenamefont {{Forero-Romero}}, \citenamefont
  {{Frenk}}, \citenamefont {{Garcia-Quintero}}, \citenamefont
  {{Gazta{\~n}aga}}, \citenamefont {{Gil-Mar{\'\i}n}}, \citenamefont
  {{Gontcho}}, \citenamefont {{Gonzalez-Morales}}, \citenamefont
  {{Gonzalez-Perez}}, \citenamefont {{Gordon}}, \citenamefont {{Green}},
  \citenamefont {{Gruen}}, \citenamefont {{Gsponer}}, \citenamefont
  {{Gutierrez}}, \citenamefont {{Guy}}, \citenamefont {{Hadzhiyska}},
  \citenamefont {{Hahn}}, \citenamefont {{Hanif}}, \citenamefont
  {{Herrera-Alcantar}}, \citenamefont {{Honscheid}}, \citenamefont {{Howlett}},
  \citenamefont {{Huterer}}, \citenamefont {{Ir{\v{s}}i{\v{c}}}}, \citenamefont
  {{Ishak}}, \citenamefont {{Juneau}}, \citenamefont {{Kara{\c{c}}ayl{\i}}},
  \citenamefont {{Kehoe}}, \citenamefont {{Kent}}, \citenamefont {{Kirkby}},
  \citenamefont {{Kremin}}, \citenamefont {{Krolewski}}, \citenamefont {{Lai}},
  \citenamefont {{Lan}}, \citenamefont {{Landriau}}, \citenamefont {{Lang}},
  \citenamefont {{Lasker}}, \citenamefont {{Le Goff}}, \citenamefont {{Le
  Guillou}}, \citenamefont {{Leauthaud}}, \citenamefont {{Levi}}, \citenamefont
  {{Li}}, \citenamefont {{Linder}}, \citenamefont {{Lodha}}, \citenamefont
  {{Magneville}}, \citenamefont {{Manera}}, \citenamefont {{Margala}},
  \citenamefont {{Martini}}, \citenamefont {{Maus}}, \citenamefont
  {{McDonald}}, \citenamefont {{Medina-Varela}}, \citenamefont {{Meisner}},
  \citenamefont {{Mena-Fern{\'a}ndez}}, \citenamefont {{Miquel}}, \citenamefont
  {{Moon}}, \citenamefont {{Moore}}, \citenamefont {{Moustakas}}, \citenamefont
  {{Mudur}}, \citenamefont {{Mueller}}, \citenamefont
  {{Mu{\~n}oz-Guti{\'e}rrez}}, \citenamefont {{Myers}}, \citenamefont
  {{Nadathur}}, \citenamefont {{Napolitano}}, \citenamefont {{Neveux}},
  \citenamefont {{Newman}}, \citenamefont {{Nguyen}}, \citenamefont {{Nie}},
  \citenamefont {{Niz}}, \citenamefont {{Noriega}}, \citenamefont
  {{Padmanabhan}}, \citenamefont {{Paillas}}, \citenamefont
  {{Palanque-Delabrouille}}, \citenamefont {{Pan}}, \citenamefont {{Penmetsa}},
  \citenamefont {{Percival}}, \citenamefont {{Pieri}}, \citenamefont {{Pinon}},
  \citenamefont {{Poppett}}, \citenamefont {{Porredon}}, \citenamefont
  {{Prada}}, \citenamefont {{P{\'e}rez-Fern{\'a}ndez}}, \citenamefont
  {{P{\'e}rez-R{\`a}fols}}, \citenamefont {{Rabinowitz}}, \citenamefont
  {{Raichoor}}, \citenamefont {{Ram{\'\i}rez-P{\'e}rez}}, \citenamefont
  {{Ramirez-Solano}}, \citenamefont {{Ravoux}}, \citenamefont
  {{Rashkovetskyi}}, \citenamefont {{Rezaie}}, \citenamefont {{Rich}},
  \citenamefont {{Rocher}}, \citenamefont {{Rockosi}}, \citenamefont {{Roe}},
  \citenamefont {{Rosado-Marin}}, \citenamefont {{Ross}}, \citenamefont
  {{Rossi}}, \citenamefont {{Ruggeri}}, \citenamefont {{Ruhlmann-Kleider}},
  \citenamefont {{Samushia}}, \citenamefont {{Sanchez}}, \citenamefont
  {{Saulder}}, \citenamefont {{Schlafly}}, \citenamefont {{Schlegel}},
  \citenamefont {{Schubnell}}, \citenamefont {{Seo}}, \citenamefont
  {{Shafieloo}}, \citenamefont {{Sharples}}, \citenamefont {{Silber}},
  \citenamefont {{Slosar}}, \citenamefont {{Smith}}, \citenamefont
  {{Sprayberry}}, \citenamefont {{Tan}}, \citenamefont {{Tarl{\'e}}},
  \citenamefont {{Taylor}}, \citenamefont {{Trusov}}, \citenamefont
  {{Ure{\~n}a-L{\'o}pez}}, \citenamefont {{Vaisakh}}, \citenamefont {{Valcin}},
  \citenamefont {{Valdes}}, \citenamefont {{Vargas-Maga{\~n}a}}, \citenamefont
  {{Verde}}, \citenamefont {{Walther}}, \citenamefont {{Wang}}, \citenamefont
  {{Wang}}, \citenamefont {{Weaver}}, \citenamefont {{Weaverdyck}},
  \citenamefont {{Wechsler}}, \citenamefont {{Weinberg}}, \citenamefont
  {{White}}, \citenamefont {{Yu}}, \citenamefont {{Yu}}, \citenamefont
  {{Yuan}}, \citenamefont {{Y{\`e}che}}, \citenamefont {{Zaborowski}},
  \citenamefont {{Zarrouk}}, \citenamefont {{Zhang}}, \citenamefont {{Zhao}},
  \citenamefont {{Zhao}}, \citenamefont {{Zhou}}, \citenamefont {{Zhuang}},\
  and\ \citenamefont {{Zou}}}]{DESI:2024mwx}%
  \BibitemOpen
  \bibfield  {author} {\bibinfo {author} {\bibnamefont {{DESI Collaboration}}},
  \bibinfo {author} {\bibfnamefont {A.~G.}\ \bibnamefont {{Adame}}}, \bibinfo
  {author} {\bibfnamefont {J.}~\bibnamefont {{Aguilar}}}, \bibinfo {author}
  {\bibfnamefont {S.}~\bibnamefont {{Ahlen}}}, \bibinfo {author} {\bibfnamefont
  {S.}~\bibnamefont {{Alam}}}, \bibinfo {author} {\bibfnamefont {D.~M.}\
  \bibnamefont {{Alexander}}}, \bibinfo {author} {\bibfnamefont
  {M.}~\bibnamefont {{Alvarez}}}, \bibinfo {author} {\bibfnamefont
  {O.}~\bibnamefont {{Alves}}}, \bibinfo {author} {\bibfnamefont
  {A.}~\bibnamefont {{Anand}}}, \bibinfo {author} {\bibfnamefont
  {U.}~\bibnamefont {{Andrade}}}, \bibinfo {author} {\bibfnamefont
  {E.}~\bibnamefont {{Armengaud}}}, \bibinfo {author} {\bibfnamefont
  {S.}~\bibnamefont {{Avila}}}, \bibinfo {author} {\bibfnamefont
  {A.}~\bibnamefont {{Aviles}}}, \bibinfo {author} {\bibfnamefont
  {H.}~\bibnamefont {{Awan}}}, \bibinfo {author} {\bibfnamefont
  {B.}~\bibnamefont {{Bahr-Kalus}}}, \bibinfo {author} {\bibfnamefont
  {S.}~\bibnamefont {{Bailey}}}, \bibinfo {author} {\bibfnamefont
  {C.}~\bibnamefont {{Baltay}}}, \bibinfo {author} {\bibfnamefont
  {A.}~\bibnamefont {{Bault}}}, \bibinfo {author} {\bibfnamefont
  {J.}~\bibnamefont {{Behera}}}, \bibinfo {author} {\bibfnamefont
  {S.}~\bibnamefont {{BenZvi}}}, \bibinfo {author} {\bibfnamefont
  {A.}~\bibnamefont {{Bera}}}, \bibinfo {author} {\bibfnamefont
  {F.}~\bibnamefont {{Beutler}}}, \bibinfo {author} {\bibfnamefont
  {D.}~\bibnamefont {{Bianchi}}}, \bibinfo {author} {\bibfnamefont
  {C.}~\bibnamefont {{Blake}}}, \bibinfo {author} {\bibfnamefont
  {R.}~\bibnamefont {{Blum}}}, \bibinfo {author} {\bibfnamefont
  {S.}~\bibnamefont {{Brieden}}}, \bibinfo {author} {\bibfnamefont
  {A.}~\bibnamefont {{Brodzeller}}}, \bibinfo {author} {\bibfnamefont
  {D.}~\bibnamefont {{Brooks}}}, \bibinfo {author} {\bibfnamefont
  {E.}~\bibnamefont {{Buckley-Geer}}}, \bibinfo {author} {\bibfnamefont
  {E.}~\bibnamefont {{Burtin}}}, \bibinfo {author} {\bibfnamefont
  {R.}~\bibnamefont {{Calderon}}}, \bibinfo {author} {\bibfnamefont
  {R.}~\bibnamefont {{Canning}}}, \bibinfo {author} {\bibfnamefont
  {A.}~\bibnamefont {{Carnero Rosell}}}, \bibinfo {author} {\bibfnamefont
  {R.}~\bibnamefont {{Cereskaite}}}, \bibinfo {author} {\bibfnamefont {J.~L.}\
  \bibnamefont {{Cervantes-Cota}}}, \bibinfo {author} {\bibfnamefont
  {S.}~\bibnamefont {{Chabanier}}}, \bibinfo {author} {\bibfnamefont
  {E.}~\bibnamefont {{Chaussidon}}}, \bibinfo {author} {\bibfnamefont
  {J.}~\bibnamefont {{Chaves-Montero}}}, \bibinfo {author} {\bibfnamefont
  {S.}~\bibnamefont {{Chen}}}, \bibinfo {author} {\bibfnamefont
  {X.}~\bibnamefont {{Chen}}}, \bibinfo {author} {\bibfnamefont
  {T.}~\bibnamefont {{Claybaugh}}}, \bibinfo {author} {\bibfnamefont
  {S.}~\bibnamefont {{Cole}}}, \bibinfo {author} {\bibfnamefont
  {A.}~\bibnamefont {{Cuceu}}}, \bibinfo {author} {\bibfnamefont {T.~M.}\
  \bibnamefont {{Davis}}}, \bibinfo {author} {\bibfnamefont {K.}~\bibnamefont
  {{Dawson}}}, \bibinfo {author} {\bibfnamefont {A.}~\bibnamefont {{de la
  Macorra}}}, \bibinfo {author} {\bibfnamefont {A.}~\bibnamefont {{de
  Mattia}}}, \bibinfo {author} {\bibfnamefont {N.}~\bibnamefont {{Deiosso}}},
  \bibinfo {author} {\bibfnamefont {A.}~\bibnamefont {{Dey}}}, \bibinfo
  {author} {\bibfnamefont {B.}~\bibnamefont {{Dey}}}, \bibinfo {author}
  {\bibfnamefont {Z.}~\bibnamefont {{Ding}}}, \bibinfo {author} {\bibfnamefont
  {P.}~\bibnamefont {{Doel}}}, \bibinfo {author} {\bibfnamefont
  {J.}~\bibnamefont {{Edelstein}}}, \bibinfo {author} {\bibfnamefont
  {S.}~\bibnamefont {{Eftekharzadeh}}}, \bibinfo {author} {\bibfnamefont
  {D.~J.}\ \bibnamefont {{Eisenstein}}}, \bibinfo {author} {\bibfnamefont
  {A.}~\bibnamefont {{Elliott}}}, \bibinfo {author} {\bibfnamefont
  {P.}~\bibnamefont {{Fagrelius}}}, \bibinfo {author} {\bibfnamefont
  {K.}~\bibnamefont {{Fanning}}}, \bibinfo {author} {\bibfnamefont
  {S.}~\bibnamefont {{Ferraro}}}, \bibinfo {author} {\bibfnamefont
  {J.}~\bibnamefont {{Ereza}}}, \bibinfo {author} {\bibfnamefont
  {N.}~\bibnamefont {{Findlay}}}, \bibinfo {author} {\bibfnamefont
  {B.}~\bibnamefont {{Flaugher}}}, \bibinfo {author} {\bibfnamefont
  {A.}~\bibnamefont {{Font-Ribera}}}, \bibinfo {author} {\bibfnamefont
  {D.}~\bibnamefont {{Forero-S{\'a}nchez}}}, \bibinfo {author} {\bibfnamefont
  {J.~E.}\ \bibnamefont {{Forero-Romero}}}, \bibinfo {author} {\bibfnamefont
  {C.~S.}\ \bibnamefont {{Frenk}}}, \bibinfo {author} {\bibfnamefont
  {C.}~\bibnamefont {{Garcia-Quintero}}}, \bibinfo {author} {\bibfnamefont
  {E.}~\bibnamefont {{Gazta{\~n}aga}}}, \bibinfo {author} {\bibfnamefont
  {H.}~\bibnamefont {{Gil-Mar{\'\i}n}}}, \bibinfo {author} {\bibfnamefont
  {S.~G.~A.}\ \bibnamefont {{Gontcho}}}, \bibinfo {author} {\bibfnamefont
  {A.~X.}\ \bibnamefont {{Gonzalez-Morales}}}, \bibinfo {author} {\bibfnamefont
  {V.}~\bibnamefont {{Gonzalez-Perez}}}, \bibinfo {author} {\bibfnamefont
  {C.}~\bibnamefont {{Gordon}}}, \bibinfo {author} {\bibfnamefont
  {D.}~\bibnamefont {{Green}}}, \bibinfo {author} {\bibfnamefont
  {D.}~\bibnamefont {{Gruen}}}, \bibinfo {author} {\bibfnamefont
  {R.}~\bibnamefont {{Gsponer}}}, \bibinfo {author} {\bibfnamefont
  {G.}~\bibnamefont {{Gutierrez}}}, \bibinfo {author} {\bibfnamefont
  {J.}~\bibnamefont {{Guy}}}, \bibinfo {author} {\bibfnamefont
  {B.}~\bibnamefont {{Hadzhiyska}}}, \bibinfo {author} {\bibfnamefont
  {C.}~\bibnamefont {{Hahn}}}, \bibinfo {author} {\bibfnamefont {M.~M.~S.}\
  \bibnamefont {{Hanif}}}, \bibinfo {author} {\bibfnamefont {H.~K.}\
  \bibnamefont {{Herrera-Alcantar}}}, \bibinfo {author} {\bibfnamefont
  {K.}~\bibnamefont {{Honscheid}}}, \bibinfo {author} {\bibfnamefont
  {C.}~\bibnamefont {{Howlett}}}, \bibinfo {author} {\bibfnamefont
  {D.}~\bibnamefont {{Huterer}}}, \bibinfo {author} {\bibfnamefont
  {V.}~\bibnamefont {{Ir{\v{s}}i{\v{c}}}}}, \bibinfo {author} {\bibfnamefont
  {M.}~\bibnamefont {{Ishak}}}, \bibinfo {author} {\bibfnamefont
  {S.}~\bibnamefont {{Juneau}}}, \bibinfo {author} {\bibfnamefont {N.~G.}\
  \bibnamefont {{Kara{\c{c}}ayl{\i}}}}, \bibinfo {author} {\bibfnamefont
  {R.}~\bibnamefont {{Kehoe}}}, \bibinfo {author} {\bibfnamefont
  {S.}~\bibnamefont {{Kent}}}, \bibinfo {author} {\bibfnamefont
  {D.}~\bibnamefont {{Kirkby}}}, \bibinfo {author} {\bibfnamefont
  {A.}~\bibnamefont {{Kremin}}}, \bibinfo {author} {\bibfnamefont
  {A.}~\bibnamefont {{Krolewski}}}, \bibinfo {author} {\bibfnamefont
  {Y.}~\bibnamefont {{Lai}}}, \bibinfo {author} {\bibfnamefont {T.~W.}\
  \bibnamefont {{Lan}}}, \bibinfo {author} {\bibfnamefont {M.}~\bibnamefont
  {{Landriau}}}, \bibinfo {author} {\bibfnamefont {D.}~\bibnamefont {{Lang}}},
  \bibinfo {author} {\bibfnamefont {J.}~\bibnamefont {{Lasker}}}, \bibinfo
  {author} {\bibfnamefont {J.~M.}\ \bibnamefont {{Le Goff}}}, \bibinfo {author}
  {\bibfnamefont {L.}~\bibnamefont {{Le Guillou}}}, \bibinfo {author}
  {\bibfnamefont {A.}~\bibnamefont {{Leauthaud}}}, \bibinfo {author}
  {\bibfnamefont {M.~E.}\ \bibnamefont {{Levi}}}, \bibinfo {author}
  {\bibfnamefont {T.~S.}\ \bibnamefont {{Li}}}, \bibinfo {author}
  {\bibfnamefont {E.}~\bibnamefont {{Linder}}}, \bibinfo {author}
  {\bibfnamefont {K.}~\bibnamefont {{Lodha}}}, \bibinfo {author} {\bibfnamefont
  {C.}~\bibnamefont {{Magneville}}}, \bibinfo {author} {\bibfnamefont
  {M.}~\bibnamefont {{Manera}}}, \bibinfo {author} {\bibfnamefont
  {D.}~\bibnamefont {{Margala}}}, \bibinfo {author} {\bibfnamefont
  {P.}~\bibnamefont {{Martini}}}, \bibinfo {author} {\bibfnamefont
  {M.}~\bibnamefont {{Maus}}}, \bibinfo {author} {\bibfnamefont
  {P.}~\bibnamefont {{McDonald}}}, \bibinfo {author} {\bibfnamefont
  {L.}~\bibnamefont {{Medina-Varela}}}, \bibinfo {author} {\bibfnamefont
  {A.}~\bibnamefont {{Meisner}}}, \bibinfo {author} {\bibfnamefont
  {J.}~\bibnamefont {{Mena-Fern{\'a}ndez}}}, \bibinfo {author} {\bibfnamefont
  {R.}~\bibnamefont {{Miquel}}}, \bibinfo {author} {\bibfnamefont
  {J.}~\bibnamefont {{Moon}}}, \bibinfo {author} {\bibfnamefont
  {S.}~\bibnamefont {{Moore}}}, \bibinfo {author} {\bibfnamefont
  {J.}~\bibnamefont {{Moustakas}}}, \bibinfo {author} {\bibfnamefont
  {N.}~\bibnamefont {{Mudur}}}, \bibinfo {author} {\bibfnamefont
  {E.}~\bibnamefont {{Mueller}}}, \bibinfo {author} {\bibfnamefont
  {A.}~\bibnamefont {{Mu{\~n}oz-Guti{\'e}rrez}}}, \bibinfo {author}
  {\bibfnamefont {A.~D.}\ \bibnamefont {{Myers}}}, \bibinfo {author}
  {\bibfnamefont {S.}~\bibnamefont {{Nadathur}}}, \bibinfo {author}
  {\bibfnamefont {L.}~\bibnamefont {{Napolitano}}}, \bibinfo {author}
  {\bibfnamefont {R.}~\bibnamefont {{Neveux}}}, \bibinfo {author}
  {\bibfnamefont {J.~A.}\ \bibnamefont {{Newman}}}, \bibinfo {author}
  {\bibfnamefont {N.~M.}\ \bibnamefont {{Nguyen}}}, \bibinfo {author}
  {\bibfnamefont {J.}~\bibnamefont {{Nie}}}, \bibinfo {author} {\bibfnamefont
  {G.}~\bibnamefont {{Niz}}}, \bibinfo {author} {\bibfnamefont {H.~E.}\
  \bibnamefont {{Noriega}}}, \bibinfo {author} {\bibfnamefont {N.}~\bibnamefont
  {{Padmanabhan}}}, \bibinfo {author} {\bibfnamefont {E.}~\bibnamefont
  {{Paillas}}}, \bibinfo {author} {\bibfnamefont {N.}~\bibnamefont
  {{Palanque-Delabrouille}}}, \bibinfo {author} {\bibfnamefont
  {J.}~\bibnamefont {{Pan}}}, \bibinfo {author} {\bibfnamefont
  {S.}~\bibnamefont {{Penmetsa}}}, \bibinfo {author} {\bibfnamefont {W.~J.}\
  \bibnamefont {{Percival}}}, \bibinfo {author} {\bibfnamefont {M.~M.}\
  \bibnamefont {{Pieri}}}, \bibinfo {author} {\bibfnamefont {M.}~\bibnamefont
  {{Pinon}}}, \bibinfo {author} {\bibfnamefont {C.}~\bibnamefont {{Poppett}}},
  \bibinfo {author} {\bibfnamefont {A.}~\bibnamefont {{Porredon}}}, \bibinfo
  {author} {\bibfnamefont {F.}~\bibnamefont {{Prada}}}, \bibinfo {author}
  {\bibfnamefont {A.}~\bibnamefont {{P{\'e}rez-Fern{\'a}ndez}}}, \bibinfo
  {author} {\bibfnamefont {I.}~\bibnamefont {{P{\'e}rez-R{\`a}fols}}}, \bibinfo
  {author} {\bibfnamefont {D.}~\bibnamefont {{Rabinowitz}}}, \bibinfo {author}
  {\bibfnamefont {A.}~\bibnamefont {{Raichoor}}}, \bibinfo {author}
  {\bibfnamefont {C.}~\bibnamefont {{Ram{\'\i}rez-P{\'e}rez}}}, \bibinfo
  {author} {\bibfnamefont {S.}~\bibnamefont {{Ramirez-Solano}}}, \bibinfo
  {author} {\bibfnamefont {C.}~\bibnamefont {{Ravoux}}}, \bibinfo {author}
  {\bibfnamefont {M.}~\bibnamefont {{Rashkovetskyi}}}, \bibinfo {author}
  {\bibfnamefont {M.}~\bibnamefont {{Rezaie}}}, \bibinfo {author}
  {\bibfnamefont {J.}~\bibnamefont {{Rich}}}, \bibinfo {author} {\bibfnamefont
  {A.}~\bibnamefont {{Rocher}}}, \bibinfo {author} {\bibfnamefont
  {C.}~\bibnamefont {{Rockosi}}}, \bibinfo {author} {\bibfnamefont {N.~A.}\
  \bibnamefont {{Roe}}}, \bibinfo {author} {\bibfnamefont {A.}~\bibnamefont
  {{Rosado-Marin}}}, \bibinfo {author} {\bibfnamefont {A.~J.}\ \bibnamefont
  {{Ross}}}, \bibinfo {author} {\bibfnamefont {G.}~\bibnamefont {{Rossi}}},
  \bibinfo {author} {\bibfnamefont {R.}~\bibnamefont {{Ruggeri}}}, \bibinfo
  {author} {\bibfnamefont {V.}~\bibnamefont {{Ruhlmann-Kleider}}}, \bibinfo
  {author} {\bibfnamefont {L.}~\bibnamefont {{Samushia}}}, \bibinfo {author}
  {\bibfnamefont {E.}~\bibnamefont {{Sanchez}}}, \bibinfo {author}
  {\bibfnamefont {C.}~\bibnamefont {{Saulder}}}, \bibinfo {author}
  {\bibfnamefont {E.~F.}\ \bibnamefont {{Schlafly}}}, \bibinfo {author}
  {\bibfnamefont {D.}~\bibnamefont {{Schlegel}}}, \bibinfo {author}
  {\bibfnamefont {M.}~\bibnamefont {{Schubnell}}}, \bibinfo {author}
  {\bibfnamefont {H.}~\bibnamefont {{Seo}}}, \bibinfo {author} {\bibfnamefont
  {A.}~\bibnamefont {{Shafieloo}}}, \bibinfo {author} {\bibfnamefont
  {R.}~\bibnamefont {{Sharples}}}, \bibinfo {author} {\bibfnamefont
  {J.}~\bibnamefont {{Silber}}}, \bibinfo {author} {\bibfnamefont
  {A.}~\bibnamefont {{Slosar}}}, \bibinfo {author} {\bibfnamefont
  {A.}~\bibnamefont {{Smith}}}, \bibinfo {author} {\bibfnamefont
  {D.}~\bibnamefont {{Sprayberry}}}, \bibinfo {author} {\bibfnamefont
  {T.}~\bibnamefont {{Tan}}}, \bibinfo {author} {\bibfnamefont
  {G.}~\bibnamefont {{Tarl{\'e}}}}, \bibinfo {author} {\bibfnamefont
  {P.}~\bibnamefont {{Taylor}}}, \bibinfo {author} {\bibfnamefont
  {S.}~\bibnamefont {{Trusov}}}, \bibinfo {author} {\bibfnamefont {L.~A.}\
  \bibnamefont {{Ure{\~n}a-L{\'o}pez}}}, \bibinfo {author} {\bibfnamefont
  {R.}~\bibnamefont {{Vaisakh}}}, \bibinfo {author} {\bibfnamefont
  {D.}~\bibnamefont {{Valcin}}}, \bibinfo {author} {\bibfnamefont
  {F.}~\bibnamefont {{Valdes}}}, \bibinfo {author} {\bibfnamefont
  {M.}~\bibnamefont {{Vargas-Maga{\~n}a}}}, \bibinfo {author} {\bibfnamefont
  {L.}~\bibnamefont {{Verde}}}, \bibinfo {author} {\bibfnamefont
  {M.}~\bibnamefont {{Walther}}}, \bibinfo {author} {\bibfnamefont
  {B.}~\bibnamefont {{Wang}}}, \bibinfo {author} {\bibfnamefont {M.~S.}\
  \bibnamefont {{Wang}}}, \bibinfo {author} {\bibfnamefont {B.~A.}\
  \bibnamefont {{Weaver}}}, \bibinfo {author} {\bibfnamefont {N.}~\bibnamefont
  {{Weaverdyck}}}, \bibinfo {author} {\bibfnamefont {R.~H.}\ \bibnamefont
  {{Wechsler}}}, \bibinfo {author} {\bibfnamefont {D.~H.}\ \bibnamefont
  {{Weinberg}}}, \bibinfo {author} {\bibfnamefont {M.}~\bibnamefont {{White}}},
  \bibinfo {author} {\bibfnamefont {J.}~\bibnamefont {{Yu}}}, \bibinfo {author}
  {\bibfnamefont {Y.}~\bibnamefont {{Yu}}}, \bibinfo {author} {\bibfnamefont
  {S.}~\bibnamefont {{Yuan}}}, \bibinfo {author} {\bibfnamefont
  {C.}~\bibnamefont {{Y{\`e}che}}}, \bibinfo {author} {\bibfnamefont {E.~A.}\
  \bibnamefont {{Zaborowski}}}, \bibinfo {author} {\bibfnamefont
  {P.}~\bibnamefont {{Zarrouk}}}, \bibinfo {author} {\bibfnamefont
  {H.}~\bibnamefont {{Zhang}}}, \bibinfo {author} {\bibfnamefont
  {C.}~\bibnamefont {{Zhao}}}, \bibinfo {author} {\bibfnamefont
  {R.}~\bibnamefont {{Zhao}}}, \bibinfo {author} {\bibfnamefont
  {R.}~\bibnamefont {{Zhou}}}, \bibinfo {author} {\bibfnamefont
  {T.}~\bibnamefont {{Zhuang}}}, \ and\ \bibinfo {author} {\bibfnamefont
  {H.}~\bibnamefont {{Zou}}},\ }\href {\doibase 10.48550/arXiv.2404.03002}
  {\bibfield  {journal} {\bibinfo  {journal} {arXiv e-prints}\ ,\ \bibinfo
  {eid} {arXiv:2404.03002}} (\bibinfo {year} {2024})},\ \Eprint
  {http://arxiv.org/abs/2404.03002} {arXiv:2404.03002 [astro-ph.CO]}
  \BibitemShut {NoStop}%
\bibitem [{\citenamefont {Adame}\ \emph
  {et~al.}(2024{\natexlab{a}})\citenamefont {Adame} \emph
  {et~al.}}]{DESI:2024hhd}%
  \BibitemOpen
  \bibfield  {author} {\bibinfo {author} {\bibfnamefont {A.~G.}\ \bibnamefont
  {Adame}} \emph {et~al.} (\bibinfo {collaboration} {DESI}),\ }\href@noop {} {\
   (\bibinfo {year} {2024}{\natexlab{a}})},\ \Eprint
  {http://arxiv.org/abs/2411.12022} {arXiv:2411.12022 [astro-ph.CO]}
  \BibitemShut {NoStop}%
\bibitem [{\citenamefont {Adame}\ \emph
  {et~al.}(2024{\natexlab{b}})\citenamefont {Adame} \emph
  {et~al.}}]{DESI:2024jis}%
  \BibitemOpen
  \bibfield  {author} {\bibinfo {author} {\bibfnamefont {A.~G.}\ \bibnamefont
  {Adame}} \emph {et~al.} (\bibinfo {collaboration} {DESI}),\ }\href@noop {} {\
   (\bibinfo {year} {2024}{\natexlab{b}})},\ \Eprint
  {http://arxiv.org/abs/2411.12021} {arXiv:2411.12021 [astro-ph.CO]}
  \BibitemShut {NoStop}%
\bibitem [{\citenamefont {{\'O Colg\'ain}}\ \emph {et~al.}(2024)\citenamefont
  {{\'O Colg\'ain}}, \citenamefont {Dainotti}, \citenamefont {Capozziello},
  \citenamefont {Pourojaghi}, \citenamefont {Sheikh-Jabbari},\ and\
  \citenamefont {Stojkovic}}]{Colgain:2024xqj}%
  \BibitemOpen
  \bibfield  {author} {\bibinfo {author} {\bibfnamefont {E.}~\bibnamefont {{\'O
  Colg\'ain}}}, \bibinfo {author} {\bibfnamefont {M.~G.}\ \bibnamefont
  {Dainotti}}, \bibinfo {author} {\bibfnamefont {S.}~\bibnamefont
  {Capozziello}}, \bibinfo {author} {\bibfnamefont {S.}~\bibnamefont
  {Pourojaghi}}, \bibinfo {author} {\bibfnamefont {M.~M.}\ \bibnamefont
  {Sheikh-Jabbari}}, \ and\ \bibinfo {author} {\bibfnamefont {D.}~\bibnamefont
  {Stojkovic}},\ }\href@noop {} {\  (\bibinfo {year} {2024})},\ \Eprint
  {http://arxiv.org/abs/2404.08633} {arXiv:2404.08633 [astro-ph.CO]}
  \BibitemShut {NoStop}%
\bibitem [{\citenamefont {Wu}\ and\ \citenamefont
  {Huterer}(2017)}]{Wu:2017fpr}%
  \BibitemOpen
  \bibfield  {author} {\bibinfo {author} {\bibfnamefont {H.-Y.}\ \bibnamefont
  {Wu}}\ and\ \bibinfo {author} {\bibfnamefont {D.}~\bibnamefont {Huterer}},\
  }\href {\doibase 10.1093/mnras/stx1967} {\bibfield  {journal} {\bibinfo
  {journal} {Mon. Not. Roy. Astron. Soc.}\ }\textbf {\bibinfo {volume} {471}},\
  \bibinfo {pages} {4946} (\bibinfo {year} {2017})},\ \Eprint
  {http://arxiv.org/abs/1706.09723} {arXiv:1706.09723 [astro-ph.CO]}
  \BibitemShut {NoStop}%
\bibitem [{\citenamefont {Knox}\ and\ \citenamefont
  {Millea}(2020)}]{Knox:2019rjx}%
  \BibitemOpen
  \bibfield  {author} {\bibinfo {author} {\bibfnamefont {L.}~\bibnamefont
  {Knox}}\ and\ \bibinfo {author} {\bibfnamefont {M.}~\bibnamefont {Millea}},\
  }\href {\doibase 10.1103/PhysRevD.101.043533} {\bibfield  {journal} {\bibinfo
   {journal} {Phys. Rev. D}\ }\textbf {\bibinfo {volume} {101}},\ \bibinfo
  {pages} {043533} (\bibinfo {year} {2020})},\ \Eprint
  {http://arxiv.org/abs/1908.03663} {arXiv:1908.03663 [astro-ph.CO]}
  \BibitemShut {NoStop}%
\bibitem [{\citenamefont {Amon}\ and\ \citenamefont
  {Efstathiou}(2022)}]{Amon:2022azi}%
  \BibitemOpen
  \bibfield  {author} {\bibinfo {author} {\bibfnamefont {A.}~\bibnamefont
  {Amon}}\ and\ \bibinfo {author} {\bibfnamefont {G.}~\bibnamefont
  {Efstathiou}},\ }\href {\doibase 10.1093/mnras/stac2429} {\bibfield
  {journal} {\bibinfo  {journal} {Mon. Not. Roy. Astron. Soc.}\ }\textbf
  {\bibinfo {volume} {516}},\ \bibinfo {pages} {5355} (\bibinfo {year}
  {2022})},\ \Eprint {http://arxiv.org/abs/2206.11794} {arXiv:2206.11794
  [astro-ph.CO]} \BibitemShut {NoStop}%
\bibitem [{\citenamefont {Preston}\ \emph {et~al.}(2023)\citenamefont
  {Preston}, \citenamefont {Amon},\ and\ \citenamefont
  {Efstathiou}}]{Preston:2023uup}%
  \BibitemOpen
  \bibfield  {author} {\bibinfo {author} {\bibfnamefont {C.}~\bibnamefont
  {Preston}}, \bibinfo {author} {\bibfnamefont {A.}~\bibnamefont {Amon}}, \
  and\ \bibinfo {author} {\bibfnamefont {G.}~\bibnamefont {Efstathiou}},\
  }\href {\doibase 10.1093/mnras/stad2573} {\bibfield  {journal} {\bibinfo
  {journal} {Mon. Not. Roy. Astron. Soc.}\ }\textbf {\bibinfo {volume} {525}},\
  \bibinfo {pages} {5554} (\bibinfo {year} {2023})},\ \Eprint
  {http://arxiv.org/abs/2305.09827} {arXiv:2305.09827 [astro-ph.CO]}
  \BibitemShut {NoStop}%
\bibitem [{\citenamefont {Weinberg}(1989)}]{Weinberg:1988cp}%
  \BibitemOpen
  \bibfield  {author} {\bibinfo {author} {\bibfnamefont {S.}~\bibnamefont
  {Weinberg}},\ }\href {\doibase 10.1103/RevModPhys.61.1} {\bibfield  {journal}
  {\bibinfo  {journal} {Rev. Mod. Phys.}\ }\textbf {\bibinfo {volume} {61}},\
  \bibinfo {pages} {1} (\bibinfo {year} {1989})}\BibitemShut {NoStop}%
\bibitem [{\citenamefont {Weinberg}(2000)}]{Weinberg:2000yb}%
  \BibitemOpen
  \bibfield  {author} {\bibinfo {author} {\bibfnamefont {S.}~\bibnamefont
  {Weinberg}},\ }in\ \href@noop {} {\emph {\bibinfo {booktitle} {{4th
  International Symposium on Sources and Detection of Dark Matter in the
  Universe (DM 2000)}}}}\ (\bibinfo {year} {2000})\ pp.\ \bibinfo {pages}
  {18--26},\ \Eprint {http://arxiv.org/abs/astro-ph/0005265}
  {arXiv:astro-ph/0005265} \BibitemShut {NoStop}%
\bibitem [{\citenamefont {Peebles}\ and\ \citenamefont
  {Ratra}(2003)}]{Peebles:2002gy}%
  \BibitemOpen
  \bibfield  {author} {\bibinfo {author} {\bibfnamefont {P.~J.~E.}\
  \bibnamefont {Peebles}}\ and\ \bibinfo {author} {\bibfnamefont
  {B.}~\bibnamefont {Ratra}},\ }\href {\doibase 10.1103/RevModPhys.75.559}
  {\bibfield  {journal} {\bibinfo  {journal} {Rev. Mod. Phys.}\ }\textbf
  {\bibinfo {volume} {75}},\ \bibinfo {pages} {559} (\bibinfo {year} {2003})},\
  \Eprint {http://arxiv.org/abs/astro-ph/0207347} {arXiv:astro-ph/0207347}
  \BibitemShut {NoStop}%
\bibitem [{\citenamefont {Dvali}\ and\ \citenamefont
  {Gomez}(2016)}]{Dvali:2014gua}%
  \BibitemOpen
  \bibfield  {author} {\bibinfo {author} {\bibfnamefont {G.}~\bibnamefont
  {Dvali}}\ and\ \bibinfo {author} {\bibfnamefont {C.}~\bibnamefont {Gomez}},\
  }\href {\doibase 10.1002/andp.201500216} {\bibfield  {journal} {\bibinfo
  {journal} {Annalen Phys.}\ }\textbf {\bibinfo {volume} {528}},\ \bibinfo
  {pages} {68} (\bibinfo {year} {2016})},\ \Eprint
  {http://arxiv.org/abs/1412.8077} {arXiv:1412.8077 [hep-th]} \BibitemShut
  {NoStop}%
\bibitem [{\citenamefont {Dvali}\ and\ \citenamefont
  {Gomez}(2019)}]{Dvali:2018fqu}%
  \BibitemOpen
  \bibfield  {author} {\bibinfo {author} {\bibfnamefont {G.}~\bibnamefont
  {Dvali}}\ and\ \bibinfo {author} {\bibfnamefont {C.}~\bibnamefont {Gomez}},\
  }\href {\doibase 10.1002/prop.201800092} {\bibfield  {journal} {\bibinfo
  {journal} {Fortsch. Phys.}\ }\textbf {\bibinfo {volume} {67}},\ \bibinfo
  {pages} {1800092} (\bibinfo {year} {2019})},\ \Eprint
  {http://arxiv.org/abs/1806.10877} {arXiv:1806.10877 [hep-th]} \BibitemShut
  {NoStop}%
\bibitem [{\citenamefont {{Obied}}\ \emph {et~al.}(2018)\citenamefont
  {{Obied}}, \citenamefont {{Ooguri}}, \citenamefont {{Spodyneiko}},\ and\
  \citenamefont {{Vafa}}}]{Obied:2018sgi}%
  \BibitemOpen
  \bibfield  {author} {\bibinfo {author} {\bibfnamefont {G.}~\bibnamefont
  {{Obied}}}, \bibinfo {author} {\bibfnamefont {H.}~\bibnamefont {{Ooguri}}},
  \bibinfo {author} {\bibfnamefont {L.}~\bibnamefont {{Spodyneiko}}}, \ and\
  \bibinfo {author} {\bibfnamefont {C.}~\bibnamefont {{Vafa}}},\ }\href
  {\doibase 10.48550/arXiv.1806.08362} {\bibfield  {journal} {\bibinfo
  {journal} {arXiv e-prints}\ ,\ \bibinfo {eid} {arXiv:1806.08362}} (\bibinfo
  {year} {2018})},\ \Eprint {http://arxiv.org/abs/1806.08362} {arXiv:1806.08362
  [hep-th]} \BibitemShut {NoStop}%
\end{thebibliography}%

\end{document}